\documentclass[showpacs,pra,superscriptaddress,twocolumn]{revtex4-1}
\usepackage{pgf}
\usepackage[utf8]{inputenc}
\usepackage{amsmath}
\usepackage{braket}
\usepackage{amssymb}
\usepackage{amsmath}
\usepackage{enumitem} 
\usepackage{appendix}
\usepackage{graphicx}
\usepackage{dsfont}
\usepackage{tikz}
\usepackage{epstopdf}
\usepackage[pdftex,bookmarks,colorlinks]{hyperref}
\usepackage{hyperref}
\usepackage{booktabs}

\usepackage{ulem}

\newcommand{\opa}[1]{{\hat{a}^{\phantom \dagger}}_{#1}}
\newcommand{\opadag}[1]{{\hat{a}^{\dagger}}_{#1}}
\newcommand{\Tr}[1]{\operatorname{Tr} #1}
\newcommand{\mean}[1]{\left\langle #1\right\rangle}
\newcommand{\nep}{\textrm{e}}
\newcommand{\nb}{{\boldsymbol{n}}}

\newcommand{\ud}{\mathrm{d}}

\newcommand{\opbdag}[1]{{\hat{b}^{\dagger}}_{#1}}

\newcommand{\opb}[1]{{\hat{b}^{\phantom \dagger}}_{#1}}

\begin{document}
\title{Non-ergodic behaviour of {the clean Bose-Hubbard chain}}
\author{Angelo Russomanno}
\affiliation{Max-Planck-Institut f\"ur Physik Komplexer Systeme, N\"othnitzer Stra{\ss}e 38, D-01187, Dresden, Germany}
\author{Michele Fava}
\affiliation{Rudolf Peierls Centre for Theoretical Physics, Clarendon Laboratory, University of Oxford, Oxford OX1 3PU, UK}

\author{Rosario Fazio}
\affiliation{Abdus Salam ICTP, Strada Costiera 11, I-34151 Trieste, Italy}
\affiliation{Dipartimento di Fisica, Universit\`a di Napoli ``Federico II'', Monte S. Angelo, I-80126 Napoli, Italy}
\thanks{On leave}

\begin{abstract}
We study ergodicity breaking in the clean Bose-Hubbard chain for small hopping strength. We see the existence of a non-ergodic regime
by means of indicators as the half-chain entanglement entropy of the eigenstates, the average level spacing ratio, {the properties of the eigenstate-expectation distribution of the correlation and the scaling of the Inverse Participation Ratio averages.} We find that this ergodicity
breaking {is different from many-body localization} because the average half-chain entanglement entropy of the eigenstates obeys volume law.
This ergodicity breaking appears unrelated to the spectrum being organized in quasidegenerate multiplets at small hopping and finite system sizes, so in principle it can survive
also for larger system sizes. We find that some imbalance oscillations in time which could mark the existence of a glassy behaviour in space are well described by
the dynamics of a single symmetry-breaking doublet and {quantitatively} captured by a perturbative effective XXZ model. We show that the amplitude of these oscillations vanishes in
	the large-size limit. {Our findings are numerically obtained for systems with $L < 12$. Extrapolations of our scalings to larger system sizes should be taken with care, as discussed in the paper.}
\end{abstract}

\maketitle
\section{Introduction}

{
Thermalization in classical Hamiltonian systems is well understood in terms of chaotic dynamics and the related essentially ergodic exploration of the phase space~\cite{lichtenberg1983regular,Vulpiani,Berry_regirr78:proceeding}. From the quantum point of view the
physical mechanism is quite different, involving the eigenstates of the Hamiltonian being fully random strongly entangled states which appear
thermal from the point of view of local measurements. This is the paradigm of eigenstate thermalization (ETH), introduced in~\cite{Deutsch_PRA91,Sred_PRE94,Rigol_Nat,Prosen_PRE99}.

In many cases there is correspondence between classical and quantum thermalization. {This can be seen from a variety of different arguments, including Berry's random-wave conjecture for the energy eigenfunctions~\cite{Berry1_1977,PhysRevLett.51.943,Sred_PRE94,PhysRevA.34.591,Prosen_AJ}, the analogy between quantized chaotic systems and random matrix theory~\cite{PhysRevLett.52.1}, and the semiclassical periodic orbit expansion, assuming a certain randomness for the periodic orbits~\cite{PhysRevLett.75.2300}. In summary,} when a classically chaotic Hamiltonian is quantized, it gives rise
to a Hamiltonian matrix which is a random matrix and its eigenstates have exactly the required properties for thermalization of local observables. This correspondence is nevertheless highly
non trivial, because quantum effects can give rise to ergodicity-breaking phenomena with no analog in the classical domain {(we discuss some examples in the next paragraphs)}.
From a classical point of view, any non-linear Hamiltonian system with more than two degrees of freedom and no conservation law beyond energy gives
rise to chaos and {essentially ergodic dynamics}~\cite{lichtenberg1983regular}. From the quantum point of view the situation is different.

A striking example is dynamical localization (initially discovered for one degree of freedom~\cite{Boris:rotor} and later generalized to the many body case~\cite{QRKR,rylands,Michele_arxiv,Notarnicola_PRB18}), 
where a classical ergodic driven system shows a {regular-like behaviour} in the quantum regime, leading to a suppression
of energy absorption. Another example is many body localization. In contrast with the classical case, a generic disordered and interacting many-body quantum 
system does not thermalize and spontaneously generates the space-localized integrals of motion which forbid ergodic behaviour (see an overview of the subject
in the reviews~\cite{Bloch_2019,nandkishore2015many,imbrie2017review}).

It is therefore of the utmost interest to find new cases of complex many body systems which break ergodicity, in order to understand if there are general
properties in this lack of correspondence with the classical case. One very interesting advancement in this field has been the proposal of
many-body localized systems without disorder. Starting from the early proposals on this subject~\cite{Carleo,grover2014,schiulaz2015} there 
has been a constant interest in understanding if disorder is a necessary ingredient to achieve localization in an interacting system.  
Most of this activity concerns proposals based on gauge theories which are globally uniform and generate a different realization of an effective disorder 
in each superselection sector~\cite{Adam_prl17,Adam_prl171,PhysRevLett.120.030601,Adam_prl19,Adam_prb,PhysRevResearch.2.012003,
karpov2020disorderfree}. A uniform Josephson junction chain was studied in~\cite{Pino536} where it was shown that at high energies or small 
Josephson coupling the system breaks ergodicity and shows a many-body localized phase with {zero} conductivity. 
The systems considered in these works are not integrable, thus pointing out that non-ergodic behaviour can be found also in non-integrable 
clean models. {Very important in this context are the results of~\cite{Prosen_PRL98,Prosen_PRE99}, where a clean kicked spin-$1/2$ XXZ chain shows in the large-size limit a transition from a non-ergodic non-localized phase to an ergodic fully quantum chaotic phase.} Our interest is to further explore these questions and seek for non-ergodic behaviour of a clean quantum many-body system. 

Aim of this work is to analyze the ergodicity properties of the clean Bose-Hubbard model~\cite{fisher}. The equilibrium phase diagram and 
the dynamical properties of the model were extensively scrutinized in the last three decades also for its importance in the physics of optical 
lattices~\cite{reviewBloch}. Many-body localization in the presence of disorder was studied in~\cite{Delande_APP17,zak1,
zak2,Lukin256,Glen,hopjan2019manybody,yao2020manybody}. The non-ergodic behaviour of the clean version of this model was already discovered and studied in the two papers~\cite{kollath,kollath1}.
Here we confirm and extend the results of these papers. We inquire first of all the relation of the non-ergodicity with the spectral structure of the system.
Second, we study if this ergodicity breaking can be interpreted in a many-body localization paradigm.
Finally, another goal will be to understand how the glassy behaviour discovered for this model in~\cite{Carleo} is related to non ergodicity. 
%

The Bose-Hubbard model, describing a system of interacting bosons hopping on a $d$-dimensional lattice, is characterized by two energy scales, 
the hopping amplitude $J$ and the on-site repulsion $U$. Here we will consider $d=1$ with $J$ and $U$ playing their role as given in the Hamiltonian Eq.~\eqref{Hamour:eqn} (notice the factor $1/2$ in the hopping term). {For $J/U\gtrsim 0.5$ (in our notation) and $L=10$ sites this model is known to thermalize: the long-time dynamics is well described by the thermal canonical ensemble~\cite{PhysRevA.90.033606,note_cruz}. For smaller $J$ there are deviations from the canonical ensemble, and it is not known if this is an effect of the vicinity of the integrable point at $J=0$ on the dynamics of the finite system, or is an evidence for a non-ergodic regime. Here we make progress in this direction and extend this analysis. We
consider several different indicators and make finite-size scalings in order to examine the 
(non-)ergodicity as a function of the ratio $J/U$.} 

First of all we consider the half-chain entanglement entropies of the eigenstates of the system. We find that their average always obeys a volume law. For large $J$ this volume law is the 
same obeyed by a fully random state (Page value): In this regime the system obeys eigenstate thermalization~\cite{Deutsch_PRA91,
Sred_PRE94,Rigol_Nat,ikeda} and is fully quantum chaotic and thermalizing. For small $J$ ergodicity is broken and the pre-factor  of the 
volume law is significantly smaller than the Page value. For small values of $J$, this model breaks ergodicity in a way remarkably different from many-body localization, where 
the eigenstates show instead an area-law behaviour and the averaged entanglement entropy is constant with the system size.

{Then we move to} the analysis of the level spacing distribution. Keeping the focus on the spectral properties, we discover that
this non-ergodic behaviour appears unrelated to the spectrum being organized in multiplets at small $J$ and small sizes. 
{In this circumstances, in fact, the spectrum can be understood through perturbation theory in $J$. At $J=0$ the spectrum displays large degenerate subspaces. At a finite $L$, as a small $J$ is switched on, these multiplets acquire a bandwidth of order $LJ$, remaining well separated between each other by energies of order $U$. However, we expect the multiplet structure to ultimately disappear for any $J$, when $L\gtrsim U/J$. {In some cases this coincides with the transition to an ergodic behaviour~\cite{santos2010localization} but in our model we remarkably} find signatures of non-ergodic behavior even in cases where the multiplet structure breaks down. This is a crucial result as it disentangles the multiplet structure from non-ergodicity, hinting that non-ergodicity might survive to large system sizes (while the multiplet structure will eventually disappear).}


Some words of caution are necessary at this stage. We see a non-ergodic regime at small hoppings with properties that are clearly 
different from a many-body localized phase. In this way we confirm and extend the results of~\cite{kollath,kollath1}. {Moreover, a many-body system showing in the thermodynamic limit a non-ergodic non-localized phase and an ergodic phase has been reported in literature~\cite{Prosen_PRL98,Prosen_PRE99}, and the evidences we provide point towards this paradigm.} However, given 
the system sizes that we are able to reach, we should keep in mind the possibility that this non-ergodic regime is a finite-size effect and we do not 
know if it can be extrapolated towards the thermodynamic limit. {Some hint in this direction is provided by the behaviour of the entanglement-entropy averages restricted to the high-entropy states (see Sec.~\ref{sec:entro}), but the system sizes are too small for a definitive statement. }The same limitation does not allow to make a clear-cut observation of the value of the 
transition point to the ergodic phase. 

{In the physics of this ergodicity breaking a crucial role is played by the ``rare'' states introduced in~\cite{PhysRevLett.105.250401}. These states do not obey ETH and at finite system size they are frequent enough to forbid thermalization at small $J$. We call them non-thermal states and in this context, in analogy with regular trajectories in classical Hamiltonian systems~\cite{lichtenberg1983regular}, we talk about regular-like behaviour. The question is if at larger system sizes these states become rare enough to allow thermalization or give rise to an extended non-thermal phase, as it occurs for instance in the Rosenzweig{\textendash}Porter model~\cite{pinotto}. For now we have not enough evidences to clearly decide, that's why we talk about a non-ergodic regime (and not a phase). {Nevertheless, results coming from the distribution of the eigenstate expectations of the correlation and from the scaling of the averages of the Inverse Participation Ratio point towards an extended non-ergodic phase, as we show in Sec.~\ref{other}. Independently of the thermal nature, a very important finding of ours} is that the eigenstates are spatially extended in contrast to a somewhat related model where they are many-body localized~\cite{Pino536}.}

{The absence of {real-space} localization can be seen also in the behaviour of the imbalance. This analysis is connected to}
 the relation of non ergodicity with the glassy behaviour found in~\cite{Carleo}. We prepare 
the system in a number-imbalanced state and we see that, for $J$ small enough, the imbalance shows oscillations with a frequency decreasing with the system size, confirming 
to longer times the behaviour observed in~\cite{Carleo}. We find that the frequency decreases exponentially with the system size. Using perturbative arguments, we show that this phenomenon is {quantitatively} described by Rabi 
oscillations in the symmetry-breaking doublet of an effective XXZ model. 
The perturbative scheme itself depends
on the multiplet structure of the spectrum and both disappear for large system sizes.


The paper is organized as follows. In Section~\ref{model:sec} we introduce the model  and in Section~\ref{quantob:sec} the quantities and the 
observables we use to describe it. In Section~\ref{sec:entro} we study the ergodicity breaking by means of the behaviour of the half-chain entanglement entropy
whose volume-law behaviour is in sharp contrast with the area-law behaviour in many-body localized systems. In Section~\ref{dos:sec} we study the ergodicity breaking
by means of the spectral properties and we see that it is unrelated to the multiplet structure of the spectrum. {In Section~\ref{other} we study the behaviour of the eigenstate expectations of the correlation and of the average of the Inverse Participation Ratio.}
In Section~\ref{imbdyn:sec} we study the imbalance dynamics and we interpret the oscillations we find as Rabi oscillations involving a single symmetry-breaking doublet. In Appendixes~\ref{perturb:sec} and~\ref{breaking} we discuss in detail the perturbative analysis allowing us to construct an effective XXZ model well describing this phenomenon. In Section~\ref{conc:sec} we draw our conclusions.


\section{Model and methods}

\subsection{The model} \label{model:sec}
The one-dimensional Bose-Hubbard model~\cite{fisher} is characterized by the Hamiltonian
\begin{align} \label{Hamour:eqn}
  \hat{H}= \frac{U}{2}\sum_{j=1}^L\hat{n}_j(\hat{n}_j-1) -  \frac{J}{2}\sum_{j=1}^L\left(\hat{a}_j^\dagger\hat{a}_{j+1}+{\rm H.~c.}\right)\,,
\end{align}
the first term describes the on-site repulsion while the second the hopping between neighboring sites.  In Eq.(\ref{Hamour:eqn})
$L$ is the system size, $\hat{a}_j$ are bosonic operators and $\hat{n}_j\equiv \hat{a}_j^\dagger\hat{a}_{j}$ are number operators, and $j$ labels 
the site. In the rest of the paper we will fix the interaction strength $U=1$; we will explore the different dynamical behaviours modifying the hopping 
strength $J$.
}
The total-boson-number operator $\hat{N}\equiv\sum_j\hat{n}_j$ is conserved. Defining the filling factor $\nu\equiv \mean{\hat{N}}/L$, it is possible 
to see that the Hilbert space dimension is $\dim\mathcal{H}(L)=\binom{L(\nu+1)-1}{L-1}$ {(if $L\nu$ is an integer)}. 
In our analysis we will restrict to the case $\nu=1$ (one boson per site) and label the state using 
the Fock basis, i.e. the basis of the simultaneous eigenstates of all the $\hat{n}_j$ operators (we will denote it as $\left\lbrace\ket{\nb}\right\rbrace$). 

Throughout the work we will consider periodic boundary conditions (unless otherwise specified), so the system has the translation and the inversion symmetries. 
Therefore, choosing initial states invariant under these symmetry operations, the dynamics restricts to the Hilbert subspace $\mathcal{H}_S(L)$ fully symmetric 
under these symmetry operations. {More explicitly, $\mathcal{H}_S(L)$ is defined as the subspace corresponding to the zero-momentum sector and even with respect to inversion. In this way we restrict} the dimension of the interesting Hilbert subspace, so we can perform full exact diagonalizations for 
system sizes up to $L= 11$. In particular, the dimension of the fully-symmetric subspace is smaller than the full Hilbert space dimension by a factor $\sim 2L$.


\subsection{Quantities and observables} \label{quantob:sec}
{
Our analysis will include both observing the dynamics of the system after an initial preparation in a given non-equilibrium state and by a statistical 
analysis of the properties of the spectrum. Correspondingly we will consider different quantities.
}
\subsubsection{Eigenvalues and eigenstates}

We define the Hamiltonian eigenvalues as $E_\alpha$ and the corresponding eigenvectors $\ket{\varphi_\alpha}$. In order to see if there is eigenstate thermalization 
we focus on the properties of eigenstates. 

We consider an important basis independent quantity, the half-chain entanglement entropy, which is defined in the following way. We divide the system in two partitions 
$A$ and $B$. When $L$ is even they are both long $L/2$. When $L$ is odd, one of them is long $L/2-1$ and the other $L/2+1$. So the {\it full} Hilbert space 
(not the symmetrized one) has a tensor product structure $\mathcal{H}(L)=\mathcal{H}_A\otimes\mathcal{H}_B$. Once we have done that, we define for each eigenstate
\begin{equation} \label{entropy:eqn}
  S_{L/2}^{(\alpha)}=-\Tr_A[\hat{\rho}_A^{(\alpha)}\log\hat{\rho}_A^{(\alpha)}]\quad{\rm with}\quad \hat{\rho}_A^{(\alpha)}=\Tr_B[\ket{\varphi_\alpha}\bra{\varphi_\alpha}]\,,
\end{equation}
where $\Tr_B$ is the partial trace over $\mathcal{H}_B$. Notice that now the states are taken living in the full Hilbert space. In order to pass from fully-symmetric subspace to the full Hilbert
space one must apply a linear transformation (an isometry). We will see that the entanglement entropy is a very precious quantity to probe if there is ETH 
and thermalization.
We also study the properties of the energy eigenvalues, density of states Eq.~\eqref{density:eqn} and average level spacing ratio Eq.~\eqref{spacing:eqn} (see Sec.~\ref{dos:sec}).
Throughout all the text we define the average over the eigenstates inside the full Hilbert space as
\begin{equation} \label{mean:eqn}
  \mean{(\cdots)}=\frac{1}{\dim\mathcal{H}_S(L)}\sum_\alpha (\cdots)_\alpha\,.
\end{equation}

\subsubsection{Dynamics}

We will study dynamics with a specific initialization. In order to explore {real-space} localization, we will follow a protocol introduced by~\cite{Schreiber842,Carleo} 
and focus on the evolution of the imbalance. In order to do that, we will initialize the system in the state
\begin{equation} \label{imbalanced_state:eqn}
  \ket{\psi_{02}}=\ket{0}\otimes\ket{2}\otimes\ket{0}\otimes\cdots\otimes\ket{2}\otimes\ket{0}\otimes\ket{2}\,,
\end{equation}
and study the evolution of the imbalance operator
\begin{equation} \label{I:eqn}
  \hat{\mathcal{I}}=\frac{ \sum_{j\;{\rm even}}\hat{n}_j - \sum_{j\;{\rm odd}}\hat{n}_j }{L}\,.
\end{equation}
In this analysis we can restrict to a portion of the Hilbert space (the one fully symmetric under translations of two sites) which is larger than the fully symmetric one, 
but still numerically affordable. For $L\leq 10$ we will use full exact diagonalization. For $L=12$, on the opposite, we will resort to Krylov technique (implemented in 
{\sc Expokit}~\cite{EXPOKIT}) and we will truncate the Hilbert space so that the number of bosons per site will be smaller than some threshold $m_{\rm th}$ (we will 
always consider $m_{\rm th}\leq 8$). We will take a Krylov subspace of dimension $M_{\rm K}\leq32$. With this technique we can address only values of 
$J\lesssim0.25$: in this case, for the sizes we consider, the time-evolved state does not deviate too much from the initial one. In order to understand if the Krylov 
technique is applicable with our numerical resources, for each of the considered cases we consider different values of $m_{\rm th}$ and $M_{\rm K}$ and verify 
that the result converges if we increase these values.

%
%
%
%
It is important to state the definition of the infinite-time average of a time-dependent quantity $\mathcal{O}(t)$,
\begin{equation} \label{medo:eqn}
  \overline{\mathcal{O}}=\frac{1}{\mathcal{T}}\int_0^\mathcal{T}\mathcal{O}(t)\ud t\,.
\end{equation}

%
\section{Behaviour of the entanglement entropy} \label{sec:entro}
\begin{figure*}
  \begin{center}
   \begin{tabular}{cc}
     \includegraphics[width=8cm]{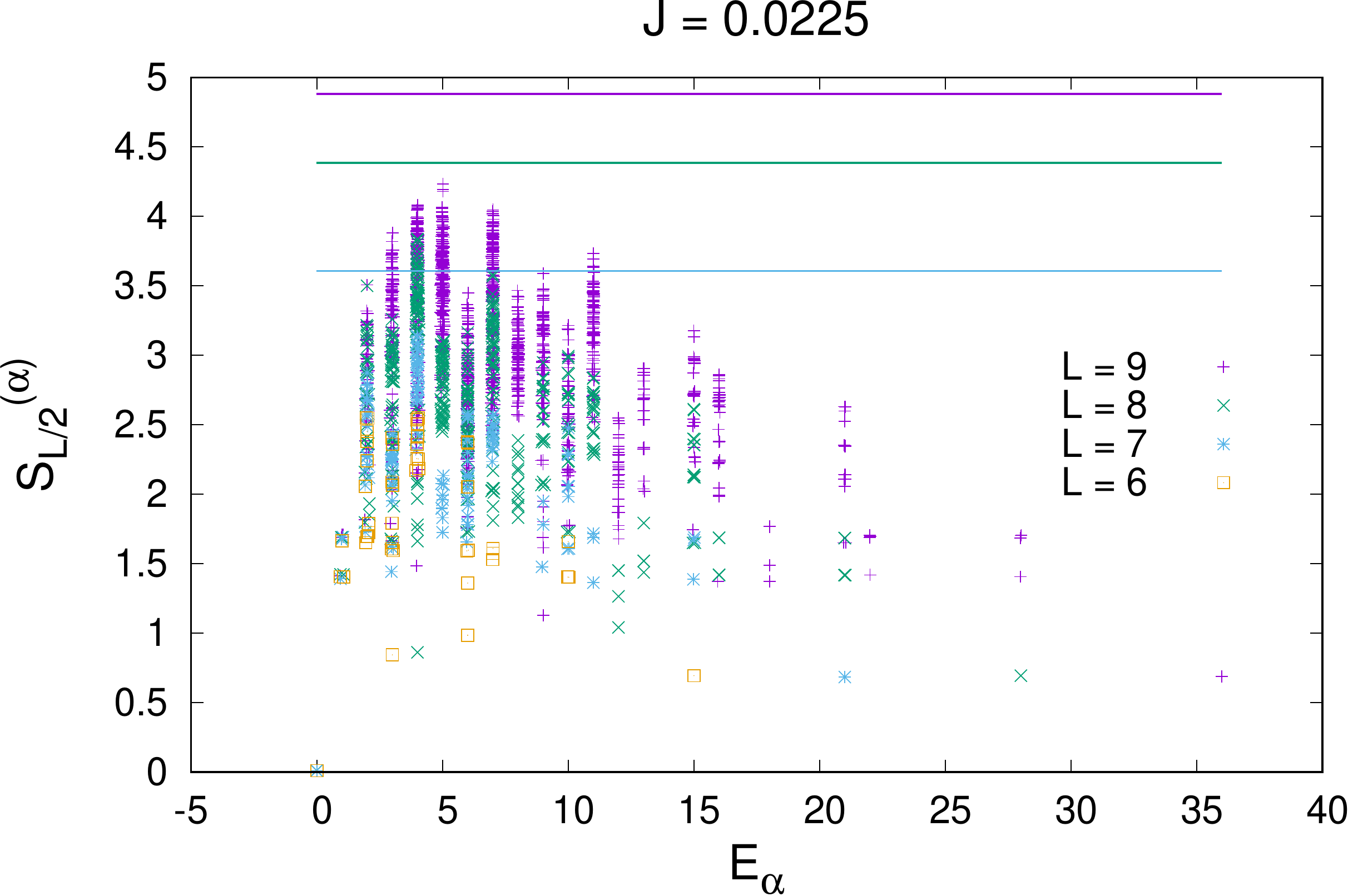}&
     \includegraphics[width=8cm]{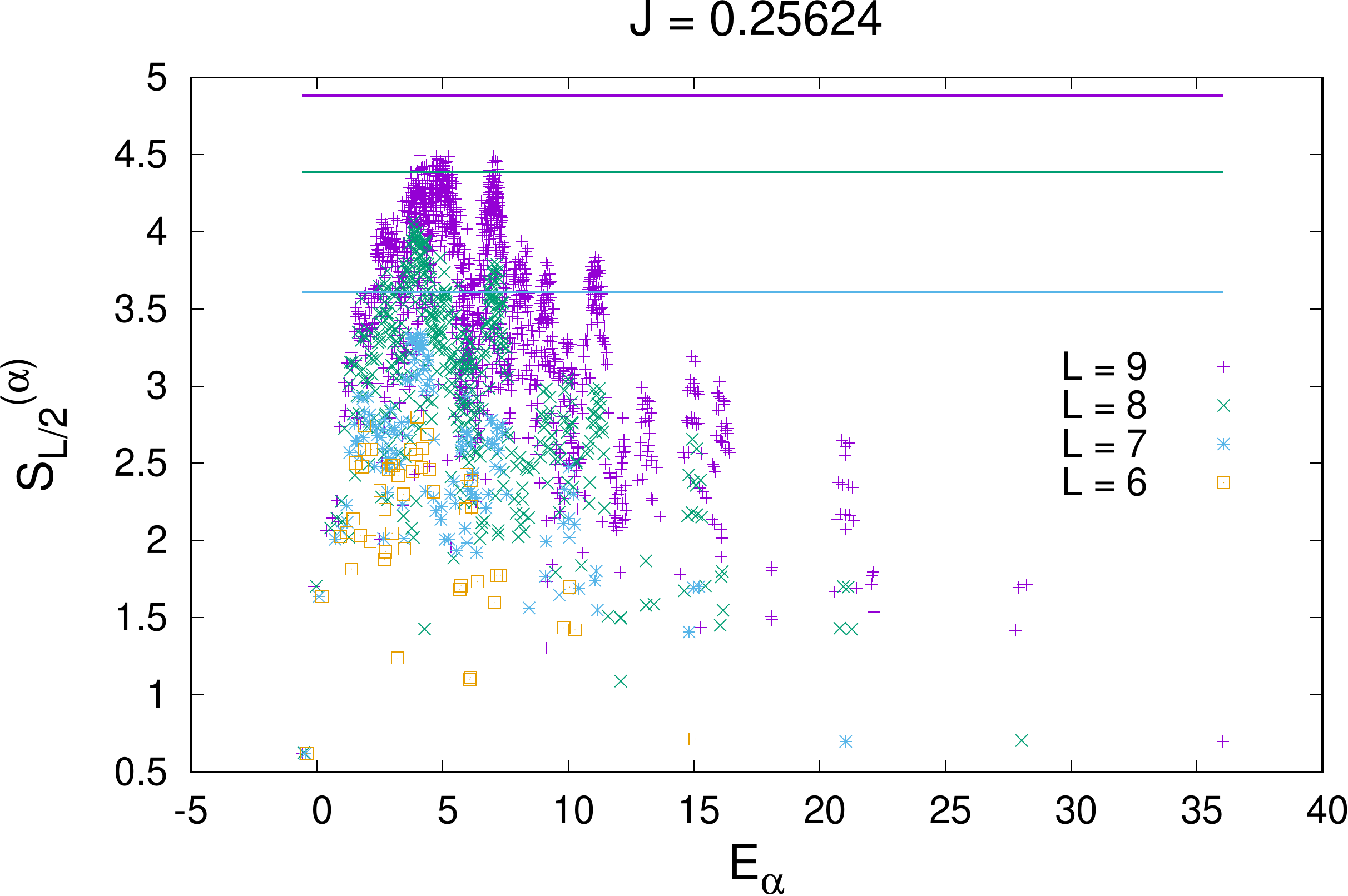}\\
     \includegraphics[width=8cm]{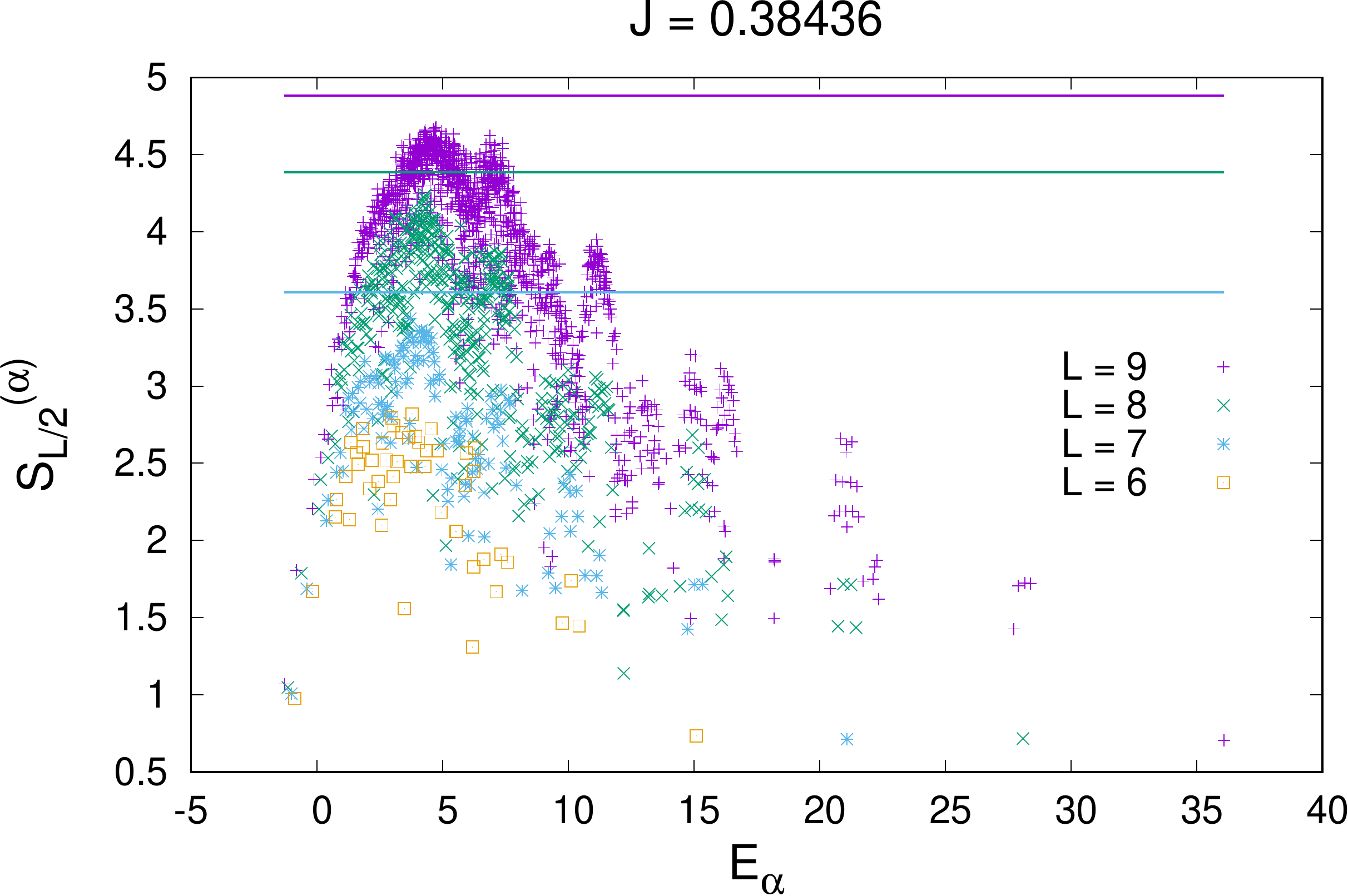}&
     \includegraphics[width=8cm]{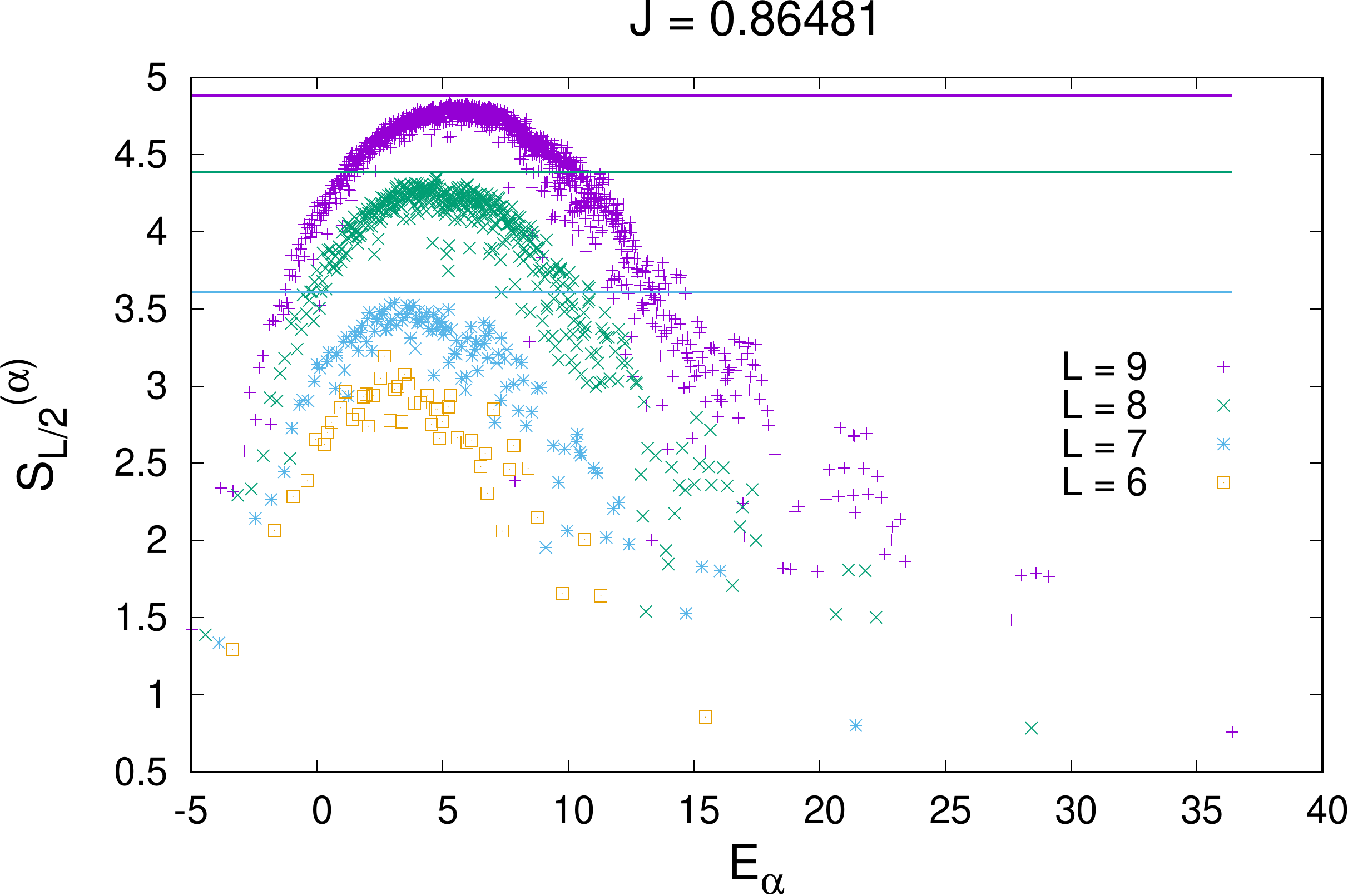}\\
     \includegraphics[width=8cm]{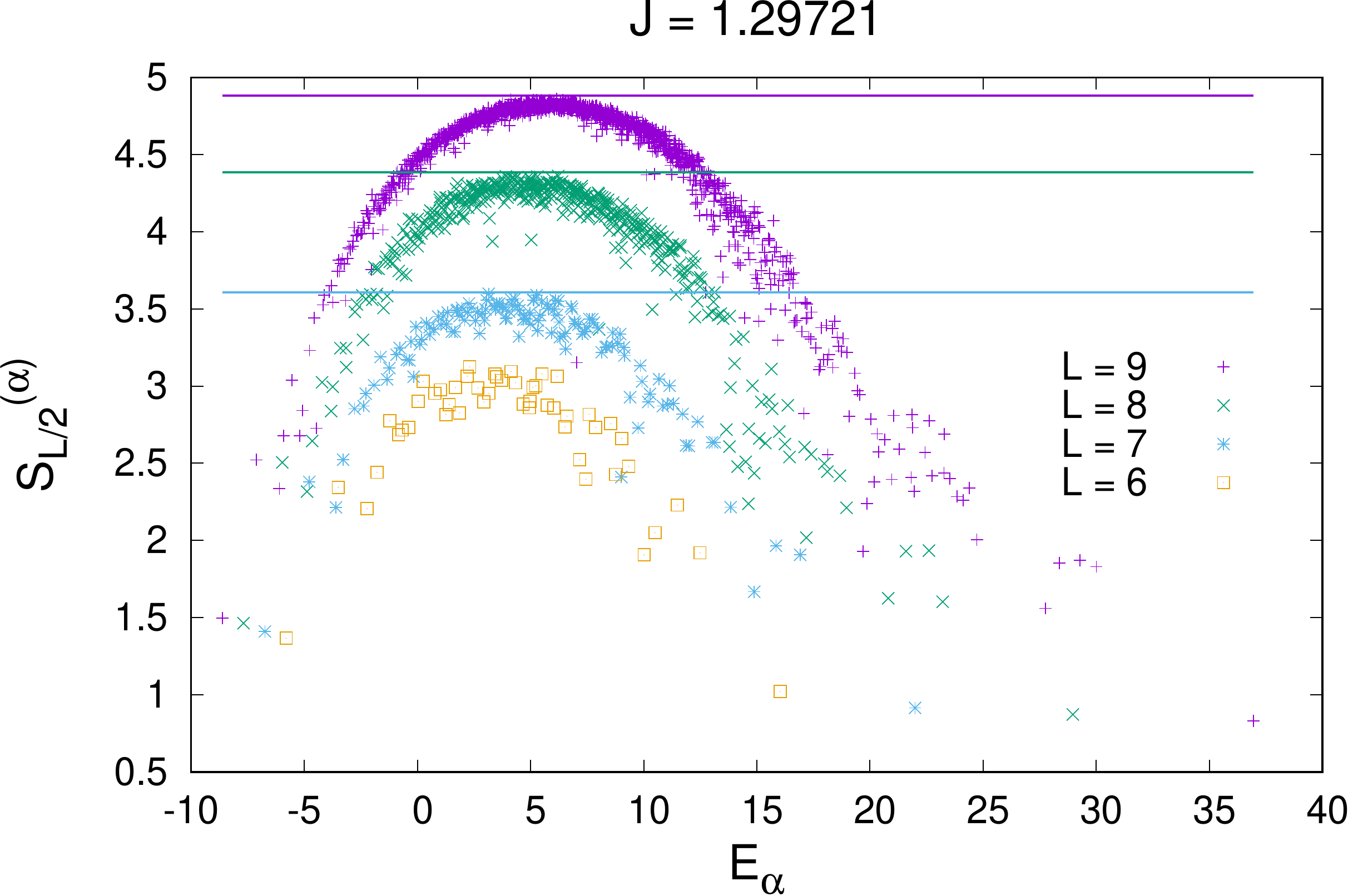}&
     \includegraphics[width=8cm]{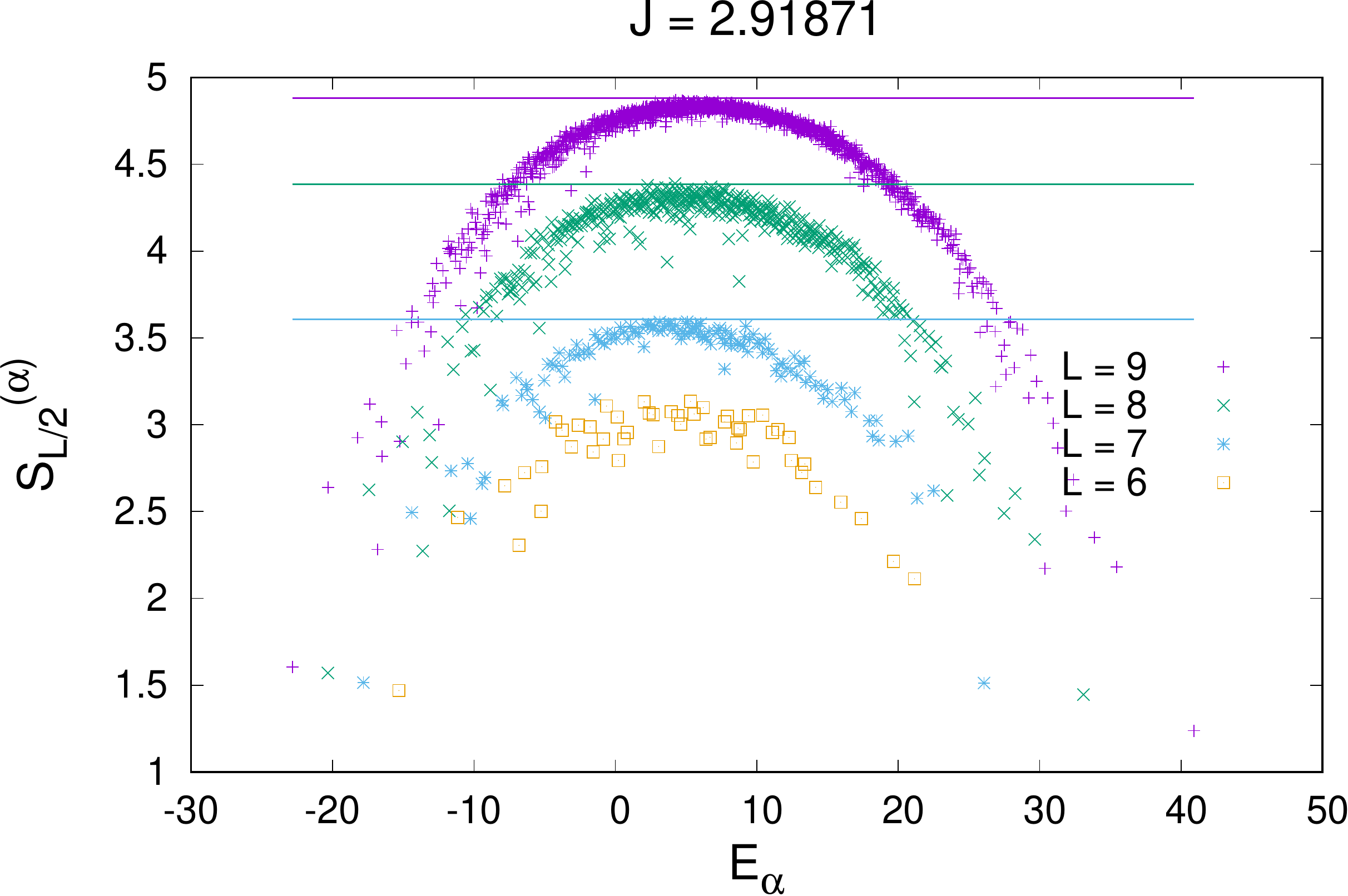}
   \end{tabular}
  \end{center}
 \caption{Examples of scatter plots of $S_{L/2}^{(\alpha)}$ versus $E_\alpha$ for different values of $L$ and $J$. The horizontal lines are the Pages values at the system-size value of the corresponding color. The Page values are evaluated for $N_{\rm rand}=2000$.}
    \label{plots_entro:fig}
\end{figure*}

We start our analysis by considering the half-chain entanglement entropy of the eigenstates Eq.~\eqref{entropy:eqn}. We plot some examples of scatter plots of $S_{L/2}^{(\alpha)}$ versus $E_\alpha$ for different values of $L$ and $J$ in Fig.~\ref{plots_entro:fig}. As a reference we plot also the line corresponding to the ``Page value'' $S_{L/2}^{\rm (Page)}$. With that phrase we mean the value obtained with the following construction, performed {analytically and in the case of a generic bipartite quantum system} by Don N. Page~\cite{Page_PRL}. We take a state in the fully symmetric Hilbert space such that 

\begin{equation} \label{stato_random:eqn}
 _S\Braket{\nb|\psi}_S=\frac{\alpha_{\nb,\psi}+i\beta_{\nb,\psi}}{\mathcal{N}_\psi}
\end{equation}
%
{
with $\alpha_{\nb,\psi}$ and $\beta_{\nb,\psi}$ drawn from from a normal distribution with zero mean and $\mathcal{N}_\psi$ the appropriate normalization constant~\cite{Page_note}.}
We project it in the full Hilbert space and we evaluate the entanglement entropy. We repeat this procedure over $N_{\rm rand}$ realizations and we take the average. We see in Fig.~\ref{plots_entro:fig} that for $J$ above some threshold, the graph becomes a more-or-less continuous curve which touches the Page value. This marks the setting-on of the ETH. Therefore we can qualitatively see that the system breaks ergodicity at small $J$ and obeys ETH at large $J$. Now we are going to discuss this feature more quantitatively.

\begin{figure}
  \begin{center}
   \begin{tabular}{c}
     \includegraphics[width=8cm]{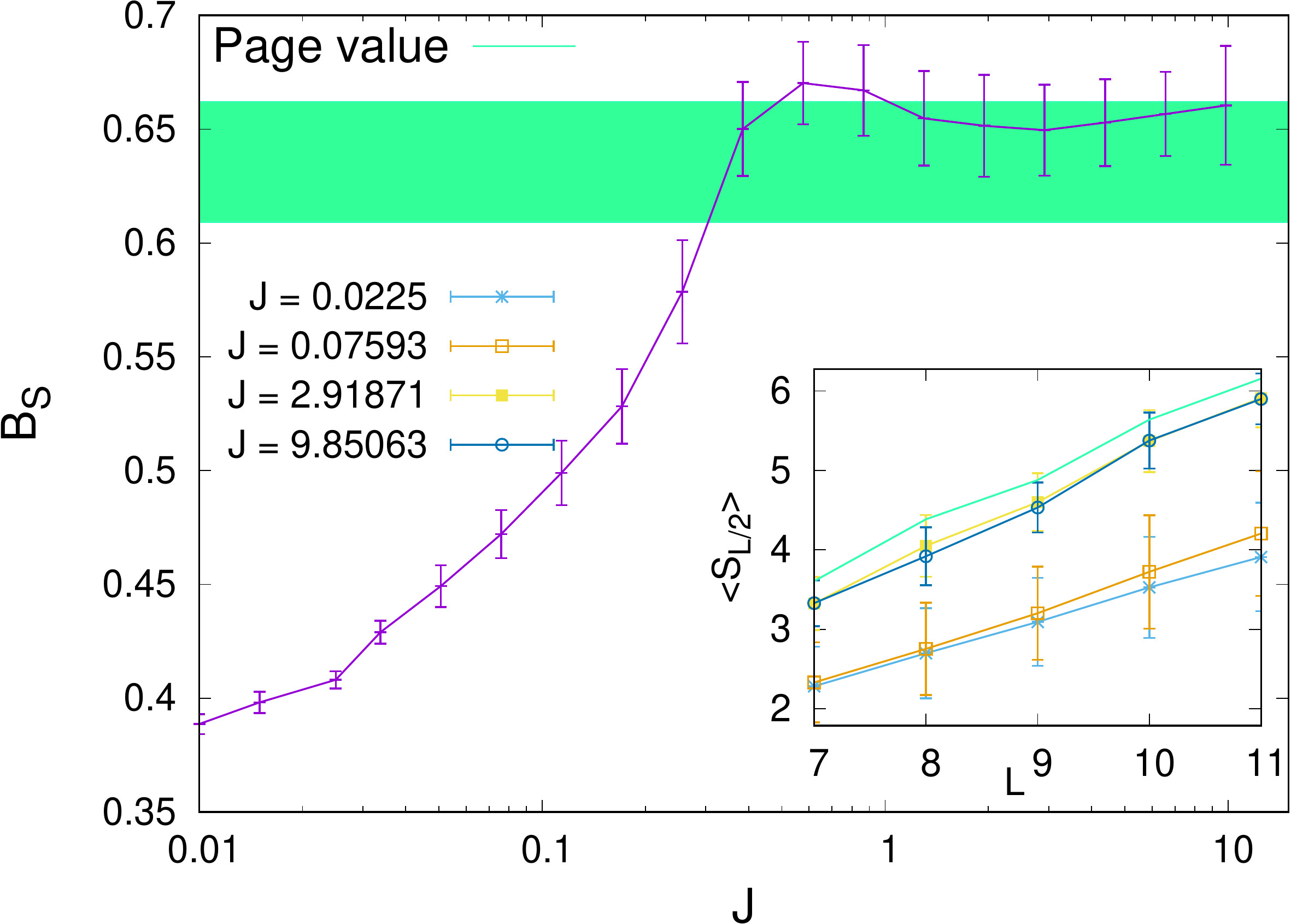}
   \end{tabular}
  \end{center}
 \caption{(Inset)  $\mean{S_{L/2}}$ versus $L$ for different values of $J$, in the ergodicity-breaking ($J=0.0225$, $J\simeq 0.17$) and in the ergodic ($J\simeq 2.92$, $J\simeq 9.8$) regime. Notice the linear dependence in all the cases and that in the ergodic regime the average closely follows the Page value of a fully random state. (Main figure) Slope of the linear fit $\mean{S_{L/2}}\sim A + B_S L$ versus $J$. Notice that it attains the Page value around $J\sim 0.9$: from there on the system behaves quantum chaotically as a random-matrix model. The green region represents the Page value of $B_S$ with uncertainty. For $L\leq 10$ we average over all the states, while for $L=11$ we average over $1200\leq N_s\leq 3800$ randomly chosen states (over a total number $\dim\mathcal{H}_S(11)=16159$). For the evaluation of the Page value, we have chosen $N_{\rm rand}=10000$ for $L\leq 10$ and $N_{\rm rand}=500$ for $L=11$.}
    \label{ententro:fig}
\end{figure}

In order to do that, we move to study the average over the eigenstates of the entanglement entropy. We show some examples of scaling with the system size $L$ in the inset of Fig.~\ref{ententro:fig}. We notice that there is always a linear increase, no matter the value of $J$, but {there is a} difference in slope between small and large $J$. In the former case ($J=0.0225$, $J\simeq 0.075$) there is ergodicity breaking and the slope is significantly smaller than the Page value. 
In the ergodic cases ($J\simeq 2.92$, $J\simeq 9.8$) we show  there is still linear increase but with a larger slope{, very near to the one of the fully-random Page value}.

{Indeed, if ETH holds, then the slope of the average entropy should coincide with the slope of the Page entropy. To show this, {we follow an argument introduced in its most formal way in~\cite{Huang_NPB19}} and recall that, assuming ETH and considering a subsystem of $l$ consecutive sites with $l \ll L$, the reduced density matrix of an eigenstate is equivalent to a thermal density matrix. In particular, it follows that the entanglement entropy of the reduced density matrix is equal to the thermal entropy computed over the appropriate ensemble. This fact has been shown to be valid up to $l=L/2$ (with better approximation for increasing $L$) for a spin $1/2$ Heisenberg chain in Ref.~\cite{PhysRevB.93.134201}.}

{
	We \textit{assume} the reduced density matrix being thermal for $l=L/2$ also in our case (we will see that it leads to conclusions confirmed by our numerical analysis). With this assumption, we can compute the $S_{L/2}$ for eigenstates at energy $E$ as the corresponding thermal entropy $S(E)$ in the microcanonical ensemble. In the definition of $S(E)$ an average over an energy shell of width $\Delta E$ is implicit. We can also average $S_{L/2}$ over eigenstates in the energy shell $[E-\Delta E/2,E+\Delta E/2]$ in order to get more regular $S_{L/2}(E)$ curves. In the microcanonical ensemble $S(E) = \log\left( \rho_S(E)\right)$, where $\rho_S$ is the density of states with energy $E$
\begin{equation}
	\rho_S(E) = \sum_\alpha \delta(E_\alpha -E)
\end{equation}
with $E_\alpha$ running over all the energy eigenvalues. (Also $\rho_S(E)$ is regularized by averaging over the energy shells).
At this point it is convenient to express  entropies $S$ and energies $E$ in terms of the corresponding energy densities $\varepsilon=E/L$ and entropy densities $s=S/L$. If we take an energy-shell width $\Delta E$ much smaller than the width of the energy spectrum, we can write the entanglement entropy averaged over the eigenstates as
\begin{equation}
	\mean{S_{L/2}} = \frac{L}{2}\frac{\int d\varepsilon\, \rho_S(\varepsilon) s(\varepsilon)}{\int d\varepsilon\, \rho_S(\varepsilon)}\,.
\end{equation}
Since $\rho_S(\varepsilon)=\exp(Ls(\varepsilon))$, when $L\gg 1$, we are justified in computing the integrals using a saddle-point approximation. Thus replacing 
\begin{equation}
	s(\varepsilon) = s(\varepsilon_\infty) - \frac{(\varepsilon-\varepsilon_\infty)^2}{2W},
\end{equation}
where $\varepsilon_\infty$ is the energy density of the maximum entropy states (corresponding to $T=\infty$), $s(\varepsilon_\infty)$ coincides with the Page value, and $W$ denotes a non-universal many-body bandwidth. The saddle point calculation then gives
\begin{equation}
	\mean{S_{L/2}} = \frac{L}{2} s(\varepsilon_\infty)+ O(1),
\end{equation}
with the $O(1)$ correction being non-universal, as they depend on $W$ {(this formula in its most general form was first found in~\cite{Huang_NPB19})}.
}


We can confirm {the analysis above} 
by doing a linear fit $\mean{S_{L/2}}\sim A + B_S L$. We show the dependence of the slope $B_S$ on $J$ in the main panel of Fig.~\ref{ententro:fig}. We can notice that the slope increases until it reaches a {value consistent with Page} around $J^*\sim 0.4$. {Here the slope} coincides with the Page value inside the errorbars. 

So, there is a clear crossover from non-ergodicity to ergodicity, but in both regimes the half-chain entanglement entropy obeys a volume law with the system size. This is a relevant result and clarifies that the system in the ergodicity-breaking regime {is space extended and }behaves in a way different from many-body localization where the half-chain entanglement entropy obeys area law. Of course these statements apply only to the system sizes we can access numerically and no statement can be done for larger system sizes. 

{It is important to perform the analysis of the entanglement entropy restricted to the highest-entropy states. In case of eigenstates thermalization entanglement entropy and thermodynamic entropy coincide, as we have elucidated above, so the highest entropy states are the ones at $T=\infty$ in this context. In order to study the properties of these states we introduce two quantities. The first one is defined as
\begin{equation} \label{lambda:eqn}
  \Lambda_S(L) = \frac{1}{\dim\mathcal{H}_S(L)}\sum_\alpha\log\left(|S^{\rm (Page)}_{L/2}-S^{(\alpha)}_{L/2}|\right)\,.
\end{equation}
The rationale is that the logarithm overweights the smallest values of the argument. Because the highest-entropy states correspond to the smallest values of the difference in the argument, they give the strongest contribution to this average.}

{In order to define the second quantity, we need to first define the integer number $1\leq\alpha^*\leq\dim\mathcal{H}_S(L)$ as the value of $\alpha$ such that the quantity $|S^{\rm (Page)}_{L/2}-S^{(\alpha^*)}_{L/2}|$ is minimum over $\alpha$. In order to consider only states in the bulk of the spectrum we restrict the average of the entanglement entropy to states around the energy $E_{\alpha^*}$. More formally, if we term the width of the energy spectrum as $D=E_{\rm max}-E_{\rm min}$, we take $0<f<1$ and restrict the sum to the states with eigenenergy $E_\alpha\in[E_{\alpha^*}-fD/2,E_{\alpha^*}+fD/2]$ (call their number $\mathcal{N}$). In this way we can define
\begin{equation} \label{S:eqn}
  \mean{S_{L/2}}_f=\frac{1}{\mathcal{N}}\sum_{\alpha\,{\rm s.t.}\,E_\alpha\in[E_{\alpha^*}-fD/2,E_{\alpha^*}+fD/2]}S^{(\alpha)}_{L/2}\,.
\end{equation}
We choose $f=0.1$, so that the sum is restricted around the state with entropy nearest to the Page value, that's to say to the highest-entropy states (in absence of ergodicity, the highest value of the entropy is smaller than the Page one).}

{We report the results for $\Lambda_S(L)$ versus $J$ for different values of $L$ in Fig,~\ref{plots_entro1:fig}(upper panel), and those for $(S^{\rm (Page)}_{L/2}-\mean{S_{L/2}}_f)/L$ in Fig.~\ref{plots_entro1:fig}(lower panel). We see that both quantities behave differently for small and large $J$. In particular, at large $J$ there is a clear scaling behaviour  for $(S^{\rm (Page)}_{L/2}-\mean{S_{L/2}}_f)/L$ and a decrease of $\Lambda_S(L)$ opposed to an increase for small $J$. We conclude that for large $J$ the highest-entropy states tend to become more ergodic as the system size is increased but the limited system sizes do not allow to precisely state the value of $J$ where this behaviour sets in. 
\begin{figure}
  \begin{center}
   \begin{tabular}{c}
     \includegraphics[width=7.5cm]{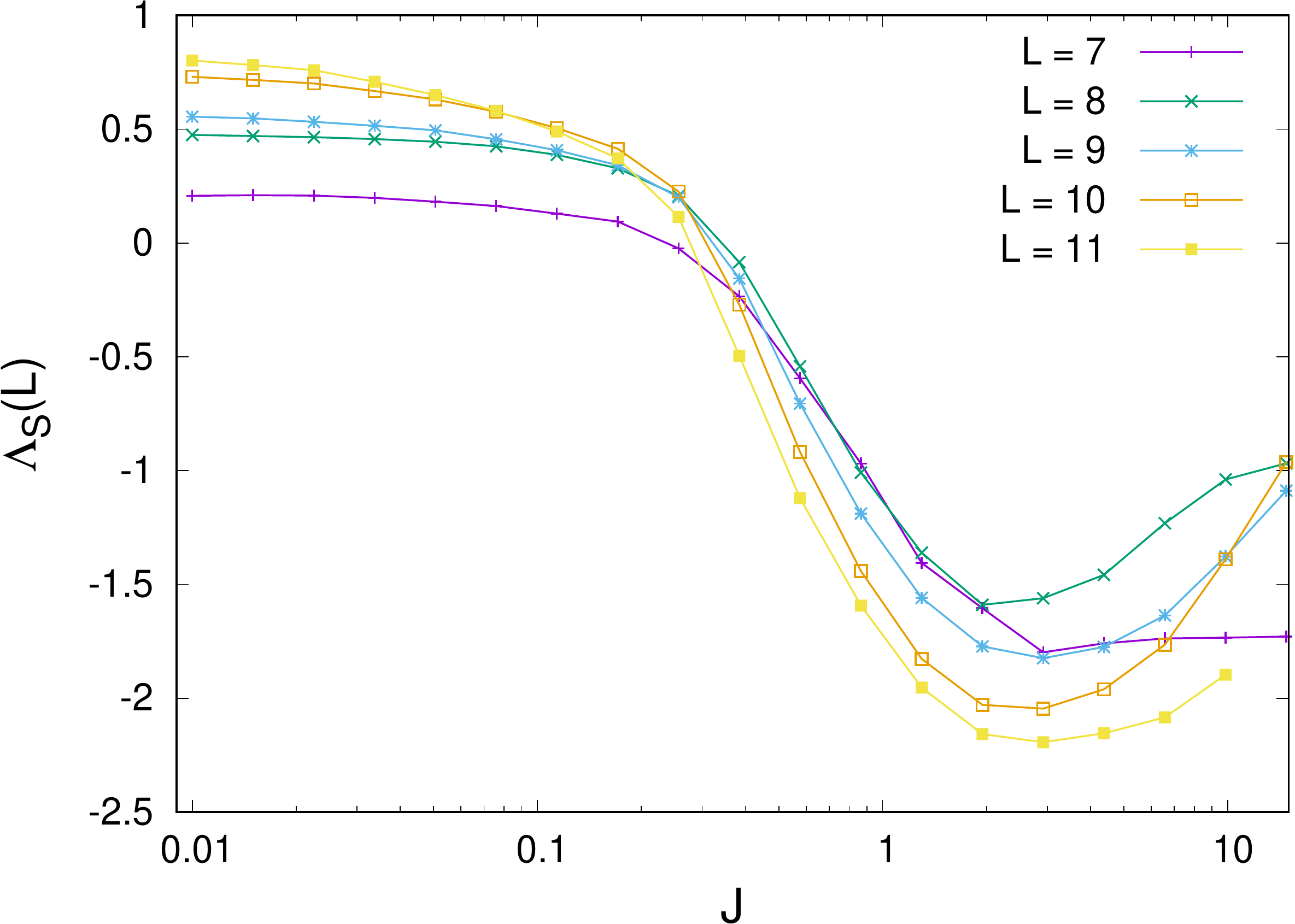}\\
     \includegraphics[width=7.5cm]{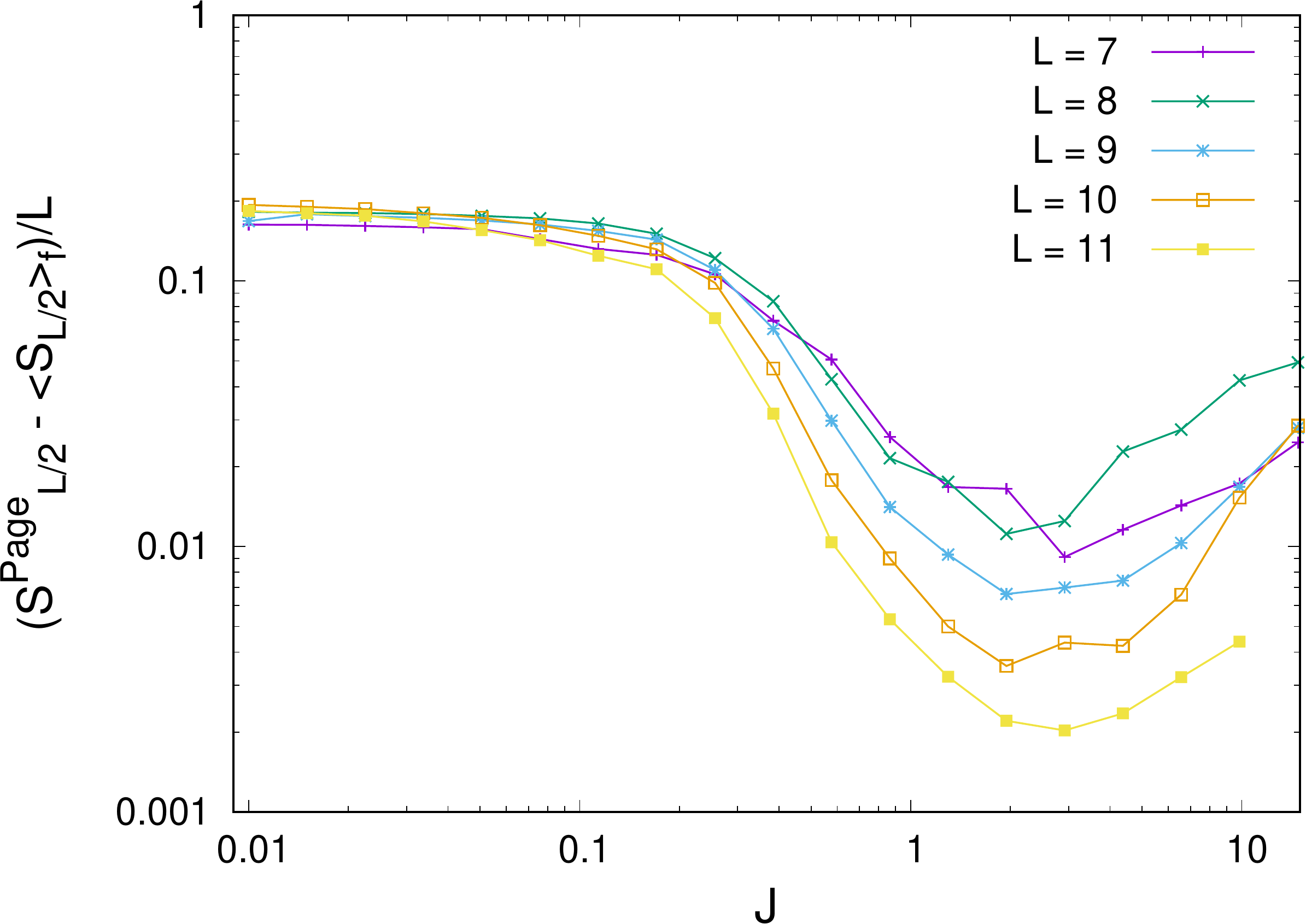}\\
   \end{tabular}
  \end{center}
 \caption{Analysis of the high-entropy states. Plot of the quantities $\Lambda_S(L)$ [Eq.~\eqref{lambda:eqn} -- upper panel ] and $\mean{S_{L/2}}_f$ [Eq.~\eqref{S:eqn} -- lower panel  -- $f=0.2$] versus $J$ for different values of $L$. For $L=11$ we average over $1200\leq N_s\leq 3800$ randomly chosen states. The behaviour for large $J$ suggests a tendency towards ergodicity for large $L$ but the limited system sizes do not allow to precisely state the value of $J$ where this behaviour sets in.}
    \label{plots_entro1:fig}
\end{figure}}
\section{Spectral properties} \label{dos:sec}
The quantum chaoticity properties of a system appear also through the properties of the spectrum of its Hamiltonian.
An important role is played by the level spacings $\lambda_\alpha=E_{\alpha+1}-E_{\alpha}$.
Indeed, the distribution of the (normalized) level spacings takes a universal form in case of quantum chaos.
If the dynamics is fully chaotic and thermalizing, the Hamiltonian behaves as
a Gaussian-Orthogonal-Ensemble random matrix~\cite{Haake,PhysRevLett.52.1,poilblanc,Berry_Les_Houches} and the level spacings obey the Wigner-Dyson (WD) distribution.
On the opposite, a classically integrable system generically shows
a Poisson distribution~\cite{Berry_PRS77} of the normalized $\lambda_\alpha$. We remark that the Poisson level spacing distribution is just a sufficient condition for integrability, not a necessary one~\cite{Berry_PRS77,Gutzwiller90}. A very convenient tool to distinguish these extreme cases and all the
intermediate ones is the average level spacing ratio~\cite{Palhuse}
\begin{equation} \label{spacing:eqn}
 r=\mean{\frac{\min(\lambda_\alpha,\lambda_{\alpha-1})}{\max(\lambda_\alpha,\lambda_{\alpha-1})}}\,.
\end{equation}
This quantity attains a value $r_{\rm WD}\simeq 0.5295$ for a fully-chaotic Wigner-Dyson level-spacing distribution and $r_{\rm P}\simeq 0.386$ for a Poisson distribution. In many-body localization phases there is a superextensive number of localized integrals of motion, a situation closely resembling classical integrability, and the level spacing ratio attains the Poisson value. In this context, the transition to ergodicity is marked by a crossover to the Wigner-Dyson value which becomes sharper and sharper as the system size increases~\cite{abanin2019distinguishing,Palhuse,Luitz15}.

\begin{figure}
  \begin{center}
    \includegraphics[width=8cm]{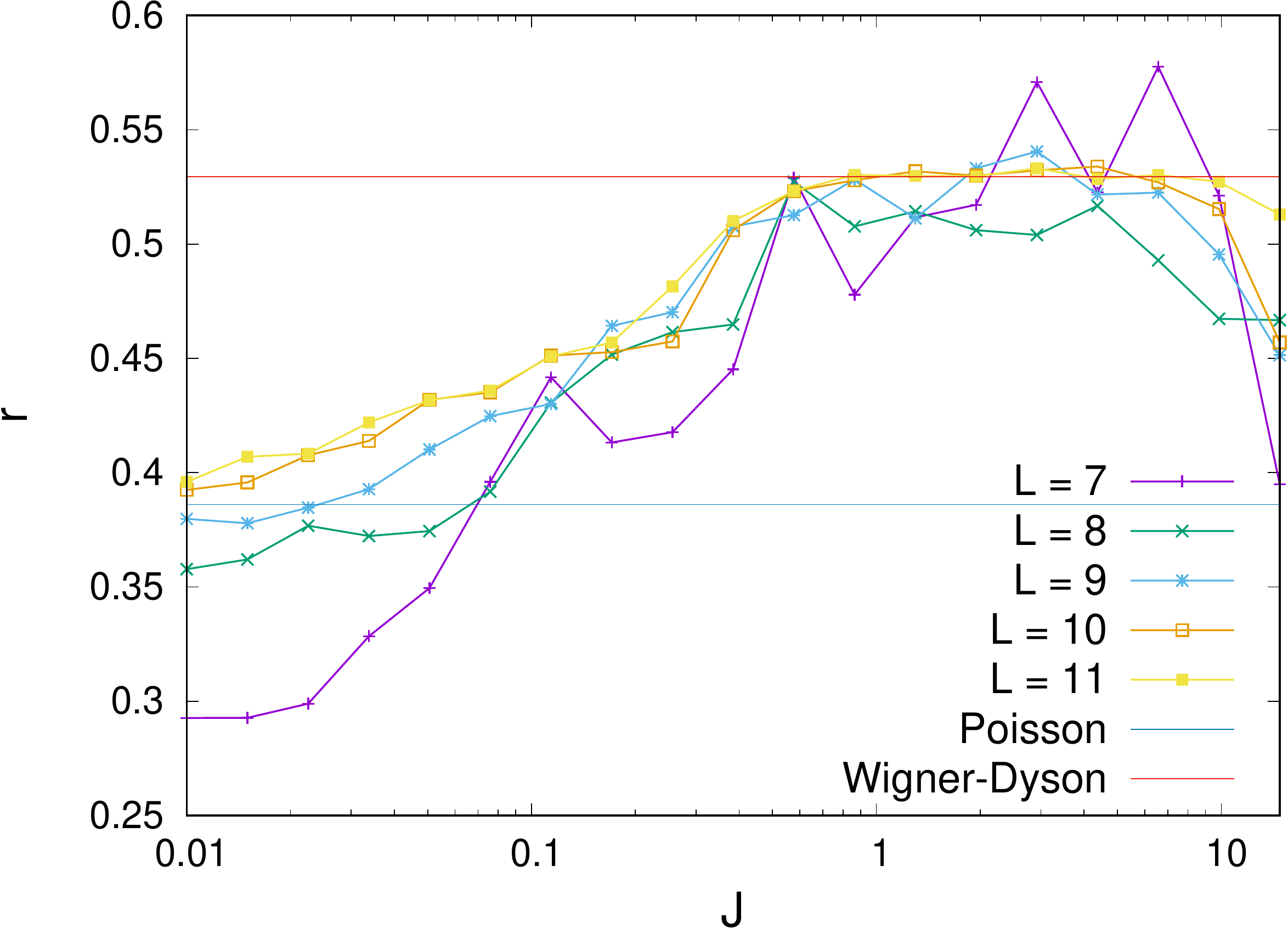}
  \end{center}
 \caption{Average level spacing ratio $r$ versus $J$ for different value of $L$.}
    \label{r_factor:fig}
\end{figure}

We evaluate the average level spacing ratio for our model
and we plot it versus $J$ for different system sizes in Fig.~\ref{r_factor:fig}. We see that it attains the Wigner-Dyson value for $J\gtrsim 0.8$ and that this feature is stable when increasing the system size. The crossover does not become sharper and sharper as the system size is increased as it instead occurs for many-body localization systems (see for instance~\cite{abanin2019distinguishing}). The value of $r$ intermediate between Poisson and Wigner-Dyson marks that the system is not ergodic (the Hamiltonian is significantly different from a random matrix) and does not thermalize.  From this fact we also know that this ergodicity breaking is most probably not associated to any {real-space} localization. Indeed, from the existing numerical evidence~\cite{Bloch_2019} we know that whenever there is a strong form of localization (Anderson or many-body localization) the average level spacing ratio gets the Poisson form.
%

%

\begin{figure*}
  \begin{center}
   \begin{tabular}{cc}
     \includegraphics[width=8cm]{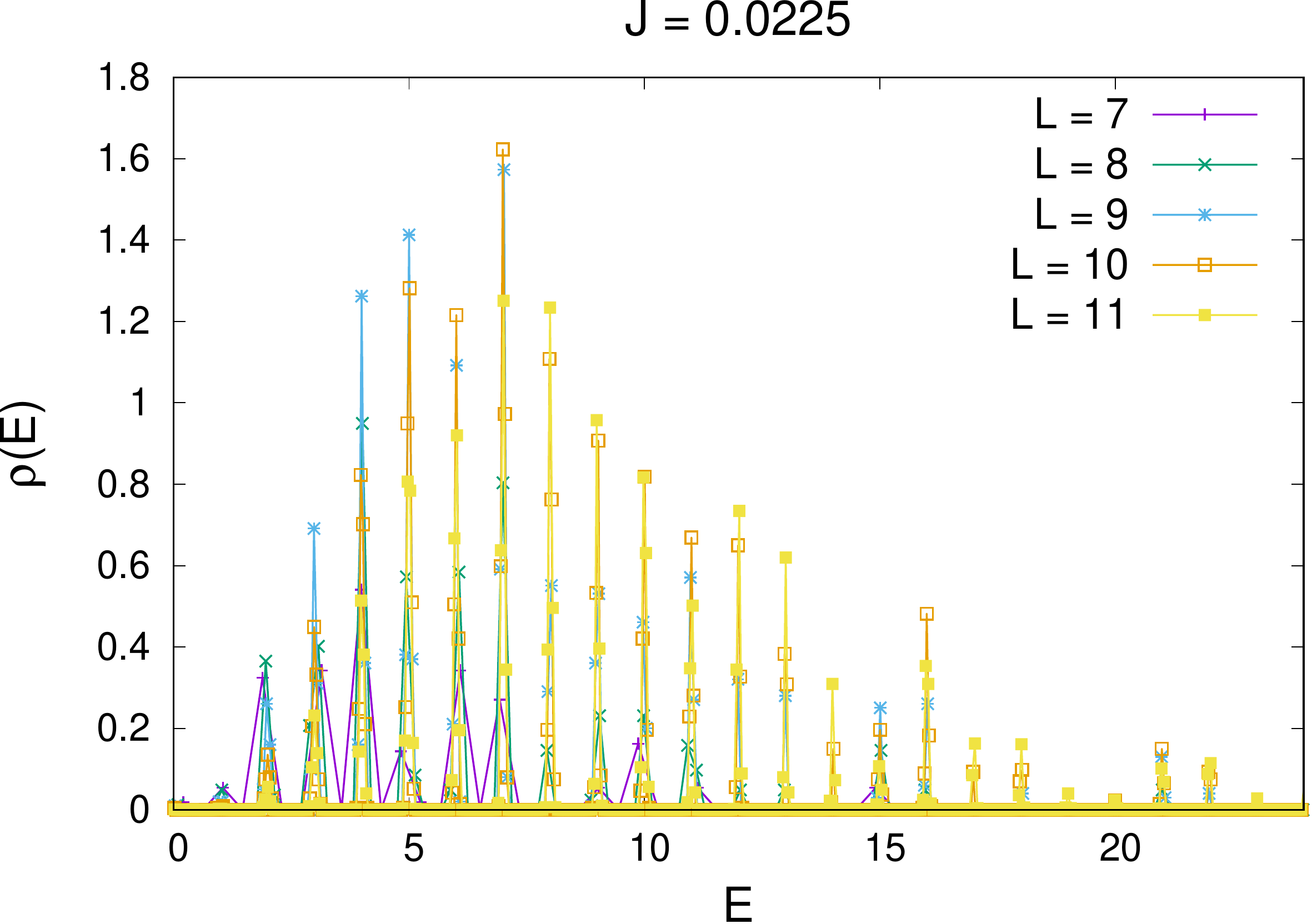}&
     \includegraphics[width=8cm]{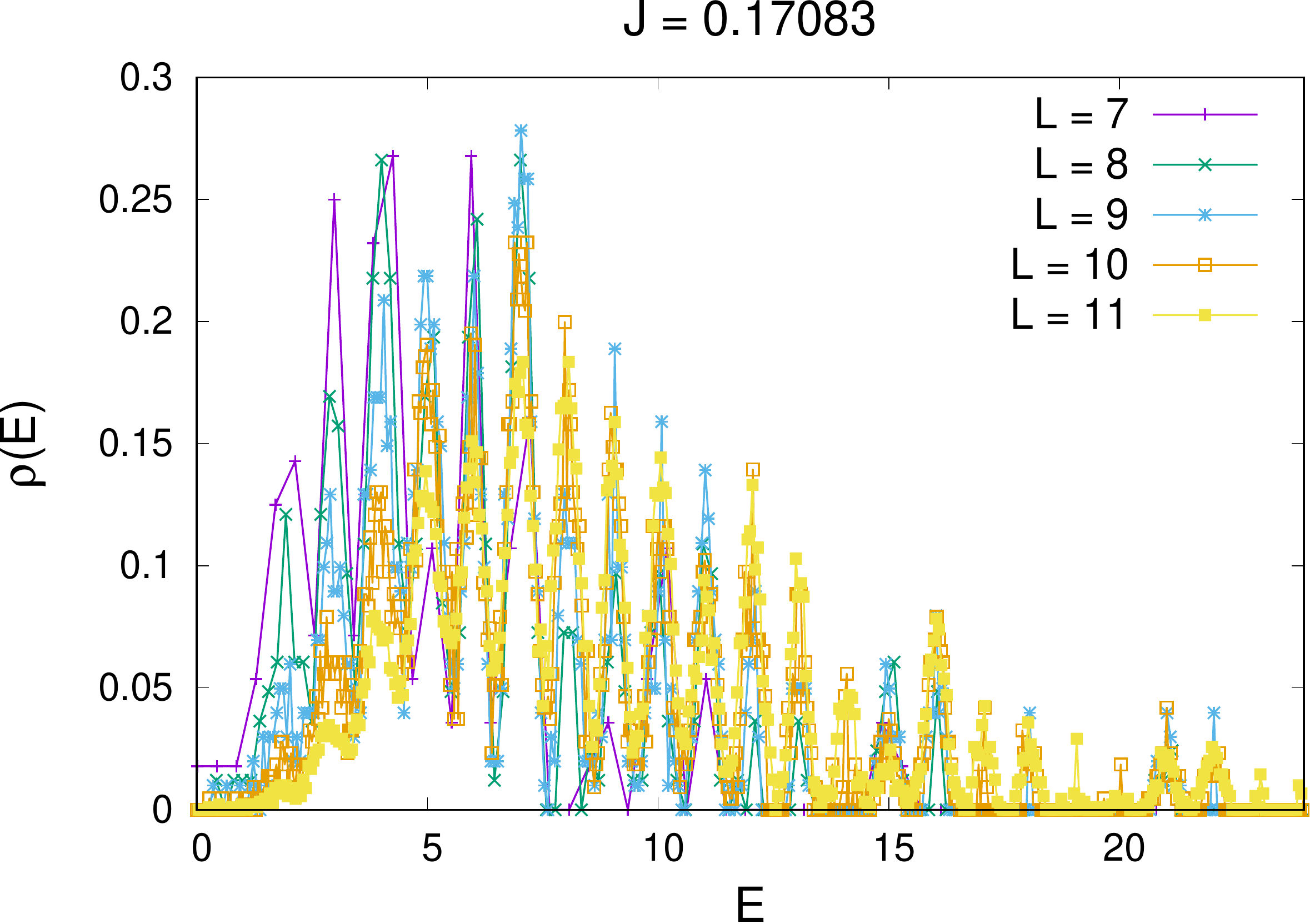}\\
     \includegraphics[width=8cm]{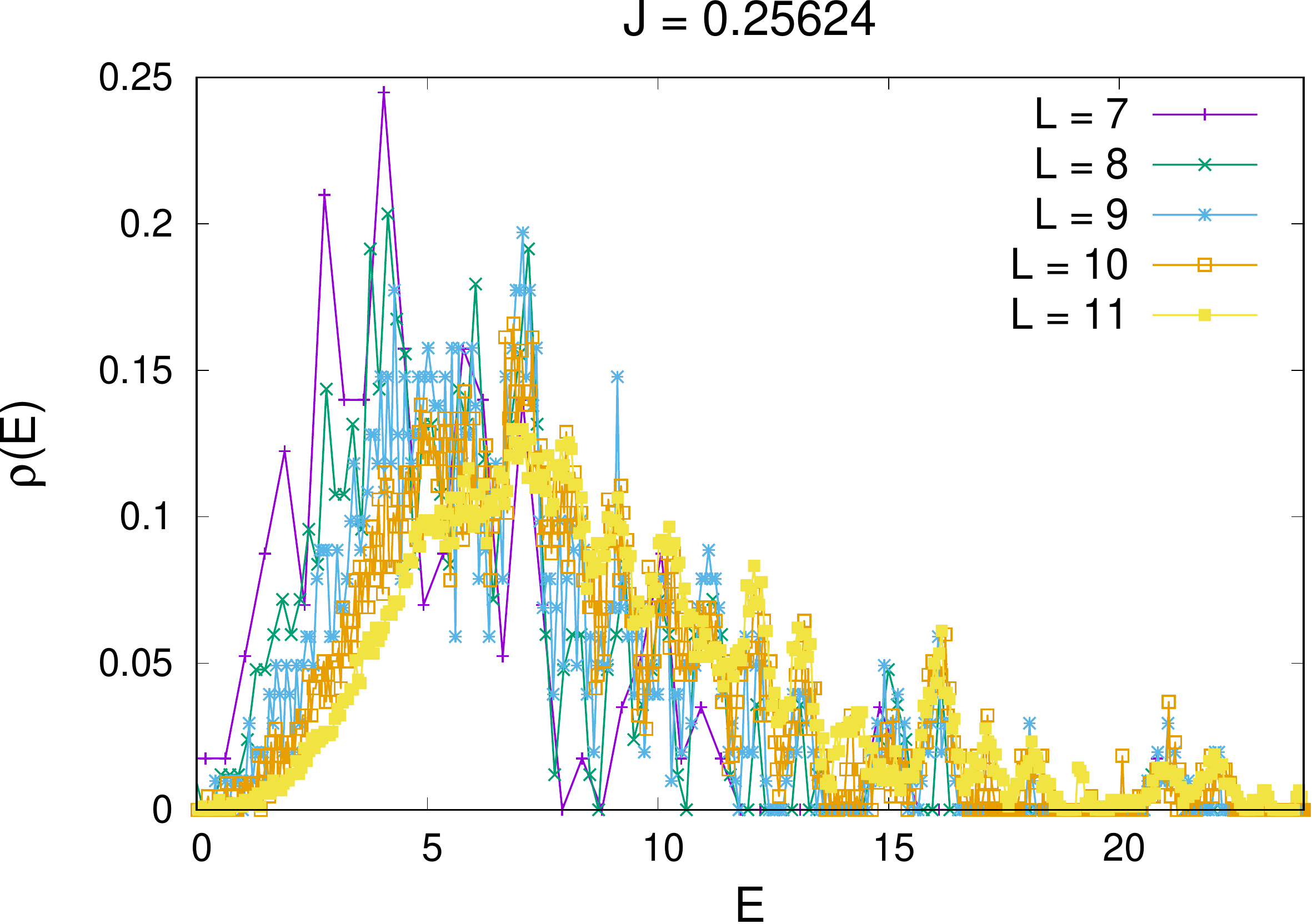}&
     \includegraphics[width=8cm]{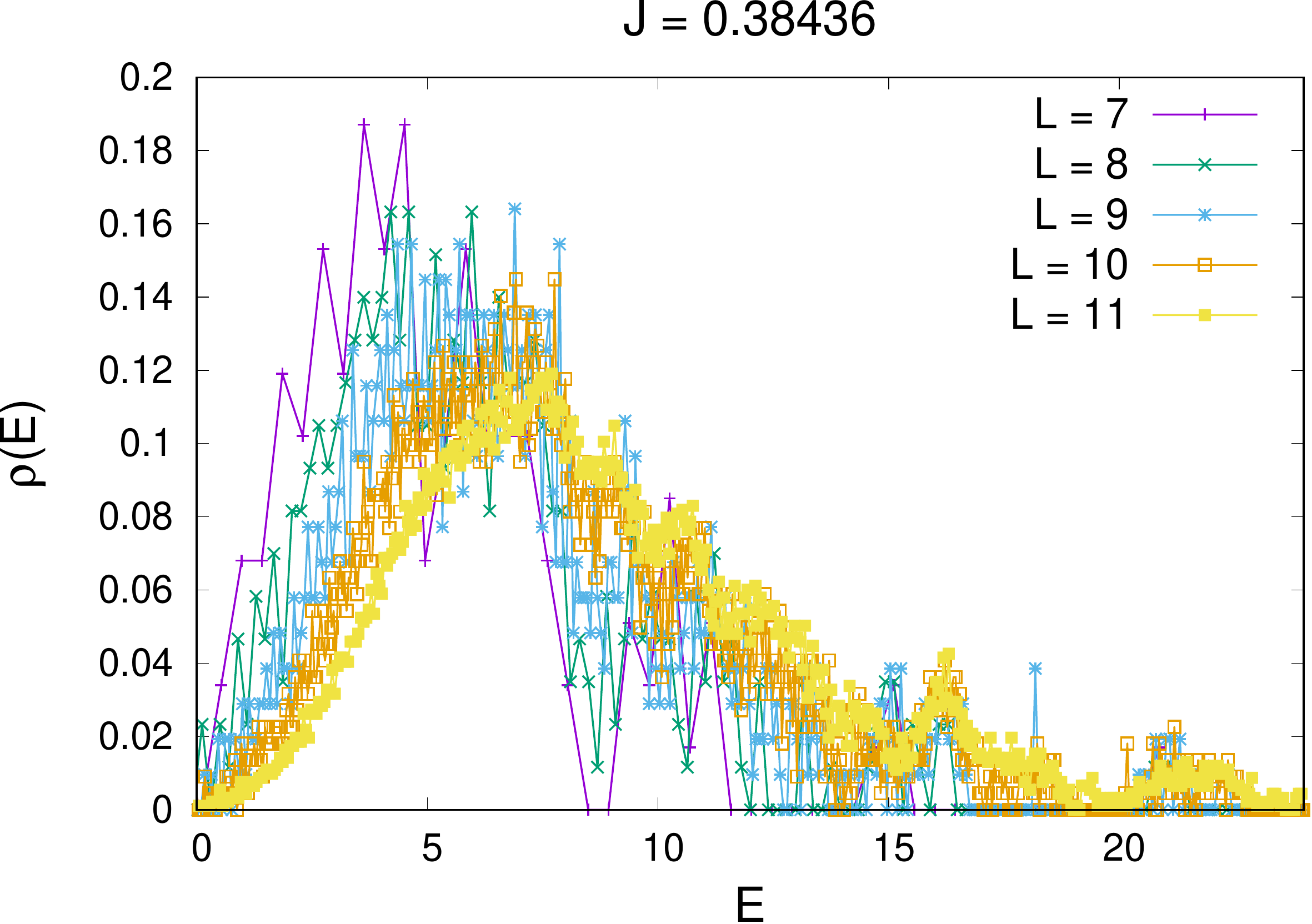}
   \end{tabular}
  \end{center}
 \caption{Examples of density of states. {(Eq.~\eqref{density:eqn} coarse-grained over a number of bins ranging from $150$ to $200$, according to the number of eigenstates). The spikes mark the multiplet structure. We see that for increasing system sizes this structure tends to be blurred. For $J\simeq0.38$ we can see it disappear for $L=9$. }}
    \label{rho:fig}
\end{figure*}

We remark that the nonergodic behaviour of the level spacing ratio is independent of the fact that the spectrum is organized in multiplets when $J$ and $L$ are small. Indeed, the spectrum is massively degenerate for $J=0$; when $J\ll U$ one can apply second order perturbation theory in $J/U$ (see~\ref{perturb:sec} for details) and see that the degenerate levels move to quasidegenerate multiplets. We can see an example of this fact in Fig.~\ref{rho:fig} (an analogous figure can be found in~\cite{lauchli2008spreading}). Here we show the density of states
\begin{equation} \label{density:eqn}
\rho(E) = \frac{1}{\dim\mathcal{H}_S(L)}\sum_\alpha\delta(E-E_\alpha)\,.
\end{equation}
for different values of $J$ and $L$.
We can see that for small $J$ the density of states shows a series of spikes, each one corresponding to a quasi-degenerate multiplet. Nevertheless we see a tendency of this structure to vanish and get diluted in a smooth continuum as the system size $L$ is increased (this fact is quite apparent for $J\simeq 0.17$ and $J\simeq 0.25$). Moreover, already at $J\simeq 0.38$ there is no trace of this structure when the system size is $L=9$. On the opposite, the average level spacing ratio attains the Wigner-Dyson value around $J\sim 0.8$. So, the organization of the spectrum in multiplets and the ergodicity breaking appear to be independent phenomena.

\begin{figure}
  \begin{center}
   \begin{tabular}{c}
     \includegraphics[width=8cm]{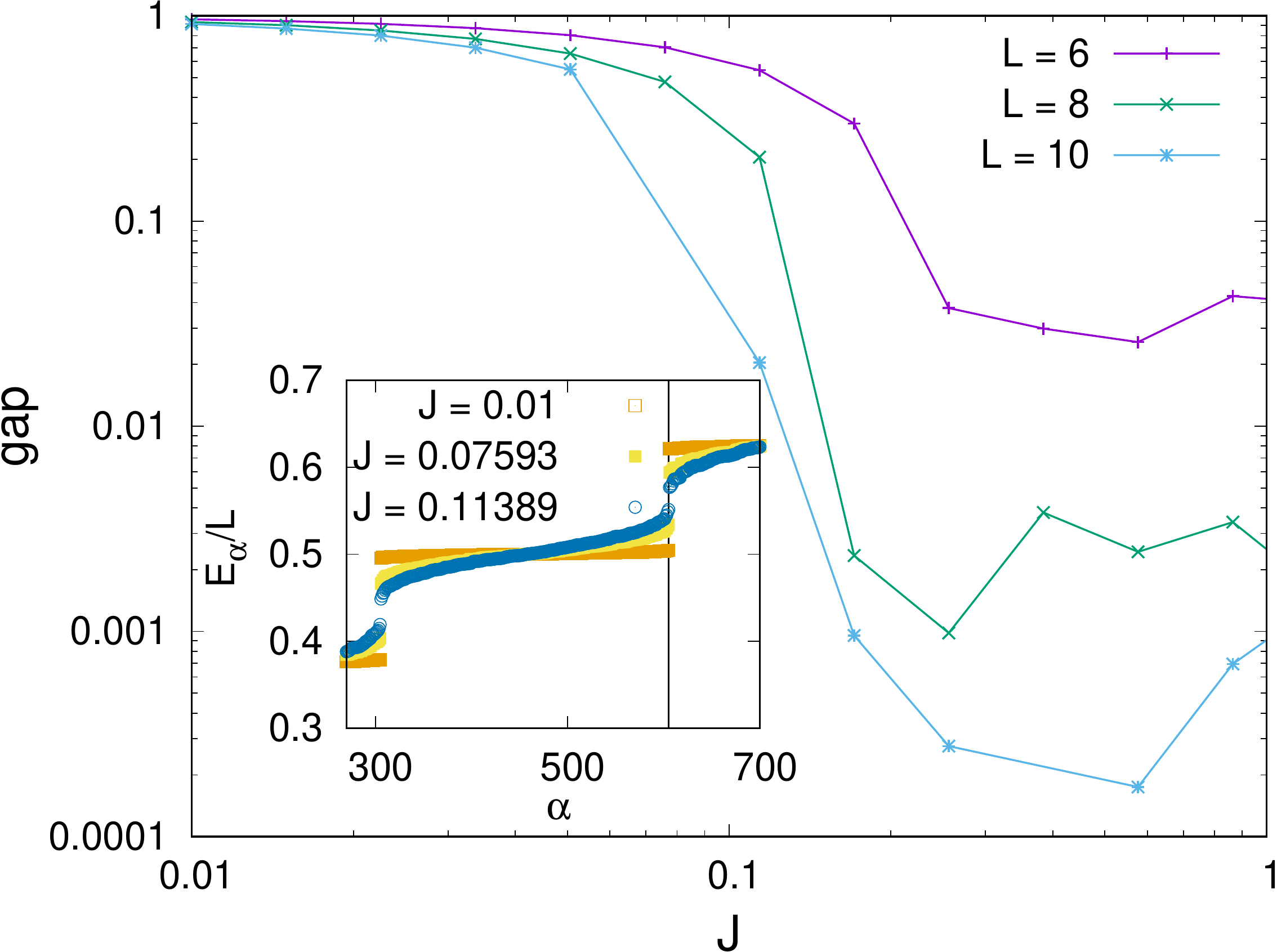}\\
   \end{tabular}
  \end{center}
 \caption{(Main panel) Gap between the multiplet around the energy density $0.5U$ and the one immediately above it. (Inset) Energy eigenvalue densities $E_\alpha/L$ in increasing order versus progressive number $\alpha$ for $L=8$. We focus on the band around energy density $0.5 U$. We see that the number of states in the band does not depend on $J$. We can evaluate the gap between the two nearby bands we are interested in as the difference of the two energy eigenvalues across the vertical black line.}
    \label{bandgap:fig}
\end{figure}

We can get a further confirmation of this fact by looking at the energy gap between two nearby multiplets. We choose in particular the gap between the multiplet around the energy density $0.5U$  and the one immediately above it. In fact, $0.5U$ is the energy density of this multiplet when $J=0$ and the spectrum is degenerate. As we show in the inset of Fig.~\ref{bandgap:fig} this degeneracy is resolved for $J\neq 0$, and for $J$ large enough nearby bands merge with each other. If we fix $L$ and we change $J$ we see that the number of states in each multiplet does not change. So, in order to evaluate the gap between the two nearby multiplets we are interested, we need to order the eigenstates in increasing energy order and compute the difference between the two eigenvalues on the two sides of the vertical black line in the inset of Fig.~\ref{bandgap:fig}. More formally, this operation defines the gap as the difference of the minimum energy of the upper multiplet and the maximum energy of the lower multiplet. We show the dependence of this gap on $J$ in the main panel of Fig.~\ref{bandgap:fig}. We see that for $L=10$ the gap decreases below $10^{-3}$ already at $J\simeq 0.17 $ and for larger values of $J$ one cannot speak about multiplets separated by a gap anymore.

Nevertheless, the results of the average level spacing ratio $r$ should be taken with caution. What we have said is certainly true for the system sizes we can numerically address, but there might be unexpected developments for larger system sizes. For instance, the slight increase with $L$ that one can see between $J\sim 0.1$ and $J\sim 0.4$ in Fig.~\ref{r_factor:fig} could lead to a convergence to the Wigner-Dyson value in the thermodynamic limit. A similar very slight increase with $L$ can be found in the average level spacing ratio resolved in energy, as one can see in~\cite{lauchli2008spreading}. What we can say for sure is that at finite system sizes there is a clear distinction in the behaviour of $r$ at large $J$ where it attains the Wigner-Dyson value and at small $J$ where it does not; the two regimes seem quite robust when one varies the system size and they seem to have no relation with the multiplet structure of the spectrum. 
{\section{Other probes of ergodicity} \label{other}
In this section we focus on two probes of ergodicity. The first one concerns the expectation on the eigenstates of a local intensive operator, the correlation, which is defined as~\cite{PhysRevLett.105.250401}
\begin{equation}
 \hat{\mathcal{G}}_1\equiv\frac{1}{L}\sum_{j=1}^L\left(\hat{a}_j^\dagger\hat{a}_{j+1}+{\rm H.~c.}\right)
\end{equation}
with periodic boundary conditions. This operator is strictly related to the hopping part of the Hamiltonian Eq.~\eqref{Hamour:eqn}. Taking inspiration from~\cite{PhysRevLett.105.250401} we study the properties of the eigenstate expectations of this operator
\begin{equation}
  (\mathcal{G}_1)_\alpha\equiv\braket{\varphi_\alpha|\hat{\mathcal{G}}_1|\varphi_\alpha}\,.
\end{equation}
If there is ergodicity and ETH, all the $(\mathcal{G}_1)_\alpha$ must be equal to the microcanonical value at energy $E_\alpha$, up to fluctuations vanishing in the thermodynamic limit~\cite{Deutsch_PRA91,Sred_PRE94,Rigol_Nat}. This implies that, restricting the $(\mathcal{G}_1)_\alpha$ to some small energy shell, their distribution should become more and more narrow as the system size is increased (ideally it should tend to a delta function in the thermodynamic limit). To probe when there is this property, we fix the energy shell around the value corresponding to the maximum of the entropy, as we have done in Eq~\eqref{S:eqn}: We restrict to the $(\mathcal{G}_1)_\alpha$ for which $E_\alpha\in[E_{\alpha^*}-fD/2,E_{\alpha^*}+fD/2]$. We use the definition of the energy-shell average used in Eq.~\eqref{S:eqn} 
\begin{equation}\label{ave:eqn}
  \mean{g(\cdots)}_{\rm Shell}\equiv\frac{1}{\mathcal{N}}\sum_{\alpha\,{\rm s.t.}\,E_\alpha\in[E_{\alpha^*}-fD/2,E_{\alpha^*}+fD/2]}g((\cdots)_\alpha)\,,
\end{equation}
(where $g(\cdots)$ is any real function) and evaluate the broadness of the distribution considering two quantities, the mean square deviation
\begin{equation}\label{del1:eqn}
\Delta_{\rm Shell}(\mathcal{G}_1) = \sqrt{\mean{\mathcal{G}_1^2}_{\rm Shell}-\mean{\mathcal{G}_1}_{\rm Shell}^2}
\end{equation}
and the $\delta$ value introduced in~\cite{Michele_arxiv}
\begin{equation}\label{del2:eqn}
 \delta_{\mathcal G}=\mean{|\mathcal{G}_1|}_{\rm Shell}-\exp\left(\mean{\log|\mathcal{G}_1|}_{\rm Shell}\right)\,.
\end{equation}
Both these quantities decrease with $L$ if the distribution of $(G_1)_\alpha$ over the chosen energy shell shrinks when the system size increases. This allows to quantitatively probe if there is a shrinking of the distribution of $(\mathcal{G}_1)_\alpha$ towards ergodicity for larger $L$. This aspect was only probed qualitatively in~\cite{PhysRevLett.105.250401}. We show numerical results for these quantities in Fig.~\ref{shrink:fig}, where we plot them versus $J$ for different values of $L$.
\begin{figure}
  \begin{center}
   \begin{tabular}{c}
     \includegraphics[width=7.5cm]{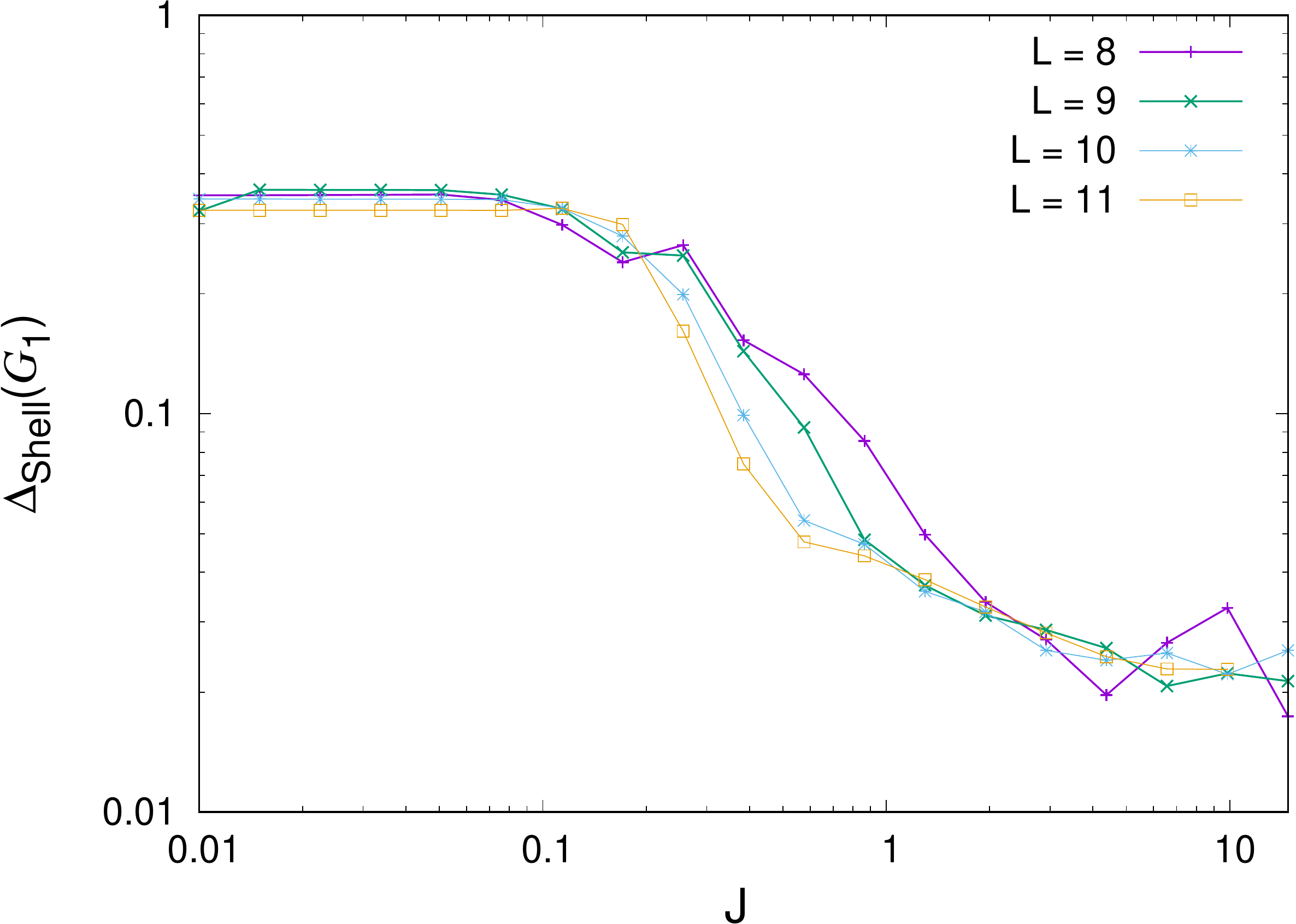}\\
     \includegraphics[width=7.5cm]{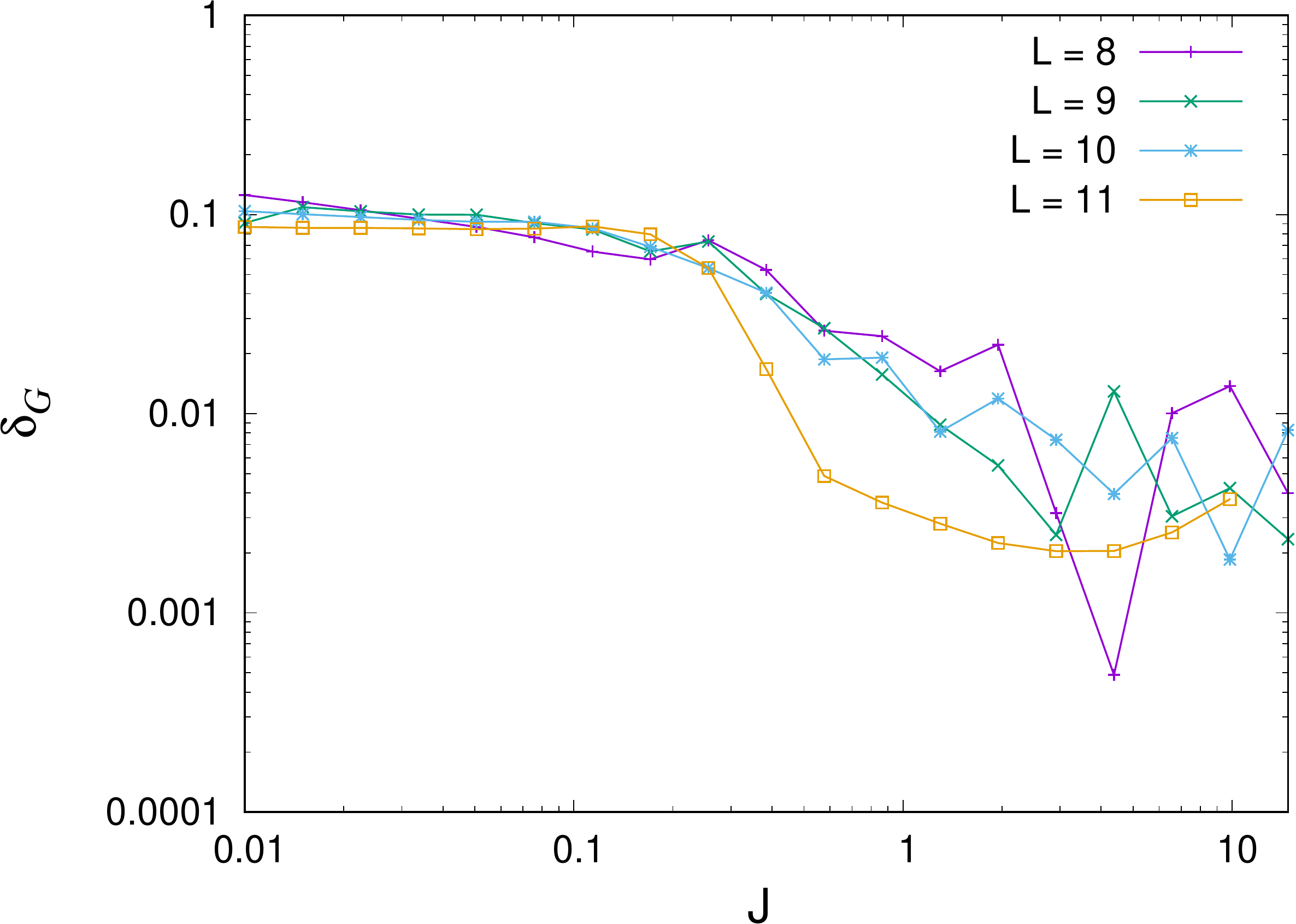}\\
   \end{tabular}
  \end{center}
 \caption{The quantities in Eqs.~\eqref{del1:eqn} (uper panel) and~\eqref{del2:eqn} (lower panel). They quantify the broadness of the distribution of $(\mathcal{G}_1)_\alpha$ over the energy shell. We consider a energy shell with $f=0.01$.}
    \label{shrink:fig}
\end{figure}
We see that the curves for $\Delta_{\rm Shell}(\mathcal{G}_1)$ (upper panel) do not change very much as $L$ increases, nevertheless they have a larger value for small $J$ and a smaller value for large $J$ and they cross around $J=0.2$. Moreover, the curves around the crossing become sharper and sharper in slope as the system size increases, suggesting that something happens there. The same thing can be observed for $\delta_{\mathcal G}$ when $L\leq 10$. Remarkably, for $L=11$ and $J>0.2$, the value of $\delta_{\mathcal G}$ significantly decreases compared to smaller $L$. On the opposite, for $J\leq 0.2$, the value of $\delta_{\mathcal G}$ for $L=11$ stays more or less the same compared to smaller $L$. This behaviour suggests that for $J>0.2$ the system tends to become more ergodic as the system size increases, while for $J\leq 0.2$ the system keeps being non thermal and not obeying ETH. Nevertheless the sizes we can attain are too small to get a ultimate conclusion.}

{Another probe for ergodicity we consider here is the Inverse Participation Ratio (IPR)~\cite{Edwards_JPC72,Wegner} of the Hamiltonian eigenstates in the basis $\{\ket{\boldsymbol{n}_S}\}$ of $\mathcal{H}_S(L)$ where $\ket{\nb_S}$ are the symmetrized simultaneous eigenstates of the operators $\hat{n}_j$
\begin{equation}
  {\rm IPR}_{\alpha}\equiv\sum_{\ket{\boldsymbol{n}_S}}|\braket{{\nb}_S|\varphi_\alpha}|^4\,.
\end{equation}
We average this quantity over the energy shell [Eq.~\eqref{ave:eqn}], in order to avoid finite-size effects related to the edges of the spectrum. The IPR is a measure of delocalization and it is useful to probe ergodicity due to the intimate relation between ergodic behaviour and delocalization in the Hilbert space. If the system is fully ergodic and the eigenstates behave as fully random states, they are fully delocalized over the Hilbert space and we should find the scaling $\mean{{\rm IPR}}_{\rm Shell}\sim 1/\dim\mathcal{H}_S(L)$; on the opposite if the system is many-body localized there can be no scaling (see for instance~\cite{ponte2015periodically}). An anomalous power-law scaling of the form $\mean{{\rm IPR}}_{\rm Shell}\sim 1/\dim\mathcal{H}_S(L)^\gamma$ with $\gamma<1$ marks the existence of a non-ergodic behaviour. There are many examples of this behaviour in the literature, and they can correspond to multifractal behaviour~\cite{PhysRevB.96.104201,PhysRevB.84.134209,PhysRevLett.79.1913} or many-body localization~\cite{Alet1}. We remark that extended non-ergodic phases can have an average level-spacing ratio near to the Wigner-Dyson one~\cite{pinotto,Notarnicola_2020}. We show our numerical results in Fig.~\ref{IPR:fig}. We see that our results are consistent with an anomalous power-law scaling of the form $\log\mean{{\rm IPR}}_{\rm Shell}=A_{IPR}-\gamma \log \dim\mathcal{H}_S(L)$ (inset). Performing the linear fit of this curve, we can evaluate the slope $-\gamma$ and we plot it in the main panel of Fig.~\ref{IPR:fig}. We plot for comparison also $-\gamma_{\rm log}$, the slope of the linear fit of $\mean{\log{\rm IPR}}_{\rm Shell}$ versus $\log \dim\mathcal{H}_S(L)$.
\begin{figure}
  \begin{center}
   \begin{tabular}{c}
     \includegraphics[width=7.5cm]{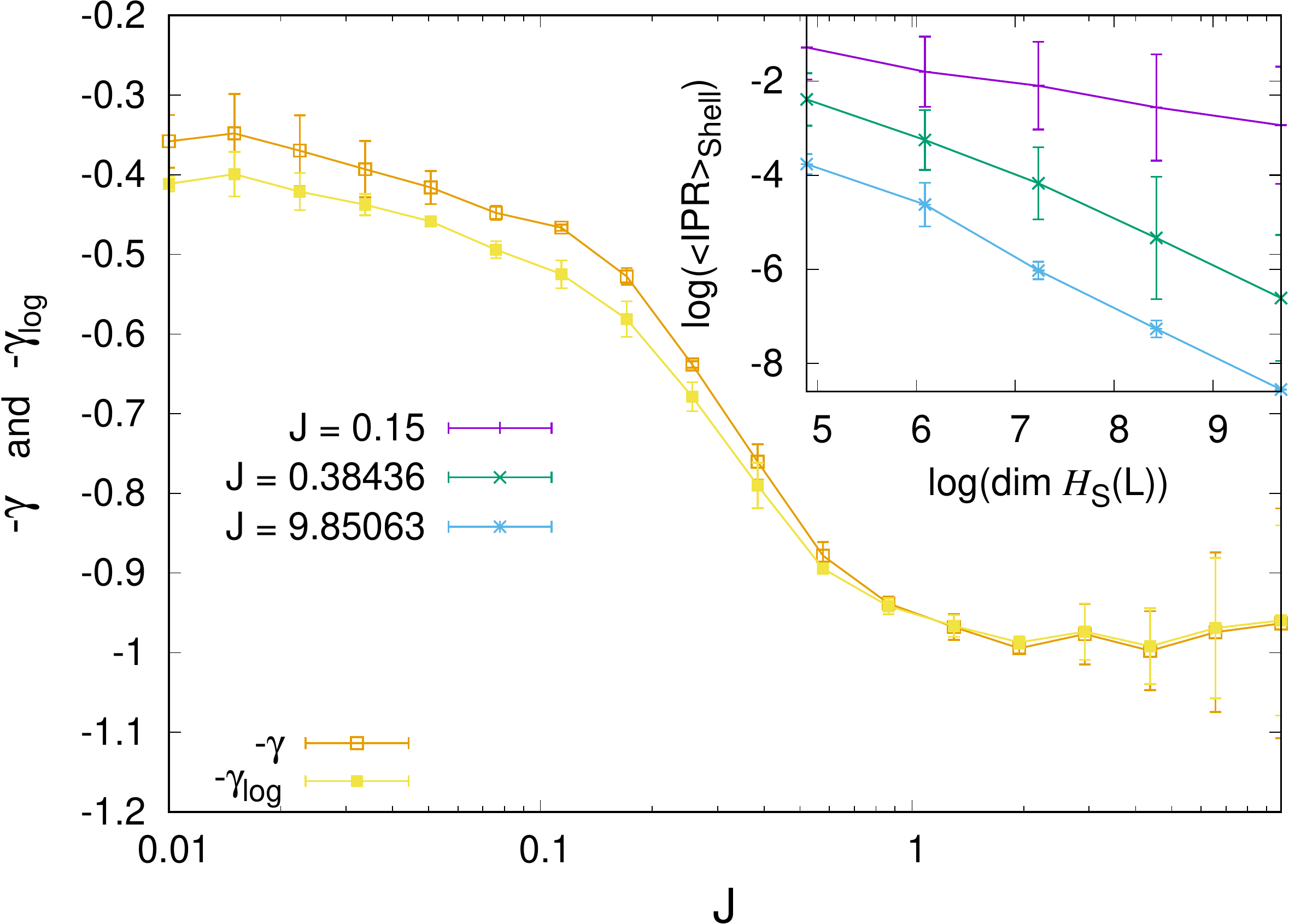}\\
   \end{tabular}
  \end{center}
 \caption{(Inset) $\log\mean{{\rm IPR}}_{\rm Shell}$ versus $\log \dim\mathcal{H}_S(L)$ (the errorbars are given by the mean square deviation evaluated for the distribution of
the ${\rm IPR}_{\alpha}$ in the energy shell). Notice the linear decay. (Main panel) Slope $-\gamma$ of the linear decay of $\log\mean{{\rm IPR}}_{\rm Shell}$ versus $J$ and slope $-\gamma_{\rm log}$ of the linear decay of $\mean{\log{\rm IPR}}_{\rm Shell}$ versus $J$ (the slopes are obtained by means of a linear least-square fit). The slope touches the full ergodic value for $J\sim 1$. We consider an energy shell with $f=0.1$.}
    \label{IPR:fig}
\end{figure}

We see that $\gamma$ is always nonvanishing. This means that there is some form of delocalization, consistently with the results we have got for the entanglement entropy in Sec.~\ref{sec:entro}. Moreover we see that $\gamma$ attains the fully-ergodic value $1$ only for $J\sim 1$. For smaller values of $J$ the system is non-ergodic. Moreover, for $J\in[0.03,0.2]$ we see that $\gamma\neq\gamma_{\rm log}$ (considering also the errorbars). This is sufficient to say that in this interval of $J$ the IPR distribution is broad, opposite to the narrow one valid in the ergodic case~\cite{Notarnicola_2020} (at least for the sizes we can numerically reach). We note that the value $J\sim 1$ where $\gamma$ attains the ergodic value is somewhat larger than the value where the average level spacing ratio attains the Wigner-Dyson value in Fig.~\ref{r_factor:fig}. This is consistent with other findings in the literature~\cite{pinotto,Notarnicola_2020}.  We are not able to say if this is a finite-size effect and futher research is needed.}
{\section{Dynamics of the Imbalance}} 
\label{imbdyn:sec}

\begin{figure}
  \begin{center}
   \begin{tabular}{c}
     \includegraphics[width=8cm]{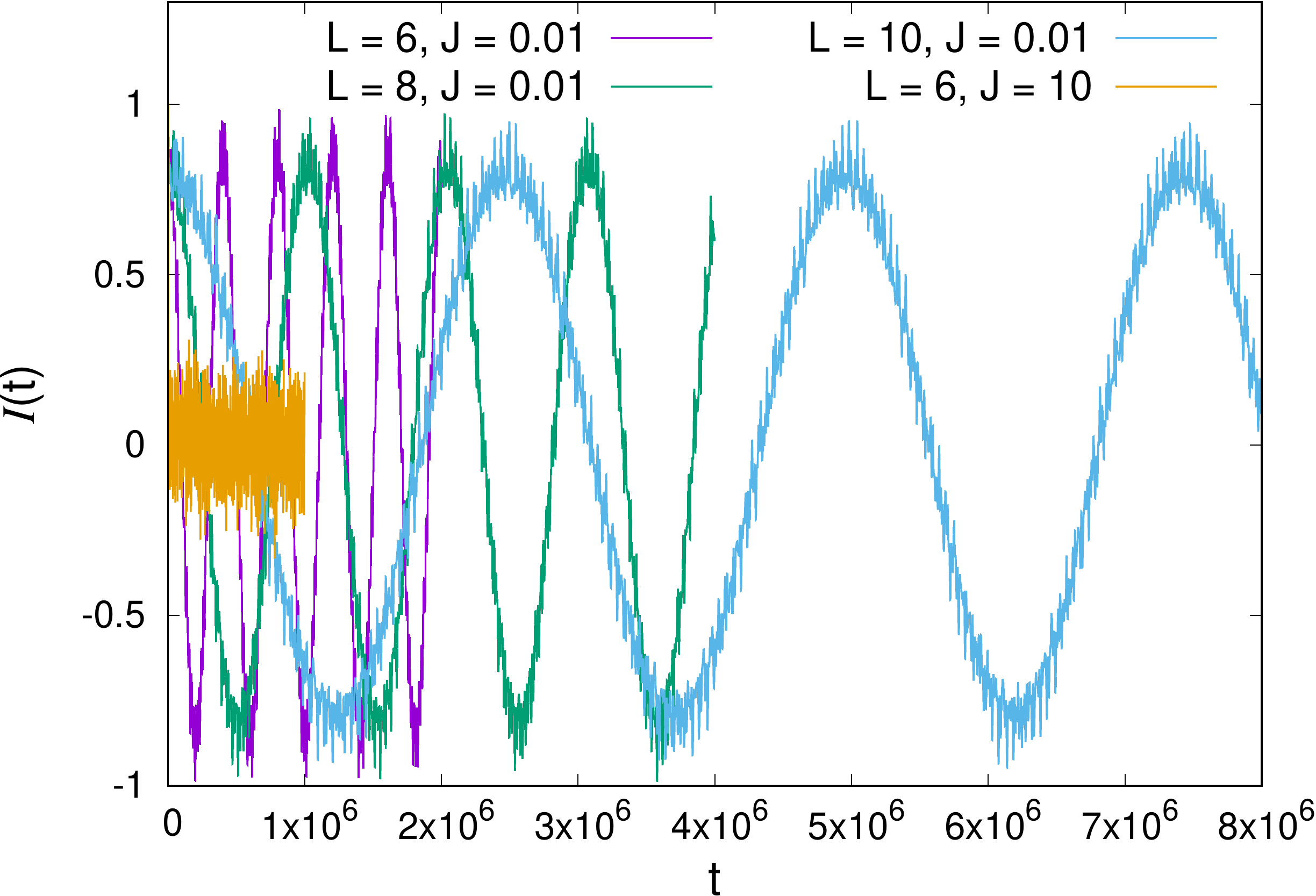}\\
     \includegraphics[width=8cm]{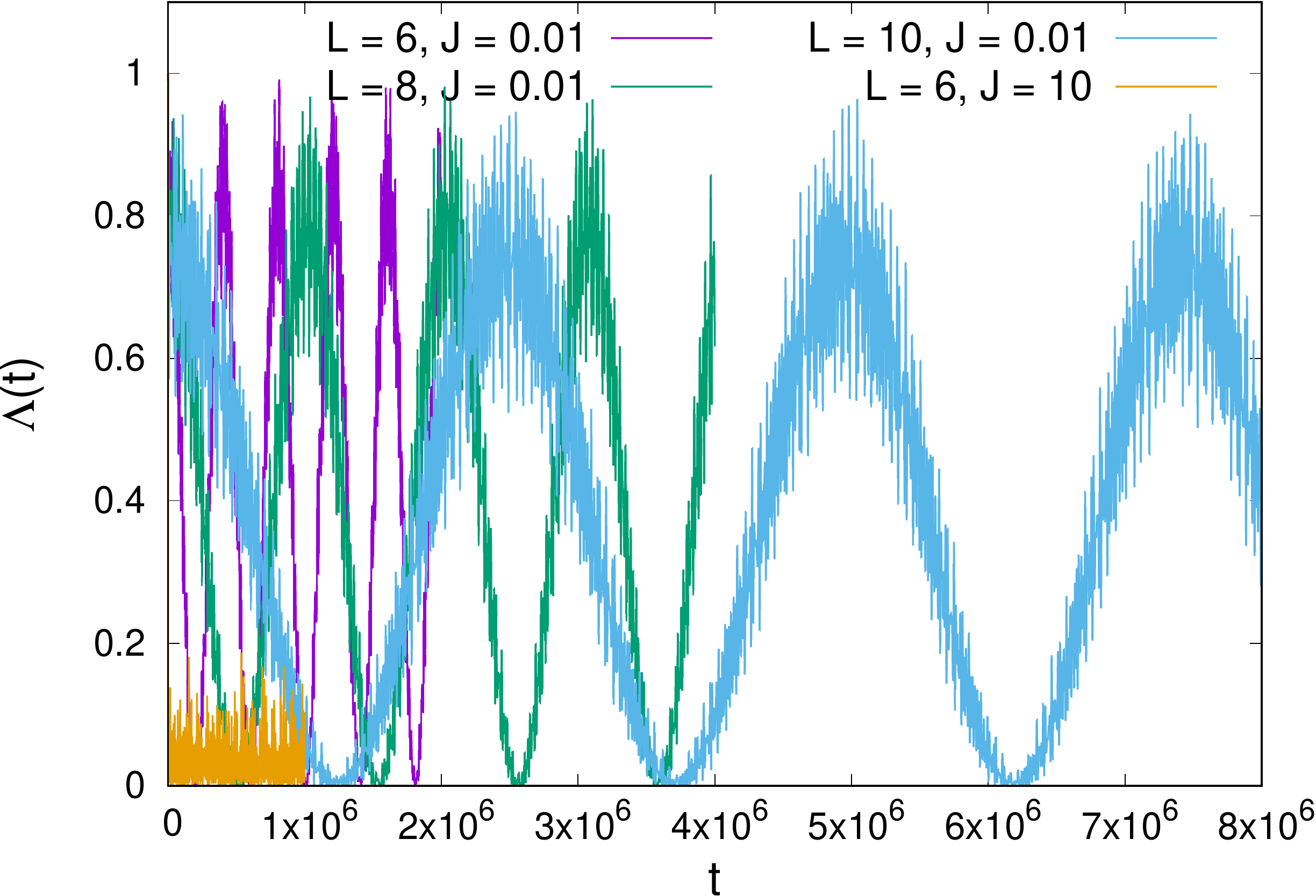}
   \end{tabular}
  \end{center}
 \caption{(Upper panel) Imbalance versus time for different values of $J$ and $L$. (Lower panel) Corresponding time evolution of the overlap. 
               Results obtained via exact diagonalization.}
    \label{imbopt:fig}
\end{figure}

We start considering some instances of evolution of the imbalance $\mathcal{I}(t)=\bra{\psi(t)}\hat{\mathcal{I}}\ket{\psi(t)}$ with $\hat{\mathcal{I}}$ 
defined in Er.~\eqref{I:eqn}. We consider also the evolution of the overlap {(known also as survival probability~\cite{cruz2020quantum})}.
\begin{equation}
  \Lambda(t) = |\bra{\psi_{02}}\left.\psi(t)\right\rangle|^2\,.
\end{equation}
in Fig.~\ref{imbopt:fig}  we plot some time traces of the imbalance (upper panel) and of the overlap  (lower panel). For $J=0.01$ we qualitatively 
see that there are large-amplitude long-period regular oscillations with a period strongly increasing with the system size (we will see that the increase 
is exponential). On the opposite, for $J=10$ there are very small irregular oscillations resembling a random noise (as appropriate for local thermalization:
As we have seen above the system obeys ETH in this regime). 

Indeed, for different values of $J$, we can see two very different regimes. In order to better understand them and their connection with the crossover to thermalization we have discussed above, let us move to a more quantitative analysis of the 
amplitude and the period of the imbalance oscillations. We quantify the amplitude of the oscillations by means of the infinite-time fluctuations of the imbalance 
\begin{equation} \label{deltaI:eqn}
  \Delta \mathcal{I}^2=\overline{\mathcal{I}^2}-\overline{\mathcal{I}}^2
\end{equation}
and of the infinite-time average of the overlap $\overline{\Lambda}$ [see definition in Eq.~\eqref{medo:eqn}]. 
If we replace in these equations the expression of the time-evolving state in terms of eigenstates and eigenvalues of the Hamiltonian
\begin{align}
  \ket{\psi(t)}&=\sum_\alpha R_\alpha \nep^{-iE_\alpha t}\ket{\varphi_\alpha}\quad{\rm with}\nonumber\\
  R_\alpha&=\bra{\varphi_\alpha}\left.\psi_{02}\right\rangle
\end{align}
and assume that there are no degeneracies in the spectrum, we easily obtain the formulae
\begin{align} \label{sommalia:eqn}
  \Delta \mathcal{I}^2&=\sum_{\alpha,\,\gamma}|R_\alpha|^2|R_\gamma|^2|\mathcal{I}_{\alpha\,\gamma}|^2\\
  \overline{\Lambda}&=\sum_\alpha|R_\alpha|^4
\end{align}
where we have defined $\mathcal{I}_{\alpha\,\gamma}\equiv\bra{\varphi_\alpha}\hat{\mathcal{I}}\ket{\varphi_\gamma}$. Notice that $\overline{\Lambda}$ is 
the IPR of the initial state $\ket{\psi_{02}}$ in the basis of the Hamiltonian eigenstates.  

\begin{figure}
  \begin{center}
   \begin{tabular}{c}
     \includegraphics[width=8cm]{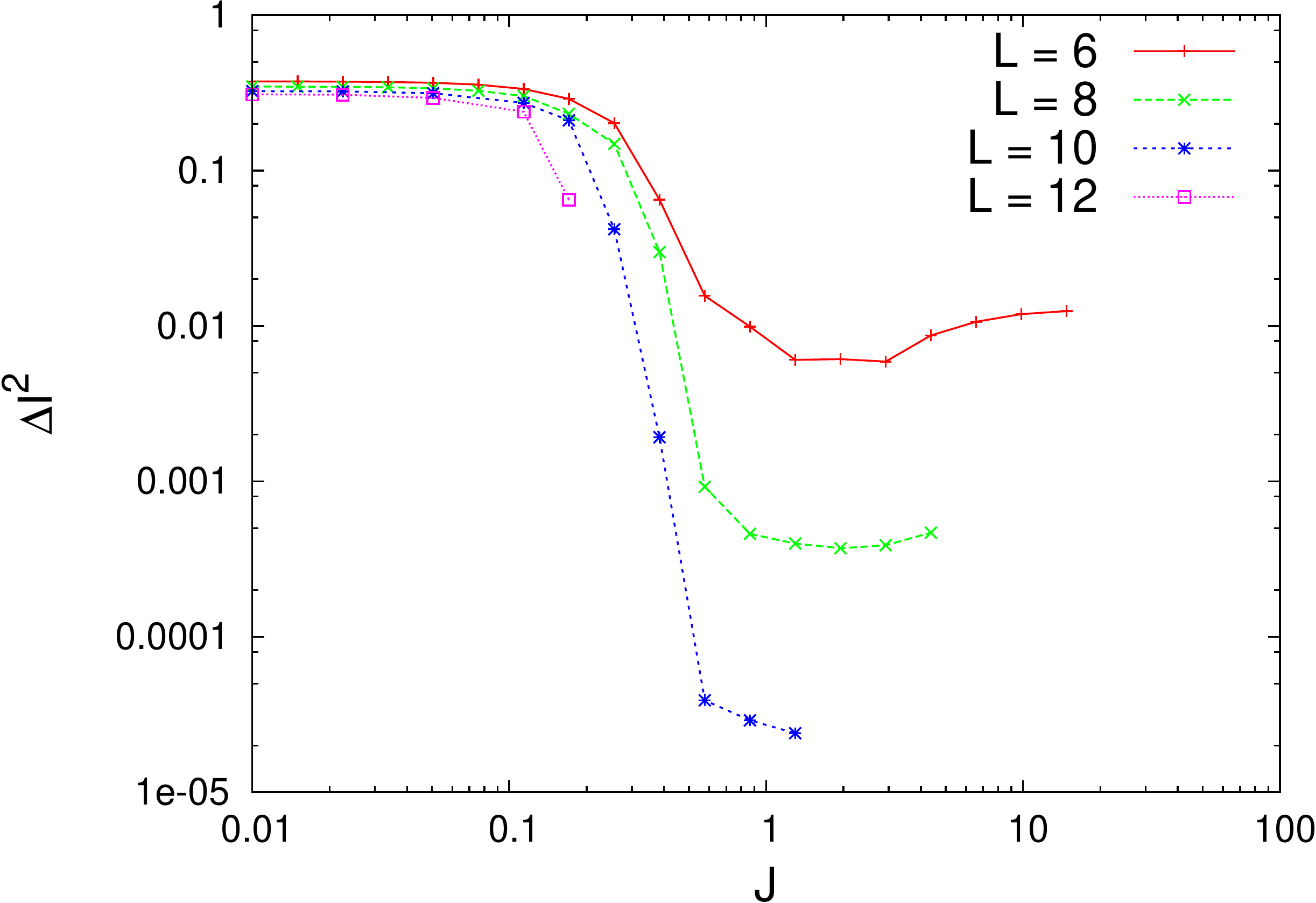}\\
     \includegraphics[width=8cm]{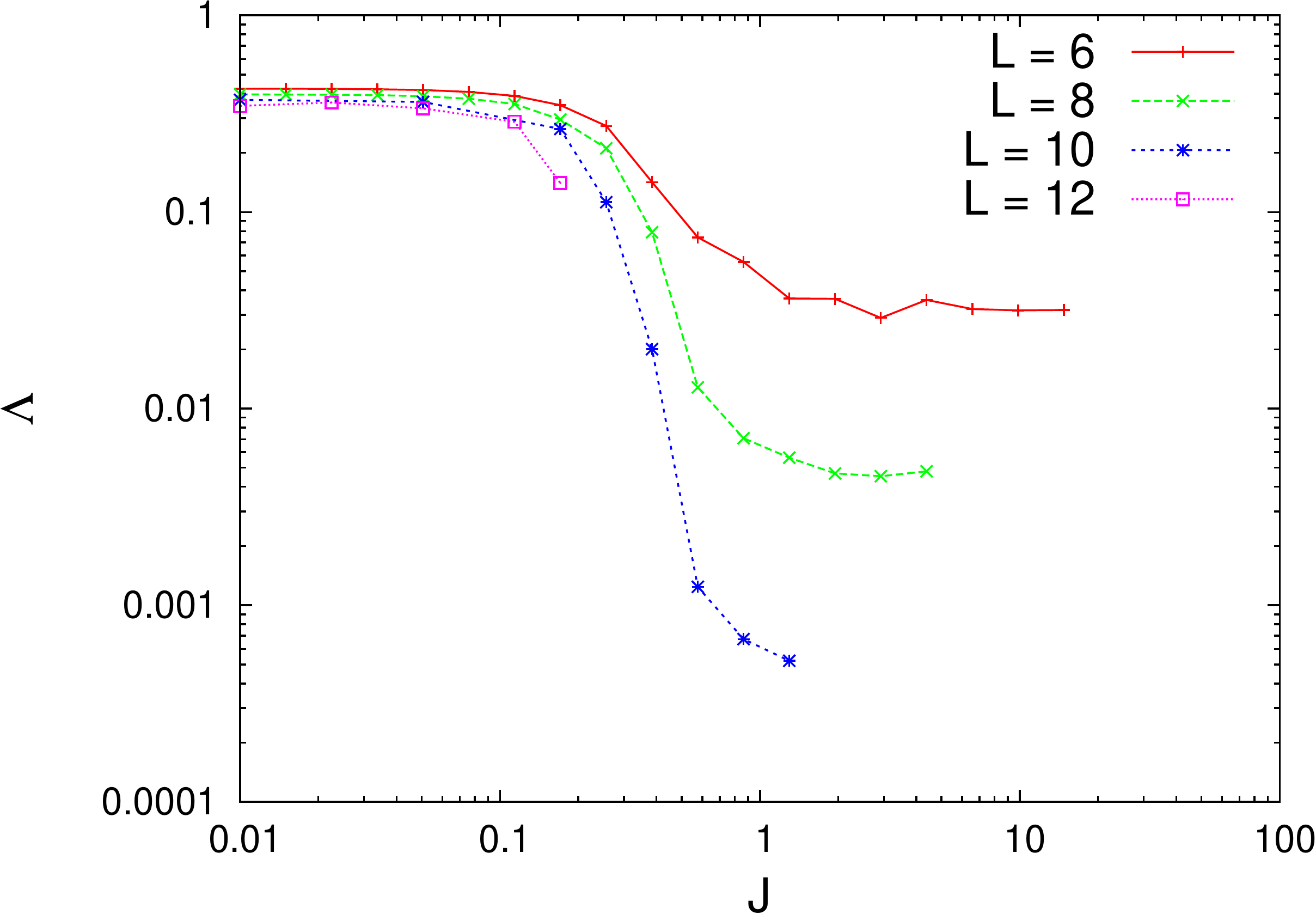}
     \pgfputat{\pgfxy(-8.2,3.25)}{\pgfbox[left,top]{$|$}}
   \end{tabular}
  \end{center}
 \caption{ (Upper panel) Time fluctuations of the imbalance versus $J$. (Lower panel) Average overlap versus $J$. (For $L=10,\,12$ averages are 
 performed over a finite time. For $L=10$, time step $Dt=10$, $t_f\ge 4.4\times 10^{6}$; for $L=10$, time step $Dt=1$, $t_f\ge 2\times 10^{6}$. 
 For $L\leq 10$ we use exact diagonalization and for $L=12$ Krylov technique with truncation.)}
    \label{timedie:fig}
\end{figure}

We show $\Delta \mathcal{I}^2$ versus $J$ in Fig.~\ref{timedie:fig} upper panel and $\overline{\Lambda}$ versus $J$ in Fig.~\ref{timedie:fig} lower panel. 
We consider different system sizes and we see that  the curves drop down towards a plateau. In this plateau we see that $\Delta \mathcal{I}^2$
and $\overline{\Lambda}$  decrease by an order of magnitude when the size is increased by two sites. Moreover, the curves drop to the plateau at a value of 
$J$ which steadily decreases as $L$ increases. We remark that the curves decrease with increasing $L$ also for small $J$ (although in a much slighter way).
%

\begin{figure}
  \begin{center}
   \begin{tabular}{c}
     \includegraphics[width=8cm]{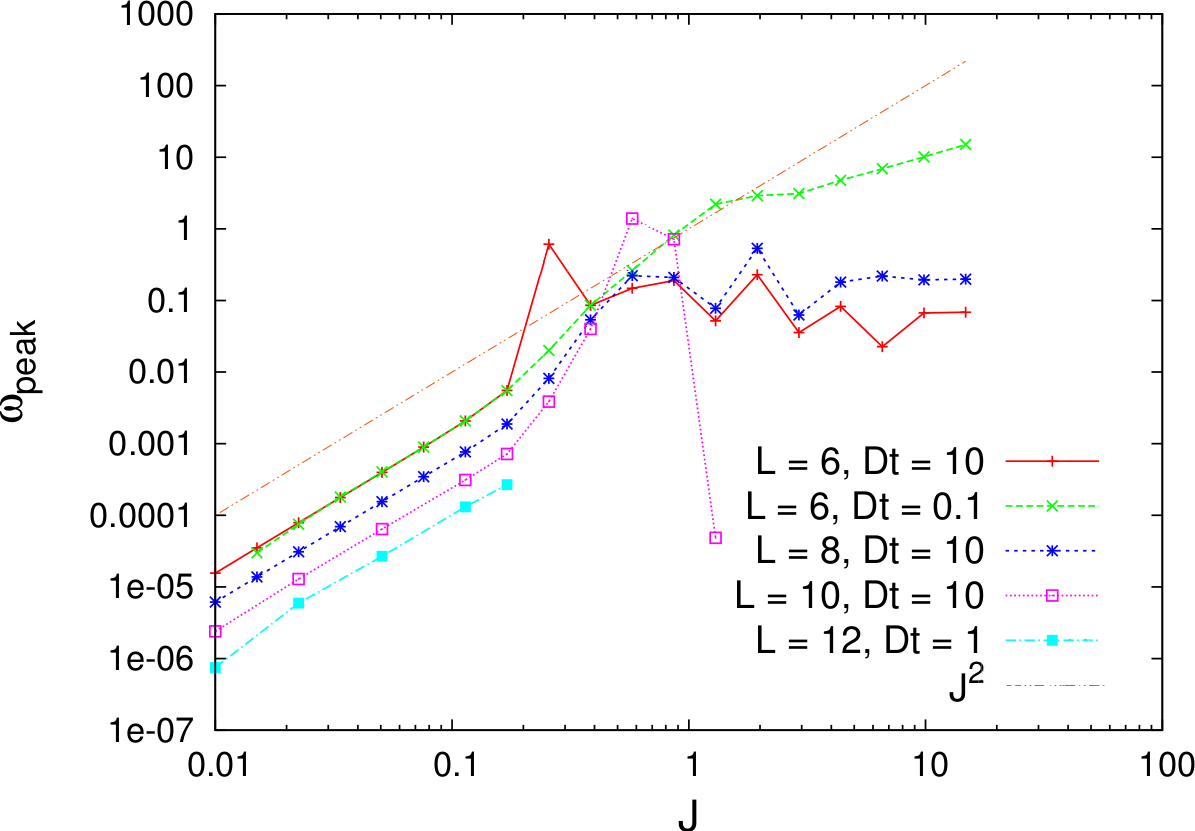}\\
     \includegraphics[width=8cm]{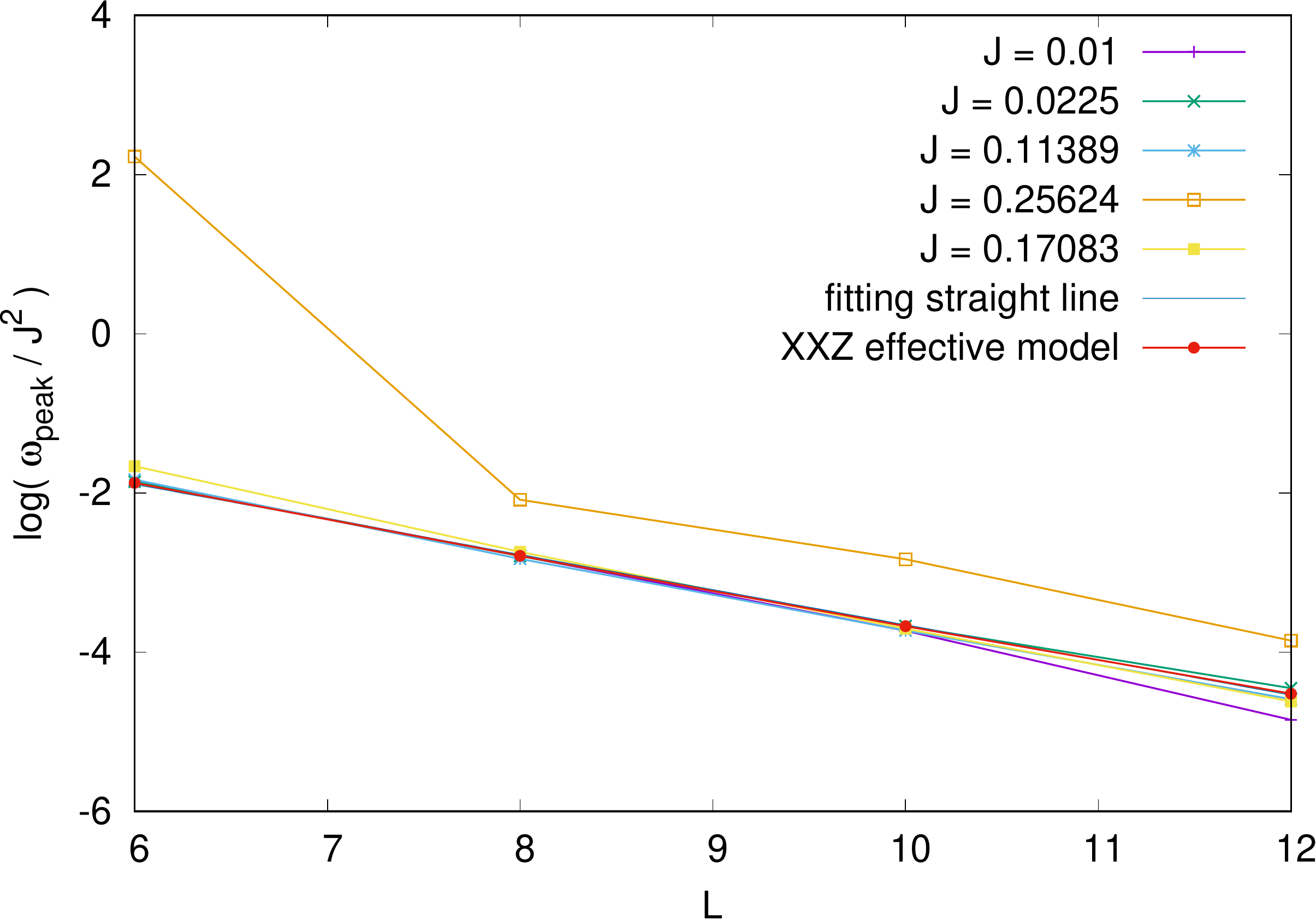}
%
   \end{tabular}
  \end{center}
 \caption{ (Upper panel) Frequency of the peak in $|\mathcal{I}(\omega)|^2$ (it coincides with the frequency of the peak in $|\Lambda(\omega)|^2$). Number of samplings $N_t=4194304$ for $L=6,\,8$ ($Dt=10$); $N_t\leq2621440$ for $L=10$ (time step $Dt=10$); $N_t\leq8388608$ for $L=12$ ($Dt=1$). (Lower panel) $\log(\omega_{\rm peak}/J^2)$ versus $L$. For $J\lesssim 0.17$ we can verify the validity of the formula Eq.~\eqref{omegaform:eqn} which we use to perform the fit of the curve at $J=0.025$. We plot also the result coming from the simulation of the XXZ effective model Eq.~\eqref{effXXZ:eqn}.}
    \label{imbpeak:fig}
\end{figure}

We can quantitatively estimate also the period of these oscillations by looking at the Fourier transform of the imbalance and the overlap. {Considering for instance the case of the imbalance, we }
define the Fourier transform as 
\begin{equation}
  \mathcal{I}(\omega)=\int_0^{t_f}\mathcal{I}(t)\nep^{i\omega t}\ud t\,,
\end{equation}
where $t_f$ is the total evolution time. Performing the Fourier transform for $\mathcal{I}(t)$ and $\Lambda(t)$, we consider the main peak at 
non-vanishing frequency~\cite{nota_freq} {corresponding to the frequency of the {dominant imbalance oscillations}. }
We find that the {peak frequencies} of the two quantities coincide, we call them $\omega_{\rm peak}$ and {we plot $\omega_{\rm peak}$ versus $J$} in the upper panel of Fig.~\ref{imbpeak:fig}.  
{We can see that $\omega_{\rm peak}$ increases quadratically with 
$J$ up to around $J\sim 0.17$ and in the lower panel of Fig.~\ref{imbpeak:fig} we can see that it decreases exponentially with the system size.} More precisely, in this interval of $J$, $\omega_{\rm peak}$ is 
well described by the formula
\begin{equation} \label{omegaform:eqn}
  \omega_{\rm peak}=BJ^2\nep^{-\alpha L}\,.
\end{equation}
We explicitly show this in the lower panel of Fig.~\ref{imbpeak:fig} where we plot $\log\left(\omega_{\rm peak}/J^2\right)$ versus $L$. We see 
that the curves are actually straight lines and the curves for different values of $J$ overlap when $J\lesssim 0.17$. We also perform a minimum 
square fit of the curve for $J=0.025$ with the formula $\log\left(\omega_{\rm peak}/J^2\right) = \log(B) - \alpha L$. We find a good agreement for 
the fit, as we show in Fig.~\ref{imbpeak:fig}, and we get as parameters $\log(B)=0.7\pm 0.1$, $\alpha=0.43\pm 0.01$. 

{In the plot we show also a curve for a {reduced model indicated as  ``XXZ effective model''.} We can use second-order perturbation theory in $J/U$ 
to interpret the dynamics at small $J$ ($J\ll U$) as the dynamics of an effective XXZ model. We describe the details of this model in the next section, for now we can notice the perfect quantitative prediction of $\omega_{\rm peak}$ given by this model, confirmed by the results of the fit of the corresponding curve with Eq.~\eqref{omegaform:eqn}, namely $\log(B_{\rm XXZ})= 0.75\pm 0.05$ and $\alpha_{\rm XXZ}=0.441\pm 0.005$ which compares well with the imbalance results stated above.}


{

\subsection{Interpretation of the imbalance oscillations}

As we show in detail in Appendix~\ref{perturb:sec}, for $J\ll 1$, whenever the system size of the system is so small that the gaps between the different quasidegenerate multiplets are still open (see Sec.~\ref{dos:sec}), we can use second-order perturbation theory in $J/U$ to understand the level structure in the different multiplets. In particular, if we focus on the multiplet to which our initial state $\ket{\psi_{02}}$ belongs, we find that the dynamics is induced by the effective Hamiltonian
\begin{equation} \label{effXXZ1:eqn}
  H^{(2)}_{eff} = \sum_{j=1}^L\left[\frac{\tilde{t}}{4}\left(\hat{\sigma}_j^+\hat{\sigma}_{j+1}^-+{\rm H.~c.}\right)-\frac{\tilde{V}}{4}\hat{\sigma}_j^z\hat{\sigma}_{j+1}^z-\frac{\tilde{V}+\tilde{\mu}}{2}\hat{\sigma}_j^z\right]\,
\end{equation}
with $\tilde{t} = (1/2)J^2/U$, $\tilde V = 4 J^2/U$ and $\tilde \mu = -J^2/U$. This is a XXZ Hamiltonian and $\ket{\psi_{02}}$ corresponds in this representation to the N\'eel state $\ket{\uparrow\downarrow\uparrow\downarrow\cdots}$.
The remarkable thing about the XXZ Hamiltonian is that it has just one single pair of eigenstates (call them $\ket{\varphi_\pm}$) which break the $\mathbb{Z}_2$ symmetry in the thermodynamic limit. As we show in detail in Appendix~\ref{breaking}, these two states are separated by a splitting $\Delta$ exponentially small in the system size and proportional to $J^2$. {This is the key leading to Rabi oscillations of a quantity called staggered magnetization and defined as~\cite{notev}
\begin{equation}
  m_S^z(t)=\frac{1}{L}\sum_{j=1}^L(-1)^{j}\braket{\psi(t)|\hat{\sigma}_j^z|\psi(t)}\,.
\end{equation}
These oscillations are very similar to the imbalance ones and we show some examples of them in Fig.~\ref{tony_stag:fig}(upper panel) Their frequency versus $L$ is reported in the lower panel of Fig.~\ref{imbpeak:fig} (labeled as ``XXZ effective model'') and we can see that they are in quantitative agreement with the frequency of the imbalance oscillations. As we have explained, also the fit with Eq.~\eqref{omegaform:eqn} gives consistent results. Moreover, as we clarify in Fig.~\ref{tony_stag:fig}(lower panel) of Appendix~\ref{breaking}, they Rabi oscillations give a good prediction also for the amplitude of the imbalance oscillations.}

{The N\'eel state has a huge overlap with the symmetry-breaking doublet and that's why the Rabi oscillations are so clearly visible. We can see the existence of this doublet and the fact that the initial state $\ket{\psi_{02}}$ has square overlap mainly with the two states in this doublet also directly in the Bose-Hubbard model. We show some example of square overlap $|\braket{\phi_\alpha|\psi_{02}}|^2$ versus $E_\alpha/L$ in Fig.~\ref{sqover:fig}. Here we can clearly see that there are two states where $|\braket{\phi_\alpha|\psi_{02}}|^2\sim 0.5$ at an energy near 1/2, the energy expectation of $\ket{\psi_{02}}$. We have also checked that the splitting $\Delta$ between these two states coincides with the frequency $\omega_{\rm peak}$ shown in Fig.~\ref{imbpeak:fig} and we show some examples of this fact in table~\ref{tabula}. So the doublet states are the ones responsible for the imbalance oscillations. One can see oscillations with a frequency exponentially small in the system size just because the splitting between $\ket{\varphi_+}$ and $\ket{\varphi_-}$ is exponentially small in the system size. We emphasize again the quantitative correctness of the prediction of the XXZ effective model for the imbalance oscillation frequency, as we can see in the lower panel of Fig.~\ref{imbpeak:fig}. }
\begin{figure}
  \begin{center}
   \begin{tabular}{c}
     \includegraphics[width=8cm]{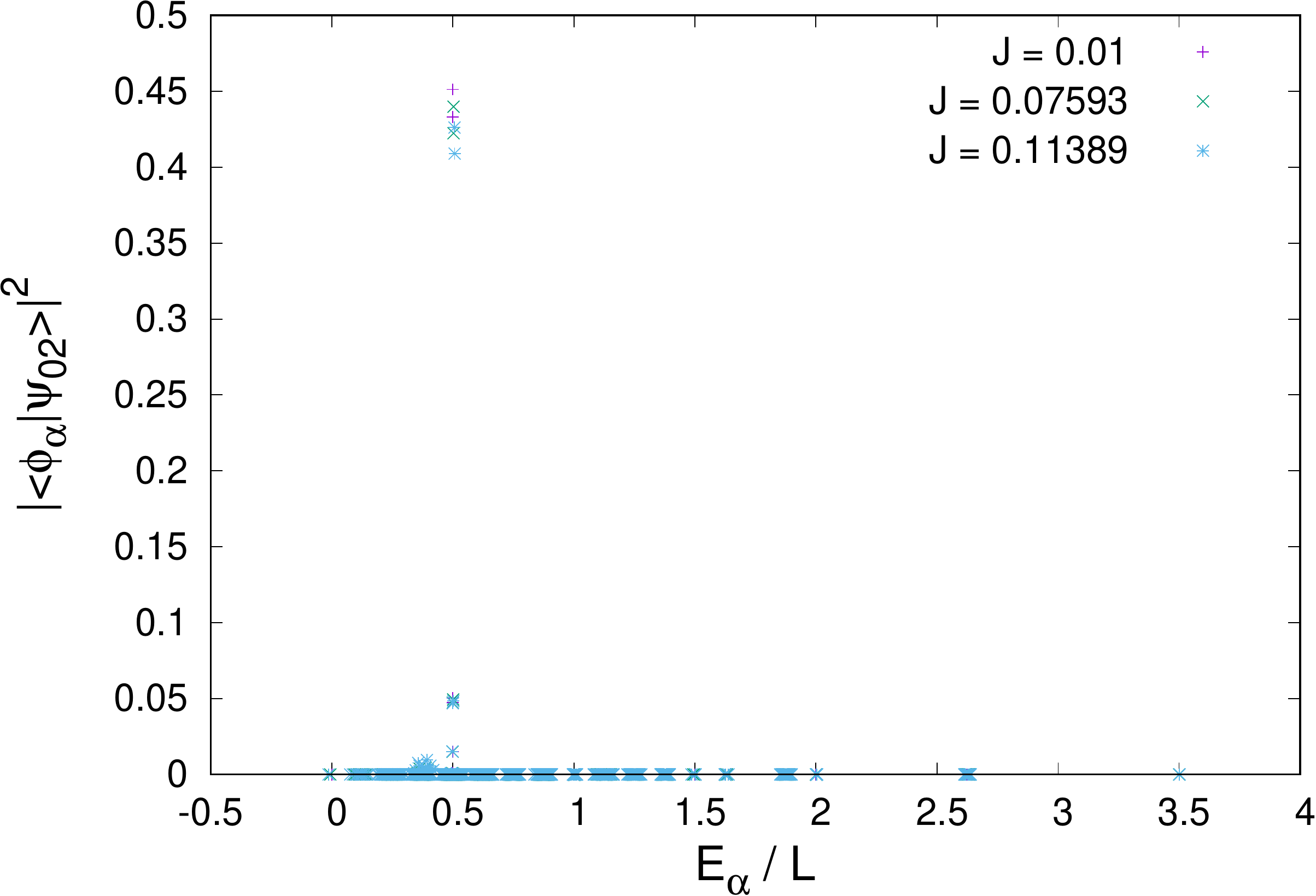}
   \end{tabular}
  \end{center}
 \caption{Square overlap of the initial state versus Hamiltonian eigenstates $|\braket{\varphi_\alpha|\psi_{02}}|^2$ versus the corresponding energy eigenvalue $E_\alpha$ divided by $L$. We take $L=8$ and different values of $J$. We see that the overlap is non-vanishing mainly for two quasidegenerate states. As we show in table~\ref{tabula}, the splitting of these two states is exponentially small in the system size and coincides with the corresponding Rabi-oscillation frequency shown in Fig.~\ref{imbpeak:fig}.}
    \label{sqover:fig}
\end{figure}
\begin{table}
\begin{tabular}{|c|c|c||c|c|}
\hline
$L$&\multicolumn{2}{c||}{$J=0.01$}&\multicolumn{2}{c|}{$J=0.11389$}\\
\hline
&$\Delta${}&$\omega_{\rm peak}$&$\Delta$&$\omega_{\rm peak}$\\
6&$1.5\cdot 10^{-5}$&$1.5\cdot 10^{-5}$&$2.08\cdot 10^{-3}$&$2.08\cdot 10^{-3}$\\
8&$6.24\cdot 10^{-6}$&$6.14\cdot 10^{-6}$&$7.68\cdot 10^{-4}$&$7.68\cdot 10^{-4}$\\
10&$2.53\cdot 10^{-6}$&$2.39\cdot 10^{-6}$&$3.12\cdot 10^{-4}$&$3.12\cdot 10^{-4}$\\
%
\hline

\end{tabular}
\caption{Comparison of the splitting $\Delta$ of the two maximum-overlap states of Fig.~\ref{sqover:fig} with the corresponding value of $\omega_{\rm peak}$ (lower panel of Fig.~\ref{imbpeak:fig}) for two values of $J$. The agreement is excellent, up to errors due to the finite time over which the Fourier transform is performed. As the lower panel of Fig.~\ref{imbpeak:fig} shows, both the quantities decay exponentially in $L$, as confirmed by the theoretical analysis of Appendixes~\ref{perturb:sec} and \ref{breaking}.}
 \label{tabula}
\end{table}

As we show in detail in Appendix~\ref{breaking}, the overlap of $\ket{\psi_{02}}$ with the symmetry-breaking doublet tends to 0 in the thermodynamic limit and so the Rabi oscillations disappear in that limit.
This can be also seen in a simpler way, considering that the gaps between the different multiplets tend to vanish in the thermodynamic limit and the perturbation theory leading to Eq.~\eqref{effXXZ1:eqn} is no more valid for $L$ larger than some threshold. In Fig.~\ref{bandgap:fig} we show the closing with the increasing system size for a gap directly relevant for our analysis. 

We remark that we can do a similar construction (with similar physical conclusions) also if we initialize our dynamics with the state $\ket{\psi_{0\,2N}}=\ket{0}\otimes\ket{2N}\otimes\ket{0}\otimes\cdots\otimes\ket{0}\otimes\ket{2N}$, as we discuss in Appendix~\ref{larger:sec}.

So, to summarize, Rabi oscillations are a phenomenon related to the behaviour of just two states in the spectrum and disappear in the thermodynamic limit. They have therefore no relation with the crossover to non ergodicity we have studied in Secs.~\ref{sec:entro},~\ref{dos:sec} and~\ref{other}. The latter phenomenon involves all the spectrum {(or at least many non-thermal states)} and we have argued it to be independent of the multiplet structure. On the opposite, the perturbative model describing Rabi oscillations strongly relies on this multiplet structure. Its disappearance at large $L$ leads to the destruction of the Rabi oscillations.


}

\section{Conclusion} \label{conc:sec}
In conclusion we have studied the ergodicity breaking in the clean Bose-Hubbard model. 

We have seen that for small hopping strength $J$ the model breaks ergodicity. We have seen this fact looking at the entanglement entropies of the eigenstates. In the non-ergodic regime the average over the eigenstates of their half-chain entanglement entropy linearly increases with the system size with a slope smaller than the fully-ergodic one. So the system is non-ergodic. The volume-law behaviour of the entanglement entropy is in strong contrast with the area law behaviour seen in many-body localization. This ergodicity breaking is confirmed by the {study of the distribution on the energy shell of the eigenstate expectations of the correlation and by the scaling of the averages of the Inverse Participation Ratio.  We get a further confirmation from the} spectral properties, {as we have seen from the results for the average level spacing ratio.} Most importantly, the ergodicity breaking appears to be independent of the spectrum being organized in quasidegenerate multiplets at small $J$ and $L$. The multiplets are doomed to disappear for the system size beyond some threshold, but the ergodicity breaking possibly survives. {One possibility is that this extended non-thermal regime gives rise to a multifractal phase as it occurs in the Rosenzweig-Porter model~\cite{pinotto} and this point will be object of further studies.} {We remark again that all our results are numerical and obtained for $L\leq 11$, and extrapolations of our scalings to larger system sizes must be taken with due care.}

Then we have moved to study the dynamics of the imbalance. We have found that it oscillates with a period exponentially large in the system size when $J\ll U$, suggesting freezing (and then some sort of {real-space} localization) in the infinite-size limit. We have interpreted this phenomenon through a perturbative theory in $J$. We have constructed an effective XXZ model describing this dynamics. We have found that this effective model shows a doublet breaking the $\mathbb{Z}_2$ symmetry and the states in this doublet have a splitting exponentially small in the system size. The imbalance oscillations with a period exponentially large in the system size were actually the Rabi oscillations in this doublet. Therefore these oscillations are a phenomenon involving just a two-dimensional subspace of the Hilbert space, opposite to the ergodicity breaking which involves all the spectrum. The overlap of the initial state with this doublet vanishes in the thermodynamic limit and so also the imbalance oscillations vanish in this limit. We have argued that releasing the approximations leading to the XXZ effective model, the decay of the imbalance-oscillations amplitude with the system size gets stronger. Beyond some value of $L$ they must disappear because the gaps between different multiplets close up. Therefore, in the thermodynamic limit we expect to see no imbalance oscillations and the apparent glassy behaviour we see occurs only at finite system sizes.

Future research will focus on the study of the relevance of our results for the many body localization occurring in the Josephson-Junction chain~\cite{Pino536}. Naively one could expect some relation because the Bose-Hubbard model at high energies maps to the Josephson junction chain~\cite{FAZIO2001235}. Nevertheless, this mapping is valid at equilibrium while here we are {considering} non-equilibrium properties. Another direction of research will be to inquire if it is possible to construct for the clean Bose-Hubbard chain an extensive number of local integrals of motion, as usually occurs for quantum integrable models~\cite{gogolin2016equilibration,essler2016quench,essler2005one}.
\acknowledgements{We acknowledge useful discussions with B.~Altshuler, M.~Heyl, I. Khaymovich, and V.~Kravtsov {and insightful comments on the manuscript from J.~De La Cruz, F.~Heidrich-Meisner, J. Hirsch, Y.~Huang, S.~Lerma and T.~Prosen}. R.~F. acknowledges partial financial support from the Google Quantum Research Award. A.~R. warmly thanks D.~Rossini for the access to computation time in the GOLDRAKE Cluster.}
\appendix
\section{Derivation of the effective XXZ model via perturbation theory} \label{perturb:sec}
%
%
We apply degenerate perturbation theory in $J$ to Eq.~\eqref{Hamour:eqn} when 
	$J\ll U\simeq 1$. Before we go on with this analysis we have to specify some details about the structure of the spectrum. When $J=0$, the Hamiltonian Eq.~\eqref{Hamour:eqn} behaves as the operator $\hat{V}$, therefore it shows massively degenerate eigenspaces, as elucidated in~\cite{Michele_arxiv}. If we switch on a small value of $J\ll U$, we find a strong mixing inside the eigenspaces, while different eigenspaces interact starting from second perturbative order in $J/U$. As a result the eigenspaces are transformed into multiplets. The different multiplets are separated by gaps of order $U$ while the levels inside each multiplet have a separation order $(J/U)^2$. We can see the existence of these multiplets for small $J$ by looking at the density of states
%
%

%
We show some examples of $\rho(E)$ for different $J$ and different $L$ in Fig.~\ref{rho:fig}. We see that for $J\sim 0.25$ it shows a series of $\delta$ peaks centered around integer values of the energy $E$ (upper left panel of Fig.~\ref{rho:fig}) to a continuum. Each point corresponds to a multiplet. Increasing the system size $L$, at some point mixing occurs also across the degenerate subspaces.  We see this fact in Fig.~\ref{bandgap:fig} where we show the closing with $L$ of the gap between the band at energy around $L/2$ (this is the energy expectation of $\ket{\psi_{02}}$) and the next one. Nevertheless, we see in this figure that for $J$ small enough the gap is open at the system sizes we can access to, and we can exploit this feature of the spectrum in order to perform a perturbative-analysis interpretation.

We are choosing as initial state of our dynamics the imbalanced state $\ket{\psi_{02}}$ [Eq.~\eqref{imbalanced_state:eqn}], so we are interested in studying the energy splitting inside the unperturbed eigenspace with $\hat{V}$-eigenvalue $V= L/2$. ($\hat{V}\equiv\frac{1}{2}\sum_j\hat{n}_{j}(\hat{n}_{j}-1)$ is the interaction part of the Hamiltonian divided by $U$.)
	
	We can show that the eigenvalues have no corrections at first order in $J$, as follows. At first order in $J$ the effective Hamiltonian is given by
	\begin{equation}
	\Braket{a|H_{eff}^{(1)}|b} = -\frac{J}{2}\delta_{V_a,V_b} \braket{a|\hat{\mathcal{G}}_1|b}\,,
	\end{equation}
	where $\delta_{V_a,V_b}$ means that states $\ket{a}$ and $\ket{b}$ are unperturbed eigenstates in the same $\hat{V}$-eigenspace. Given a state $\ket{b}=\ket{n_1 n_2 \cdots}$, we can see that $H_{eff}^{(1)}\ket{b}\neq0$ iff. $\exists$ $j$ s.t. $n_j = n_{j+1}\pm 1$. It immediately follows that the initial state $\ket{\psi_{02}}$ belongs to a subspace where $H_{eff}^{(1)}\equiv 0$, so that there still is no dynamics at first order in perturbation theory.
	
	Non-trivial dynamics starts to appear at second order in perturbation theory. We apply degenerate perturbation theory and consider coupling only within the same $\hat{V}$-eigenspace. The effective Hamiltonian at second order is then
	\begin{equation}
		\Braket{a|H_{eff}^{(2)}|b} = \frac{J^2}{4}\delta_{V_a,V_b} \sum_c \frac{\braket{a|\hat{\mathcal{G}}_1|c}\braket{c|\hat{\mathcal{G}}_1|b}}{V_a - V_c}
	\end{equation}
	So we need to hop two bosons and get back to a state with the same $V=L/2$.
	In order to describe the situation, we consider $\ket{b}=\ket{n_1 n_2 \cdots}$ and focus on a nearest neighbour pair $(i,j)$. We can reach the virtual state $\ket{c}$ with modified occupation of the pair $(i,j)$ given by $n_{i}^{(c)} = n_{i}+1$ and $n_{j}^{(c)}=n_{j}-1$ {(the discussion of the case $n_{i}^{(c)} = n_{i}-1$ and $n_{j}^{(c)}=n_{j}+1$ is essentially the same)}. We now distinguish three cases depending on which sites we act to go from $\ket{c}$ to $\ket{a}$:
	\begin{description}
		\item[(a) $(i,j)$] In this way $n_i\mapsto n_i+2$ and $n_{j}\mapsto n_{j}-2$. We obey the condition $V_a=V_b$ iff. $n_i=n_j-2$. The net result is that we moved two bosons from site $j$ to site $i$.
		\item[(b) $(j,i)$] In this way we go back to the initial state. This will produce a nearest-neighbour coupling term diagonal in the number basis.
		\item[(c) none of the above] Let's call $(l,m)$ the new pair of nearest neighbour sites on which we acted upon. The term thus produced in $H_{eff}^{(2)}$ is highly non-local. $H_{eff}^{(2)}$ can move a boson from $i$ to $j$ and from $l$ to $m$ if the overall eigenvalue of $V$ is conserved.
	\end{description}

	In order to diagonalize the effective Hamiltonian $H_{eff}^{(1)}+ H_{eff}^{(2)}$, 
we focus on the space with eigenvalue $H_{eff}^{(1)}=0$ within the $V=L/2$ $\hat{V}$-eigenspace, as we have discussed above.
	
	In an approximate way, we could say that the subspace $H_{eff}^{(1)}=0$ is given by the linear combination of product states $\ket{a}$ satisfying the condition $H_{eff}^{(1)}\ket{a}=0$. This is not exact, as e.g. $\ket{20012}$ has an overlap with states in the sector $H_{eff}^{(1)}=0$, which are however not product states. We will show {\it a posteriori} that this approximation gives a sound physical picture. Moreover, this approximation scheme is equivalent to applying perturbation theory to two nearby sites and then to extend the resulting term in the effective Hamiltonian to all the chain, as done in~\cite{Carleo,PhysRevA.76.033606,Rosch_PRL08}. Furthermore, in this approximation, non local processes of type (c) do not contribute to the Hamiltonian.

Applying this approximation, we see that the initial state $\ket{\psi_{02}}$ we are considering is connected only to number-operators eigenstates with number eigenvalues which are permutations of the initial one. Therefore, the relevant subspace is generated by all and only the number-operators eigenstates with 2 bosons in half of the sites and 0 in half of the sites.  
Considering, for instance, the state with 2 bosons in the sites $j_1,\,j_2,\,\ldots,\,j_{L/2}$ and 0 in the other sites we can write it as
        \begin{equation}
		\ket{j_1,\,j_2,\,\ldots,\,j_{L/2}}\equiv \frac{\left(\opadag{j_1}\right)^2}{2}\frac{\left(\opadag{j_2}\right)^2}{2}\cdots\frac{\left(\opadag{j_{L/2}}\right)^2}{2}\ket{0}\,.
	\end{equation}
	In this situation we can rewrite the wave function in terms of doublon creation operators. They are defined as $\opbdag{j}\equiv\frac{\left(\opadag{j}\right)^2}{2}$ and, restricting to the subspace generated by $\lbrace\ket{0}$, $\frac{\left(\opadag{j}\right)^2}{2}\ket{0}\rbrace$, $\opbdag{j}$ and $\opb{j}$ obey bosonic commutation relations~\cite{note_doublons}. 
%
%
	Moreover, the conservation of the total boson number puts the constraint that there can be exactly $L/2$ doublons, at most one per site, therefore the doublons behave as hard-core bosonic excitations.
	In terms of $b_j$ and $b_j^\dag$, $H^{(2)}_{eff}$ in the considered subspace can be expressed as
	\begin{equation}
		H^{(2)}_{eff} = \tilde{t} \sum_j (\opbdag{j} \opb{j+1} + {\rm H.~c.}) - \tilde{V}\sum_j \opbdag{j} \opb{j} \opbdag{j+1} \opb{j+1} - \tilde{\mu} \sum_j \opbdag{j} \opb{j}
	\end{equation}
	with $\tilde{t} = (1/2)J^2/U$, $\tilde V = 4 J^2/U$ and $\tilde \mu = -J^2/U$. Applying to this hard-core boson Hamiltonian a unitary transformation to Pauli matrices
\begin{equation}
  \left\{\begin{array}{ll}
     \hat{\sigma}_j^x&=\opb{j}+\opbdag{j}\\
     \hat{\sigma}_j^y&=i(\opb{j}-\opbdag{j})\\
     \hat{\sigma}_j^z&=2\opbdag{j}\opb{j}-1
  \end{array}\right.
\end{equation}
we can write it as a XXZ-model Hamiltonian
\begin{equation} \label{effXXZ:eqn}
  H^{(2)}_{eff} = \sum_{j=1}^L\left[\frac{\tilde{t}}{4}\left(\hat{\sigma}_j^+\hat{\sigma}_{j+1}^-+{\rm H.~c.}\right)-\frac{\tilde{V}}{4}\hat{\sigma}_j^z\hat{\sigma}_{j+1}^z-\frac{\tilde{V}+\tilde{\mu}}{2}\hat{\sigma}_j^z\right]\,.
\end{equation}
\section{Symmetry breaking and Rabi oscillations} \label{breaking}
The imbalance oscillations arise from the effect on the dynamics of a symmetry-breaking doublet. To show the existence of this doublet, let us perform a time-reversal transformation on the Hamiltonian Eq.~\eqref{effXXZ:eqn}. Under this transformation we get $H^{(2)}_{eff}\mapsto -H^{(2)}_{eff}$ and $t\mapsto-t$ . The Hamiltonian $-H^{(2)}_{eff}$ is in a gapped symmetry-broken phase when $\tilde V>2\tilde t$. The symmetry being broken is translation by one site which in the spin representation can be interpreted as the breaking of the $\mathbb{Z}_2$ symmetry (the reflection along the $z$ axis). In our effective model $\tilde V/(2\tilde t)=4$, so we are inside the symmetry-broken phase and we are going to discuss the dynamics of the model inside this phase.

 In order to do that, we start from the thermodynamic limit. The system breaks the $\mathbb{Z}_2$ symmetry, so in this limit there are two degenerate symmetry-breaking ground states. In the $\tilde V /\tilde t\gg1$ limit, they are the staggered $z$-magnetization states $\ket{\uparrow\downarrow\uparrow\downarrow\cdots}$ and $\ket{\downarrow\uparrow\downarrow\uparrow\cdots}$ which correspond in the Bose-Hubbard model to the bosonic states $\ket{020202\cdots}$ and $\ket{202020\cdots}$. In finite-size systems the true eigenstates will be the even and odd superposition of these states which have the form $\ket{\varphi_{\pm}}=\left(\ket{\uparrow\downarrow\uparrow\downarrow\cdots} \pm \ket{\downarrow\uparrow\downarrow\uparrow\cdots}\right)/\sqrt{2}$. These states are called cat states, being superpositions of macroscopically ordered classical states, and the elements of each superposition are related with each other by the $\mathbb{Z}_2$ symmetry. 

The states $\ket{\uparrow\downarrow\uparrow\downarrow\cdots}$ and $\ket{\downarrow\uparrow\downarrow\uparrow\cdots}$ are separated by a gap of order $J^2$ from the rest of states and are degenerate at order 0 in $\tilde t/\tilde V$. Applying degenerate perturbation theory in $\tilde t/\tilde V$ in the subspace generated by these two states, one sees that they are connected at order $L$. Therefore one sees that the two eigenstates are the cat states and that there is a splitting between them $\Delta(L)\sim J^2 (\tilde t/\tilde V)^L$, which is therefore exponentially small in the system size. 

This result is at the roots of the imbalance oscillations. Preparing the system in the imbalanced state $\ket{\psi_{02}}$ is equivalent to prepare the model in the state $\ket{\uparrow\downarrow\uparrow\downarrow\cdots}$. The evolution amounts to Rabi oscillations of the system between $\ket{\uparrow\downarrow\uparrow\downarrow\cdots}$ and $\ket{\downarrow\uparrow\downarrow\uparrow\cdots}$ with period $\Delta(L)\sim J^2\exp(-|\log(\tilde t/\tilde V)|L)$, giving rise to the imbalance oscillation frequency formula Eq.~\eqref{omegaform:eqn}
	
These conclusions are valid for the limit $\tilde V /\tilde t\gg1$ but in our case $\tilde{V}/2\tilde{t}=4$ we have a strictly similar picture. We still have a quasi-degenerate symmetry-breaking doublet at an extremum of the spectrum. These two states are separated from the rest of the spectrum by a gap of order $J^2$ and they are even and odd superposition of macroscopically ordered symmetry-breaking states, separated by a splitting $\Delta(L)\sim J^2\exp(-\alpha_{\rm XXZ}L)$, for some $\alpha_{\rm XXZ}>0$. These two states generate the so-called symmetry-breaking manifold. 

The main difference from the limit $\tilde V /\tilde t\gg1$ is the following. Before, in the limit $\tilde V /\tilde t\gg1$, the state $\ket{\psi_{02}}$ was an element of the symmetry-breaking manifold, so the dynamics was fully described by this two-level system. Now, $\ket{\psi_{02}}$ will have an overlap {\it smaller than one} with the symmetry-breaking manifold and this overlap tends to 0 in the limit $L\to\infty$~\cite{note_MPS}. In this limit the overlap with the rest of the spectrum dominates. Because the only states able to generate a Rabi-oscillation dynamics between two different symmetry sectors are the ones in the symmetry-breaking manifold, we see that the amplitude of the Rabi oscillations goes to 0 in the thermodynamic limit.

\begin{figure}
  \begin{center}
   \begin{tabular}{c}
     \includegraphics[width=8cm]{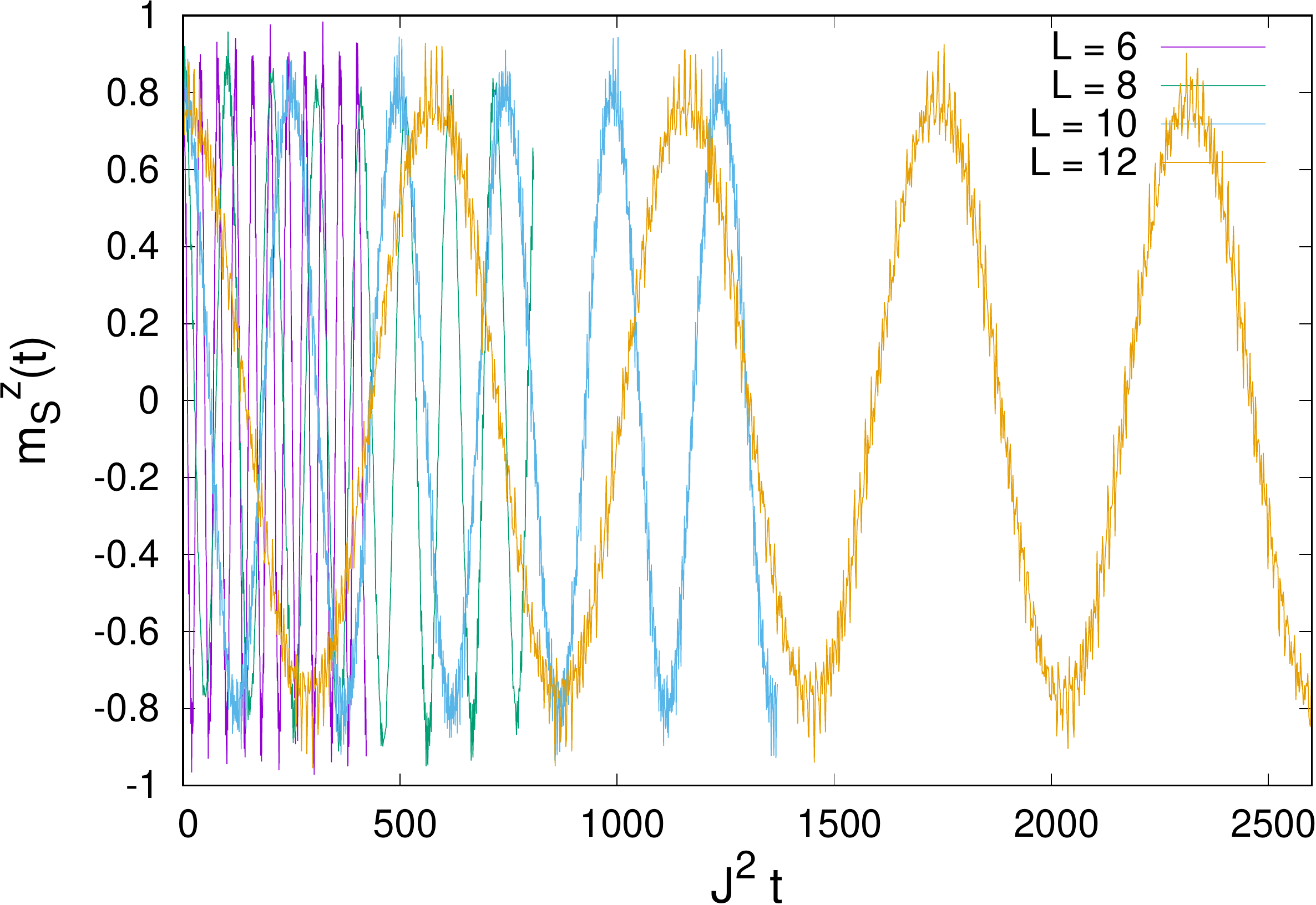}\\
     \includegraphics[width=8cm]{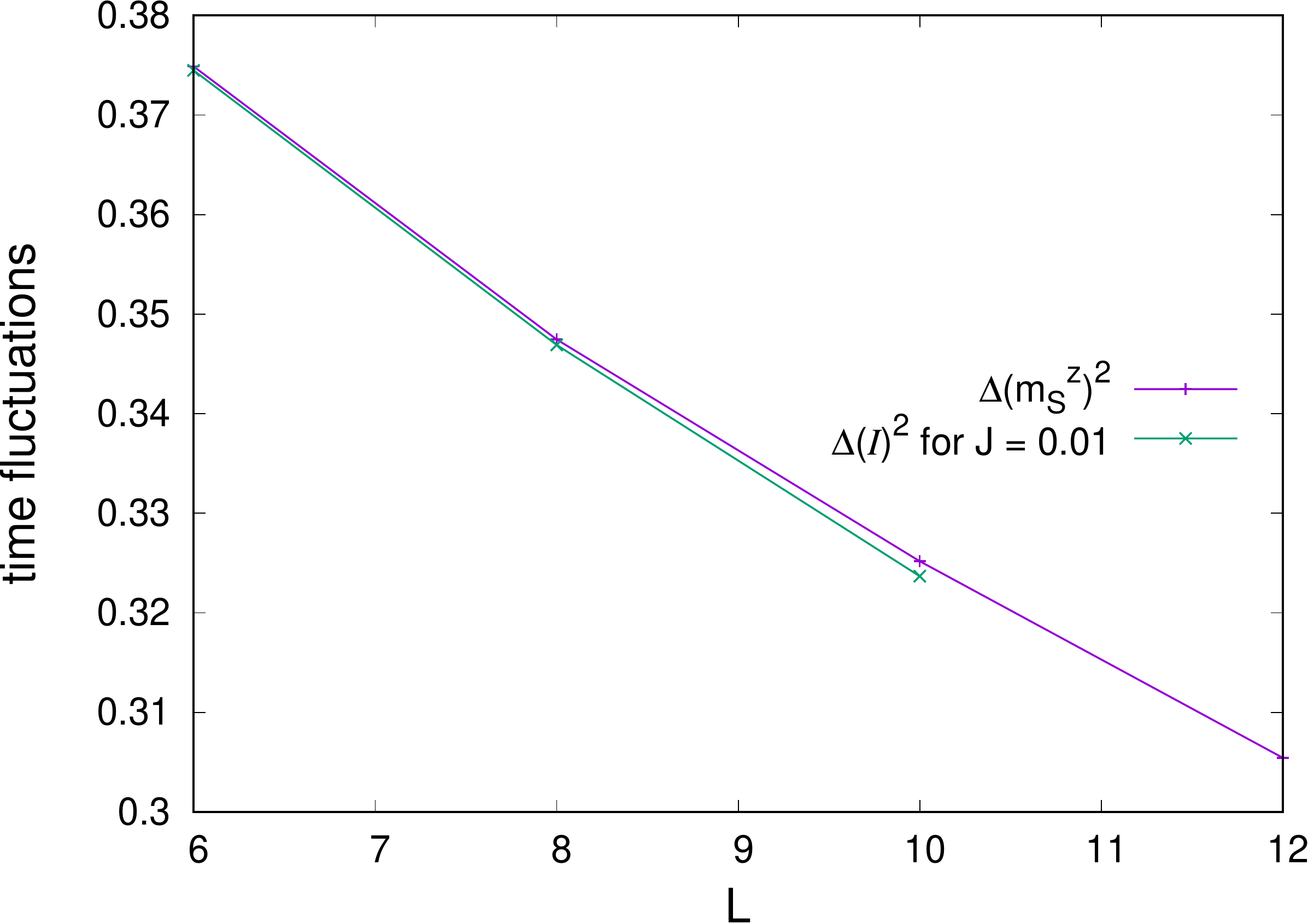}
%
   \end{tabular}
  \end{center}
 \caption{ (Upper panel) Rabi oscillations of the staggered magnetization in the XXZ effective model Eq.~\eqref{effXXZ:eqn}. (Lower panel) size dependence of the time-fluctuations in the Bose-Hubbard and in the XXZ model. (In the XXZ model time fluctuations are evaluated over the same time intervals as the ones shown in the upper panel, while in the Bose-Hubbard model we use Eq.~\eqref{sommalia:eqn}).}
    \label{tony_stag:fig}
\end{figure}

We can clearly see this effect directly in the XXZ effective model. We prepare the system in the state $\ket{\uparrow\downarrow\uparrow\downarrow\cdots}$ (which corresponds to $\ket{\psi_{02}}$ in this representation) and we study the equivalent of the imbalance for that model, that's to say the $z$-staggered magnetization
\begin{equation}
  m_S^z(t)=\frac{1}{L}\sum_{j=1}^L(-1)^{j}\braket{\psi(t)|\hat{\sigma}_j^z|\psi(t)}\,.
\end{equation}
First of all we can qualitatively see the existence of the Rabi oscillations in the time traces of Fig.~\ref{tony_stag:fig} upper panel. These plots are performed for a generic $J\ll U$ because the choice of $J$ is immaterial, being $J^2$ just a global coefficient of Eq.~\eqref{effXXZ:eqn} giving rise to an overall rescaling of the frequencies. So it is easy to see that our model predicts Rabi oscillations with a frequency proportional to $J^2$, as we observed for the Bose-Hubbard Hamiltonian in Fig.~\ref{imbpeak:fig}. Moreover, performing the Fourier transform and finding its main peak as we did above for the imbalance, we find that the Rabi oscillation frequency $\omega_{\rm peak}$ {coincides with the corresponding value for the imbalance oscillations}, as we show in the lower panel of Fig.~\ref{imbpeak:fig}. Therefore, our effective model correctly predicts for the frequency of the imbalance oscillations a dependence given by Eq.~\eqref{omegaform:eqn}, $\omega_{\rm peak}=B_{\rm XXZ}J^2\exp(-\alpha_{\rm XXZ}L)$, and the results of the Bose-Hubbard Hamiltonian obey this rule as soon as $J\ll U$ and we are in the regime where the XXZ effective model is valid. 
Looking at the curves in the lower panel of Fig.~\ref{imbpeak:fig}, we see that this prediction is in perfect agreement with the imbalance results. From the numerical fits in the effective model we get  and $\alpha_{\rm XXZ}=0.441\pm 0.005$, $\log(B_{\rm XXZ})= 0.75\pm 0.05$ which are in agreement with $\alpha=0.43\pm 0.01$, $\log(B)=1.0.7\pm 0.1$ found for the imbalance oscillations at $J=0.025$. 

In getting the effective model Eq.~\eqref{effXXZ:eqn} we neglect many terms. Neglecting these terms we have transformed a non-integrable model, like the Bose-Hubbard one, in an integrable one like the XXZ one. At least for the system sizes we can numerically reach, the terms we neglect  have almost no effect on the oscillations frequency, as we have remarked above.  The only visible effect is that of slightly accelerating the exponential decay of the oscillations frequency, thereby decreasing the splitting in the symmetry-breaking manifold.  We show the way the amplitude of the Rabi oscillations decays in the two cases -- slightly faster in the Bose-Hubbard model -- in the lower panel of Fig.~\ref{tony_stag:fig}. In both cases  we evaluate the oscillation amplitude of the relevant quantity (in one case the imbalance $I$, in the other the staggered magnetization $m_S$) as time fluctuations defined as $\Delta {(\cdots)}^2\equiv\overline{{(\cdots)}^2}-\overline{{(\cdots)}}^2$ {[see Eq.~\eqref{deltaI:eqn}].}

Nevertheless the fact that we neglect these terms does not affect our main conclusion, that the imbalance oscillations decay to 0 in the thermodynamic limit. We know that this fact is true for the XXZ model, because {the overlap of the initial state with the symmetry-breaking manifold vanishes for $L\to\infty$}. Going to the Bose-Hubbard model, we have to put in again the states we neglected in our approximation, so we have {\it more} symmetry preserving states out of the ground state manifold that are unable to support the Rabi oscillations. We expect therefore a faster decay to zero of the Rabi oscillations in the thermodynamic limit, and that is exactly what we observe in the lower panel of Fig.~\ref{tony_stag:fig}.
\subsection{Larger imbalances} \label{larger:sec}
If we prepare the system in an initial state with a larger imbalance
\begin{equation}
 \ket{\psi_{0\,2N}}=\ket{0}\otimes\ket{2N}\otimes\ket{0}\otimes\cdots\otimes\ket{0}\otimes\ket{2N}
\end{equation}
($N\in \mathbb{N}$) we can apply an analysis strictly similar to the one performed above and show that also in this case there are Rabi oscillations as in the $\ket{\psi_{02}}$ case. Applying the perturbation theory at order $2N$ between nearby sites (similarly to what is done at order 2 in~\cite{Carleo,PhysRevA.76.033606,Rosch_PRL08}) we find the effective XXZ model
\begin{equation} \label{effXXZN:eqn}
  H^{(2N)}_{eff} = \sum_{j=1}^L\left[D_{2N}J^{2N}\left(\hat{\sigma}_j^+\hat{\sigma}_{j+1}^-+{\rm H.~c.}\right)-A_{2N}J^2\hat{\sigma}_j^z\hat{\sigma}_{j+1}^z\right]\,,
\end{equation}
where $A_{2N}$ and $D_{2N}$ are positive real numbers. Also here we get a symmetry-breaking two-state manifold separated from the rest of the spectrum by a gap of order $J^2$. Similarly to before, the eigenstates in the symmetry-breaking manifold are near to $\ket{\varphi_\pm}=\frac{1}{\sqrt{2}}\left(\ket{\uparrow\downarrow\uparrow\downarrow\cdots}\pm\ket{\downarrow\uparrow\downarrow\uparrow\cdots}\right)$ which correspond in the bosonic language to $\ket{\varphi_\pm}=\frac{1}{\sqrt{2}}\left(\ket{\psi_{0\,2N}}\pm\ket{\psi_{2N\,0}}\right)$. The splitting between them now is  
$$\Delta_{2N}(L)\sim A_{2N}J^2\nep^{\left[\log(D_{2N}/ A_{2N})-(2N-2)|\log J|\right]L}\,.$$
 So we would see the same Rabi oscillations as before, with the same dependence on $L$ of $\omega_{\rm peak}$ as in Eq.~\eqref{omegaform:eqn}. The most important difference is that now the coefficient $\alpha_{\rm XXZ}(2N)$ is expected to depend on $J$ as $\alpha_{\rm XXZ}(2N)=\log(D_{2N}/ A_{2N})-(2N-2)|\log J|$. 

Moreover, we notice that the splitting $\Delta_{2N}(L)$ is smaller than before for $N$ large enough, because now it is proportional to $J^{2N}$ with $J\ll 1$. This implies that the eigenstates in the symmetry-breaking manifold are nearer than before to $\ket{\varphi_\pm}$ and the decay of the Rabi oscillations is slower.

\begin{thebibliography}{83}%
\makeatletter
\providecommand \@ifxundefined [1]{%
 \@ifx{#1\undefined}
}%
\providecommand \@ifnum [1]{%
 \ifnum #1\expandafter \@firstoftwo
 \else \expandafter \@secondoftwo
 \fi
}%
\providecommand \@ifx [1]{%
 \ifx #1\expandafter \@firstoftwo
 \else \expandafter \@secondoftwo
 \fi
}%
\providecommand \natexlab [1]{#1}%
\providecommand \enquote  [1]{``#1''}%
\providecommand \bibnamefont  [1]{#1}%
\providecommand \bibfnamefont [1]{#1}%
\providecommand \citenamefont [1]{#1}%
\providecommand \href@noop [0]{\@secondoftwo}%
\providecommand \href [0]{\begingroup \@sanitize@url \@href}%
\providecommand \@href[1]{\@@startlink{#1}\@@href}%
\providecommand \@@href[1]{\endgroup#1\@@endlink}%
\providecommand \@sanitize@url [0]{\catcode `\\12\catcode `\$12\catcode
  `\&12\catcode `\#12\catcode `\^12\catcode `\_12\catcode `\%12\relax}%
\providecommand \@@startlink[1]{}%
\providecommand \@@endlink[0]{}%
\providecommand \url  [0]{\begingroup\@sanitize@url \@url }%
\providecommand \@url [1]{\endgroup\@href {#1}{\urlprefix }}%
\providecommand \urlprefix  [0]{URL }%
\providecommand \Eprint [0]{\href }%
\providecommand \doibase [0]{http://dx.doi.org/}%
\providecommand \selectlanguage [0]{\@gobble}%
\providecommand \bibinfo  [0]{\@secondoftwo}%
\providecommand \bibfield  [0]{\@secondoftwo}%
\providecommand \translation [1]{[#1]}%
\providecommand \BibitemOpen [0]{}%
\providecommand \bibitemStop [0]{}%
\providecommand \bibitemNoStop [0]{.\EOS\space}%
\providecommand \EOS [0]{\spacefactor3000\relax}%
\providecommand \BibitemShut  [1]{\csname bibitem#1\endcsname}%
\let\auto@bib@innerbib\@empty
\bibitem [{\citenamefont {Lichtenberg}\ and\ \citenamefont
  {Lieberman}(1992)}]{lichtenberg1983regular}%
  \BibitemOpen
  \bibfield  {author} {\bibinfo {author} {\bibfnamefont {A.}~\bibnamefont
  {Lichtenberg}}\ and\ \bibinfo {author} {\bibfnamefont {M.}~\bibnamefont
  {Lieberman}},\ }\href@noop {} {\textit {\bibinfo {title} {Regular and Chaotic
  Motion}}}\ (\bibinfo  {publisher} {Springer},\ \bibinfo {year}
  {1992})\BibitemShut {NoStop}%
\bibitem [{\citenamefont {Vulpiani}\ \textit {et~al.}(2008)\citenamefont
  {Vulpiani}, \citenamefont {Falcioni},\ and\ \citenamefont
  {Castiglione}}]{Vulpiani}%
  \BibitemOpen
  \bibfield  {author} {\bibinfo {author} {\bibfnamefont {A.}~\bibnamefont
  {Vulpiani}}, \bibinfo {author} {\bibfnamefont {M.}~\bibnamefont {Falcioni}},
  \ and\ \bibinfo {author} {\bibfnamefont {P.}~\bibnamefont {Castiglione}},\
  }\href@noop {} {\textit {\bibinfo {title} {Chaos and Coarse Graining in
  Statistical Mechanics}}}\ (\bibinfo  {publisher} {Cambridge University
  Press},\ \bibinfo {year} {2008})\BibitemShut {NoStop}%
\bibitem [{\citenamefont {Berry}(1978)}]{Berry_regirr78:proceeding}%
  \BibitemOpen
  \bibfield  {author} {\bibinfo {author} {\bibfnamefont {M.~V.}\ \bibnamefont
  {Berry}},\ }in\ \href@noop {} {\textit {\bibinfo {booktitle} {{Topics in
  Nonlinear Mechanics}}}},\ Vol.~\bibinfo {volume} {46},\ \bibinfo {editor}
  {edited by\ \bibinfo {editor} {\bibfnamefont {S.}~\bibnamefont {Jorna}}}\
  (\bibinfo  {publisher} {Am.Inst.Ph.},\ \bibinfo {year} {1978})\ pp.\ \bibinfo
  {pages} {16--120}\BibitemShut {NoStop}%
\bibitem [{\citenamefont {Deutsch}(1991)}]{Deutsch_PRA91}%
  \BibitemOpen
  \bibfield  {author} {\bibinfo {author} {\bibfnamefont {J.~M.}\ \bibnamefont
  {Deutsch}},\ }\href {\doibase 10.1103/PhysRevA.43.2046} {\bibfield  {journal}
  {\bibinfo  {journal} {Phys. Rev. A}\ }\textbf {\bibinfo {volume} {43}},\
  \bibinfo {pages} {2046} (\bibinfo {year} {1991})}\BibitemShut {NoStop}%
\bibitem [{\citenamefont {Srednicki}(1994)}]{Sred_PRE94}%
  \BibitemOpen
  \bibfield  {author} {\bibinfo {author} {\bibfnamefont {M.}~\bibnamefont
  {Srednicki}},\ }\href {\doibase 10.1103/PhysRevE.50.888} {\bibfield
  {journal} {\bibinfo  {journal} {Phys. Rev. E}\ }\textbf {\bibinfo {volume}
  {50}},\ \bibinfo {pages} {888} (\bibinfo {year} {1994})}\BibitemShut
  {NoStop}%
\bibitem [{\citenamefont {Rigol}\ \textit {et~al.}(2008)\citenamefont {Rigol},
  \citenamefont {Dunjko},\ and\ \citenamefont {Olshanii}}]{Rigol_Nat}%
  \BibitemOpen
  \bibfield  {author} {\bibinfo {author} {\bibfnamefont {M.}~\bibnamefont
  {Rigol}}, \bibinfo {author} {\bibfnamefont {V.}~\bibnamefont {Dunjko}}, \
  and\ \bibinfo {author} {\bibfnamefont {M.}~\bibnamefont {Olshanii}},\ }\href
  {\doibase 10.1038/nature06838} {\bibfield  {journal} {\bibinfo  {journal}
  {Nature}\ }\textbf {\bibinfo {volume} {452}},\ \bibinfo {pages} {854}
  (\bibinfo {year} {2008})}\BibitemShut {NoStop}%
\bibitem [{\citenamefont {Prosen}(1999)}]{Prosen_PRE99}%
  \BibitemOpen
  \bibfield  {author} {\bibinfo {author} {\bibfnamefont {T.}~\
  \bibnamefont {Prosen}},\ }\href {\doibase 10.1103/PhysRevE.60.3949}
  {\bibfield  {journal} {\bibinfo  {journal} {Phys. Rev. E}\ }\textbf {\bibinfo
  {volume} {60}},\ \bibinfo {pages} {3949} (\bibinfo {year}
  {1999})}\BibitemShut {NoStop}%
\bibitem [{\citenamefont {Berry}(1977)}]{Berry1_1977}%
  \BibitemOpen
  \bibfield  {author} {\bibinfo {author} {\bibfnamefont {M.~V.}\ \bibnamefont
  {Berry}},\ }\href {\doibase 10.1088/0305-4470/10/12/016} {\bibfield
  {journal} {\bibinfo  {journal} {Journal of Physics A: Mathematical and
  General}\ }\textbf {\bibinfo {volume} {10}},\ \bibinfo {pages} {2083}
  (\bibinfo {year} {1977})}\BibitemShut {NoStop}%
\bibitem [{\citenamefont {Pechukas}(1983)}]{PhysRevLett.51.943}%
  \BibitemOpen
  \bibfield  {author} {\bibinfo {author} {\bibfnamefont {P.}~\bibnamefont
  {Pechukas}},\ }\href {\doibase 10.1103/PhysRevLett.51.943} {\bibfield
  {journal} {\bibinfo  {journal} {Phys. Rev. Lett.}\ }\textbf {\bibinfo
  {volume} {51}},\ \bibinfo {pages} {943} (\bibinfo {year} {1983})}\BibitemShut
  {NoStop}%
\bibitem [{\citenamefont {Feingold}\ and\ \citenamefont
  {Peres}(1986)}]{PhysRevA.34.591}%
  \BibitemOpen
  \bibfield  {author} {\bibinfo {author} {\bibfnamefont {M.}~\bibnamefont
  {Feingold}}\ and\ \bibinfo {author} {\bibfnamefont {A.}~\bibnamefont
  {Peres}},\ }\href {\doibase 10.1103/PhysRevA.34.591} {\bibfield  {journal}
  {\bibinfo  {journal} {Phys. Rev. A}\ }\textbf {\bibinfo {volume} {34}},\
  \bibinfo {pages} {591} (\bibinfo {year} {1986})}\BibitemShut {NoStop}%
\bibitem [{\citenamefont {Prosen}()}]{Prosen_AJ}%
  \BibitemOpen
  \bibfield  {author} {\bibinfo {author} {\bibfnamefont {T.}~\bibnamefont
  {Prosen}},\ }\href@noop {} {\bibfield  {journal} {\bibinfo  {journal} {Annals
  of Physics}\ }\textbf {\bibinfo {volume} {235}},\ \bibinfo {pages}
  {115}}\BibitemShut {NoStop}%
\bibitem [{\citenamefont {Bohigas}\ \textit {et~al.}(1984)\citenamefont
  {Bohigas}, \citenamefont {Giannoni},\ and\ \citenamefont
  {Schmit}}]{PhysRevLett.52.1}%
  \BibitemOpen
  \bibfield  {author} {\bibinfo {author} {\bibfnamefont {O.}~\bibnamefont
  {Bohigas}}, \bibinfo {author} {\bibfnamefont {M.~J.}\ \bibnamefont
  {Giannoni}}, \ and\ \bibinfo {author} {\bibfnamefont {C.}~\bibnamefont
  {Schmit}},\ }\href {\doibase 10.1103/PhysRevLett.52.1} {\bibfield  {journal}
  {\bibinfo  {journal} {Phys. Rev. Lett.}\ }\textbf {\bibinfo {volume} {52}},\
  \bibinfo {pages} {1} (\bibinfo {year} {1984})}\BibitemShut {NoStop}%
\bibitem [{\citenamefont {Eckhardt}\ and\ \citenamefont
  {Main}(1995)}]{PhysRevLett.75.2300}%
  \BibitemOpen
  \bibfield  {author} {\bibinfo {author} {\bibfnamefont {B.}~\bibnamefont
  {Eckhardt}}\ and\ \bibinfo {author} {\bibfnamefont {J.}~\bibnamefont
  {Main}},\ }\href {\doibase 10.1103/PhysRevLett.75.2300} {\bibfield  {journal}
  {\bibinfo  {journal} {Phys. Rev. Lett.}\ }\textbf {\bibinfo {volume} {75}},\
  \bibinfo {pages} {2300} (\bibinfo {year} {1995})}\BibitemShut {NoStop}%
\bibitem [{\citenamefont {Chirikov}(1991)}]{Boris:rotor}%
  \BibitemOpen
  \bibfield  {author} {\bibinfo {author} {\bibfnamefont {B.~V.}\ \bibnamefont
  {Chirikov}},\ }in\ \href@noop {} {\textit {\bibinfo {booktitle} {{Chaos and
  quantum mechanics}}}},\ \bibinfo {series} {Les Houches Lecture Series},
  Vol.~\bibinfo {volume} {52},\ \bibinfo {editor} {edited by\ \bibinfo {editor}
  {\bibnamefont {M.-J.Giannoni}}, \bibinfo {editor} {\bibfnamefont
  {A.}~\bibnamefont {Voros}}, \ and\ \bibinfo {editor} {\bibfnamefont
  {J.}~\bibnamefont {Zinn-Justin}}}\ (\bibinfo  {publisher} {Elsevier Sci.
  Publ., Amsterdam},\ \bibinfo {year} {1991})\ p.\ \bibinfo {pages}
  {443–545}\BibitemShut {NoStop}%
\bibitem [{\citenamefont {Rozenbaum}\ and\ \citenamefont
  {Galitski}(2017)}]{QRKR}%
  \BibitemOpen
  \bibfield  {author} {\bibinfo {author} {\bibfnamefont {E.~B.}\ \bibnamefont
  {Rozenbaum}}\ and\ \bibinfo {author} {\bibfnamefont {V.}~\bibnamefont
  {Galitski}},\ }\href@noop {} {\bibfield  {journal} {\bibinfo  {journal}
  {Physical Review B}\ }\textbf {\bibinfo {volume} {95}}\ \bibinfo {pages}
  {064303} (\bibinfo {year} {2017})}\BibitemShut {NoStop}%
%
\bibitem [{\citenamefont {Rylands}\ \textit {et~al.}(2019)\citenamefont
  {Rylands}, \citenamefont {Rozenbaum}, \citenamefont {Galitski},\ and\
  \citenamefont {Konik}}]{rylands}%
  \BibitemOpen
  \bibfield  {author} {\bibinfo {author} {\bibfnamefont {C.}~\bibnamefont
  {Rylands}}, \bibinfo {author} {\bibfnamefont {E.}~\bibnamefont {Rozenbaum}},
  \bibinfo {author} {\bibfnamefont {V.}~\bibnamefont {Galitski}}, \ and\
  \bibinfo {author} {\bibfnamefont {R.}~\bibnamefont {Konik}},\ }\href@noop {}
  {\bibfield  {journal} {\bibinfo  {journal} {Phys. Rev. Lett.}\ \textbf {\bibinfo {volume} {124}},\ \bibinfo {pages}
  {155302}} (\bibinfo {year} {2020})}\BibitemShut {NoStop}%
%
\bibitem [{\citenamefont {Fava}\ \textit {et~al.}(2020)\citenamefont {Fava},
  \citenamefont {Fazio},\ and\ \citenamefont {Russomanno}}]{Michele_arxiv}%
  \BibitemOpen
  \bibfield  {author} {\bibinfo {author} {\bibfnamefont {M.}~\bibnamefont
  {Fava}}, \bibinfo {author} {\bibfnamefont {R.}~\bibnamefont {Fazio}}, \ and\
  \bibinfo {author} {\bibfnamefont {A.}~\bibnamefont {Russomanno}},\
  }\href@noop {} {\bibfield  {journal} {\bibinfo  {journal} {Phys. Rev. B}\
  }\textbf {\bibinfo {volume} {101}},\ \bibinfo {pages} {064302} (\bibinfo
  {year} {2020})}\BibitemShut {NoStop}%
\bibitem [{\citenamefont {Notarnicola}\ \textit {et~al.}(2018)\citenamefont
  {Notarnicola}, \citenamefont {Iemini}, \citenamefont {Rossini}, \citenamefont
  {Fazio}, \citenamefont {Silva},\ and\ \citenamefont
  {Russomanno}}]{Notarnicola_PRB18}%
  \BibitemOpen
  \bibfield  {author} {\bibinfo {author} {\bibfnamefont {S.}~\bibnamefont
  {Notarnicola}}, \bibinfo {author} {\bibfnamefont {F.}~\bibnamefont {Iemini}},
  \bibinfo {author} {\bibfnamefont {D.}~\bibnamefont {Rossini}}, \bibinfo
  {author} {\bibfnamefont {R.}~\bibnamefont {Fazio}}, \bibinfo {author}
  {\bibfnamefont {A.}~\bibnamefont {Silva}}, \ and\ \bibinfo {author}
  {\bibfnamefont {A.}~\bibnamefont {Russomanno}},\ }\href {\doibase
  10.1103/PhysRevE.97.022202} {\bibfield  {journal} {\bibinfo  {journal} {Phys.
  Rev. E}\ }\textbf {\bibinfo {volume} {97}},\ \bibinfo {pages} {022202}
  (\bibinfo {year} {2018})}\BibitemShut {NoStop}%
\bibitem [{\citenamefont {Abanin}\ \textit
  {et~al.}(2019{\natexlab{a}})\citenamefont {Abanin}, \citenamefont {Altman},
  \citenamefont {Bloch},\ and\ \citenamefont {Serbyn}}]{Bloch_2019}%
  \BibitemOpen
  \bibfield  {author} {\bibinfo {author} {\bibfnamefont {D.~A.}\ \bibnamefont
  {Abanin}}, \bibinfo {author} {\bibfnamefont {E.}~\bibnamefont {Altman}},
  \bibinfo {author} {\bibfnamefont {I.}~\bibnamefont {Bloch}}, \ and\ \bibinfo
  {author} {\bibfnamefont {M.}~\bibnamefont {Serbyn}},\ }\href@noop {}
  {\bibfield  {journal} {\bibinfo  {journal} {Rev. Mod. Phys.}\ }\textbf
  {\bibinfo {volume} {91}},\ \bibinfo {pages} {021001} (\bibinfo {year}
  {2019}{\natexlab{a}})}\BibitemShut {NoStop}%
\bibitem [{\citenamefont {Nandkishore}\ and\ \citenamefont
  {Huse}(2015)}]{nandkishore2015many}%
  \BibitemOpen
  \bibfield  {author} {\bibinfo {author} {\bibfnamefont {R.}~\bibnamefont
  {Nandkishore}}\ and\ \bibinfo {author} {\bibfnamefont {D.~A.}\ \bibnamefont
  {Huse}},\ }\href@noop {} {\bibfield  {journal} {\bibinfo  {journal} {Annu.
  Rev. Condens. Matter Phys.}\ }\textbf {\bibinfo {volume} {6}},\ \bibinfo
  {pages} {15} (\bibinfo {year} {2015})}\BibitemShut {NoStop}%
\bibitem [{\citenamefont {Imbrie}\ \textit {et~al.}(2017)\citenamefont {Imbrie},
  \citenamefont {Ros},\ and\ \citenamefont {Scardicchio}}]{imbrie2017review}%
  \BibitemOpen
  \bibfield  {author} {\bibinfo {author} {\bibfnamefont {J.~Z.}\ \bibnamefont
  {Imbrie}}, \bibinfo {author} {\bibfnamefont {V.}~\bibnamefont {Ros}}, \ and\
  \bibinfo {author} {\bibfnamefont {A.}~\bibnamefont {Scardicchio}},\
  }\href@noop {} {\bibfield  {journal} {\bibinfo  {journal} {Annalen der
  Physik}\ }\textbf {\bibinfo {volume} {529}},\ \bibinfo {pages} {1600278}
  (\bibinfo {year} {2017})}\BibitemShut {NoStop}%
\bibitem [{\citenamefont {Carleo}\ \textit {et~al.}(2012)\citenamefont {Carleo},
  \citenamefont {Becca}, \citenamefont {Schir{\`o}},\ and\ \citenamefont
  {Fabrizio}}]{Carleo}%
  \BibitemOpen
  \bibfield  {author} {\bibinfo {author} {\bibfnamefont {G.}~\bibnamefont
  {Carleo}}, \bibinfo {author} {\bibfnamefont {F.}~\bibnamefont {Becca}},
  \bibinfo {author} {\bibfnamefont {M.}~\bibnamefont {Schir{\`o}}}, \ and\
  \bibinfo {author} {\bibfnamefont {M.}~\bibnamefont {Fabrizio}},\ }\href@noop
  {} {\bibfield  {journal} {\bibinfo  {journal} {Scientific Reports}\ }\textbf
  {\bibinfo {volume} {2}},\ \bibinfo {pages} {243} (\bibinfo {year}
  {2012})}\BibitemShut {NoStop}%
\bibitem [{\citenamefont {Grover}\ and\ \citenamefont
  {Fisher}(2014)}]{grover2014}%
  \BibitemOpen
  \bibfield  {author} {\bibinfo {author} {\bibfnamefont {T.}~\bibnamefont
  {Grover}}\ and\ \bibinfo {author} {\bibfnamefont {M.~P.~A.}\ \bibnamefont
  {Fisher}},\ }\href@noop {} {\bibfield  {journal} {\bibinfo  {journal}
  {Journal of Statistical Mechanics: Theory and Experiment}\ }\textbf {\bibinfo
  {volume} {2014}},\ \bibinfo {pages} {P10010} (\bibinfo {year}
  {2014})}\BibitemShut {NoStop}%
\bibitem [{\citenamefont {Schiulaz}\ \textit {et~al.}(2015)\citenamefont
  {Schiulaz}, \citenamefont {Silva},\ and\ \citenamefont
  {Muller}}]{schiulaz2015}%
  \BibitemOpen
  \bibfield  {author} {\bibinfo {author} {\bibfnamefont {M.}~\bibnamefont
  {Schiulaz}}, \bibinfo {author} {\bibfnamefont {A.}~\bibnamefont {Silva}}, \
  and\ \bibinfo {author} {\bibfnamefont {M.}~\bibnamefont {Muller}},\
  }\href@noop {} {\bibfield  {journal} {\bibinfo  {journal} {Physical Review
  B}\ }\textbf {\bibinfo {volume} {91}},\ \bibinfo {pages} {184202} (\bibinfo
  {year} {2015})}\BibitemShut {NoStop}%
\bibitem [{\citenamefont {Smith}\ \textit
  {et~al.}(2017{\natexlab{a}})\citenamefont {Smith}, \citenamefont {Knolle},
  \citenamefont {Kovrizhin},\ and\ \citenamefont {Moessner}}]{Adam_prl17}%
  \BibitemOpen
  \bibfield  {author} {\bibinfo {author} {\bibfnamefont {A.}~\bibnamefont
  {Smith}}, \bibinfo {author} {\bibfnamefont {J.}~\bibnamefont {Knolle}},
  \bibinfo {author} {\bibfnamefont {D.~L.}\ \bibnamefont {Kovrizhin}}, \ and\
  \bibinfo {author} {\bibfnamefont {R.}~\bibnamefont {Moessner}},\ }\href
  {\doibase 10.1103/PhysRevLett.118.266601} {\bibfield  {journal} {\bibinfo
  {journal} {Phys. Rev. Lett.}\ }\textbf {\bibinfo {volume} {118}},\ \bibinfo
  {pages} {266601} (\bibinfo {year} {2017}{\natexlab{a}})}\BibitemShut
  {NoStop}%
\bibitem [{\citenamefont {Smith}\ \textit
  {et~al.}(2017{\natexlab{b}})\citenamefont {Smith}, \citenamefont {Knolle},
  \citenamefont {Moessner},\ and\ \citenamefont {Kovrizhin}}]{Adam_prl171}%
  \BibitemOpen
  \bibfield  {author} {\bibinfo {author} {\bibfnamefont {A.}~\bibnamefont
  {Smith}}, \bibinfo {author} {\bibfnamefont {J.}~\bibnamefont {Knolle}},
  \bibinfo {author} {\bibfnamefont {R.}~\bibnamefont {Moessner}}, \ and\
  \bibinfo {author} {\bibfnamefont {D.~L.}\ \bibnamefont {Kovrizhin}},\ }\href
  {\doibase 10.1103/PhysRevLett.119.176601} {\bibfield  {journal} {\bibinfo
  {journal} {Phys. Rev. Lett.}\ }\textbf {\bibinfo {volume} {119}},\ \bibinfo
  {pages} {176601} (\bibinfo {year} {2017}{\natexlab{b}})}\BibitemShut
  {NoStop}%
\bibitem [{\citenamefont {Brenes}\ \textit {et~al.}(2018)\citenamefont {Brenes},
  \citenamefont {Dalmonte}, \citenamefont {Heyl},\ and\ \citenamefont
  {Scardicchio}}]{PhysRevLett.120.030601}%
  \BibitemOpen
  \bibfield  {author} {\bibinfo {author} {\bibfnamefont {M.}~\bibnamefont
  {Brenes}}, \bibinfo {author} {\bibfnamefont {M.}~\bibnamefont {Dalmonte}},
  \bibinfo {author} {\bibfnamefont {M.}~\bibnamefont {Heyl}}, \ and\ \bibinfo
  {author} {\bibfnamefont {A.}~\bibnamefont {Scardicchio}},\ }\href {\doibase
  10.1103/PhysRevLett.120.030601} {\bibfield  {journal} {\bibinfo  {journal}
  {Phys. Rev. Lett.}\ }\textbf {\bibinfo {volume} {120}},\ \bibinfo {pages}
  {030601} (\bibinfo {year} {2018})}\BibitemShut {NoStop}%
\bibitem [{\citenamefont {Smith}\ \textit {et~al.}(2019)\citenamefont {Smith},
  \citenamefont {Knolle}, \citenamefont {Moessner},\ and\ \citenamefont
  {Kovrizhin}}]{Adam_prl19}%
  \BibitemOpen
  \bibfield  {author} {\bibinfo {author} {\bibfnamefont {A.}~\bibnamefont
  {Smith}}, \bibinfo {author} {\bibfnamefont {J.}~\bibnamefont {Knolle}},
  \bibinfo {author} {\bibfnamefont {R.}~\bibnamefont {Moessner}}, \ and\
  \bibinfo {author} {\bibfnamefont {D.~L.}\ \bibnamefont {Kovrizhin}},\ }\href
  {\doibase 10.1103/PhysRevLett.123.086602} {\bibfield  {journal} {\bibinfo
  {journal} {Phys. Rev. Lett.}\ }\textbf {\bibinfo {volume} {123}},\ \bibinfo
  {pages} {086602} (\bibinfo {year} {2019})}\BibitemShut {NoStop}%
\bibitem [{\citenamefont {Smith}\ \textit {et~al.}(2018)\citenamefont {Smith},
  \citenamefont {Knolle}, \citenamefont {Moessner},\ and\ \citenamefont
  {Kovrizhin}}]{Adam_prb}%
  \BibitemOpen
  \bibfield  {author} {\bibinfo {author} {\bibfnamefont {A.}~\bibnamefont
  {Smith}}, \bibinfo {author} {\bibfnamefont {J.}~\bibnamefont {Knolle}},
  \bibinfo {author} {\bibfnamefont {R.}~\bibnamefont {Moessner}}, \ and\
  \bibinfo {author} {\bibfnamefont {D.~L.}\ \bibnamefont {Kovrizhin}},\ }\href
  {\doibase 10.1103/PhysRevB.97.245137} {\bibfield  {journal} {\bibinfo
  {journal} {Phys. Rev. B}\ }\textbf {\bibinfo {volume} {97}},\ \bibinfo
  {pages} {245137} (\bibinfo {year} {2018})}\BibitemShut {NoStop}%
\bibitem [{\citenamefont {Russomanno}\ \textit {et~al.}(2020)\citenamefont
  {Russomanno}, \citenamefont {Notarnicola}, \citenamefont {Surace},
  \citenamefont {Fazio}, \citenamefont {Dalmonte},\ and\ \citenamefont
  {Heyl}}]{PhysRevResearch.2.012003}%
  \BibitemOpen
  \bibfield  {author} {\bibinfo {author} {\bibfnamefont {A.}~\bibnamefont
  {Russomanno}}, \bibinfo {author} {\bibfnamefont {S.}~\bibnamefont
  {Notarnicola}}, \bibinfo {author} {\bibfnamefont {F.~M.}\ \bibnamefont
  {Surace}}, \bibinfo {author} {\bibfnamefont {R.}~\bibnamefont {Fazio}},
  \bibinfo {author} {\bibfnamefont {M.}~\bibnamefont {Dalmonte}}, \ and\
  \bibinfo {author} {\bibfnamefont {M.}~\bibnamefont {Heyl}},\ }\href {\doibase
  10.1103/PhysRevResearch.2.012003} {\bibfield  {journal} {\bibinfo  {journal}
  {Phys. Rev. Research}\ }\textbf {\bibinfo {volume} {2}},\ \bibinfo {pages}
  {012003} (\bibinfo {year} {2020})}\BibitemShut {NoStop}%
\bibitem [{\citenamefont {Karpov}\ \textit {et~al.}(2020)\citenamefont {Karpov},
  \citenamefont {Verdel}, \citenamefont {Huang}, \citenamefont {Schmitt},\ and\
  \citenamefont {Heyl}}]{karpov2020disorderfree}%
  \BibitemOpen
  \bibfield  {author} {\bibinfo {author} {\bibfnamefont {P.}~\bibnamefont
  {Karpov}}, \bibinfo {author} {\bibfnamefont {R.}~\bibnamefont {Verdel}},
  \bibinfo {author} {\bibfnamefont {Y.~P.}\ \bibnamefont {Huang}}, \bibinfo
  {author} {\bibfnamefont {M.}~\bibnamefont {Schmitt}}, \ and\ \bibinfo
  {author} {\bibfnamefont {M.}~\bibnamefont {Heyl}},\ }\href@noop {} {\enquote
  {\bibinfo {title} {Disorder-free localization in an interacting
  two-dimensional lattice gauge theory},}\ } (\bibinfo {year} {2020}),\ \Eprint
  {http://arxiv.org/abs/2003.04901} {arXiv:2003.04901 [cond-mat.str-el]}
  \BibitemShut {NoStop}%
\bibitem [{\citenamefont {Pino}\ \textit {et~al.}(2016)\citenamefont {Pino},
  \citenamefont {Ioffe},\ and\ \citenamefont {Altshuler}}]{Pino536}%
  \BibitemOpen
  \bibfield  {author} {\bibinfo {author} {\bibfnamefont {M.}~\bibnamefont
  {Pino}}, \bibinfo {author} {\bibfnamefont {L.~B.}\ \bibnamefont {Ioffe}}, \
  and\ \bibinfo {author} {\bibfnamefont {B.~L.}\ \bibnamefont {Altshuler}},\
  }\href {\doibase 10.1073/pnas.1520033113} {\bibfield  {journal} {\bibinfo
  {journal} {Proceedings of the National Academy of Sciences}\ }\textbf
  {\bibinfo {volume} {113}},\ \bibinfo {pages} {536} (\bibinfo {year}
  {2016})},\ \Eprint
  {http://arxiv.org/abs/https://www.pnas.org/content/113/3/536.full.pdf}
  {https://www.pnas.org/content/113/3/536.full.pdf} \BibitemShut {NoStop}%
\bibitem [{\citenamefont {Prosen}(1998)}]{Prosen_PRL98}%
  \BibitemOpen
  \bibfield  {author} {\bibinfo {author} {\bibfnamefont {T.~}\
  \bibnamefont {Prosen}},\ }\href {\doibase 10.1103/PhysRevLett.80.1808}
  {\bibfield  {journal} {\bibinfo  {journal} {Phys. Rev. Lett.}\ }\textbf
  {\bibinfo {volume} {80}},\ \bibinfo {pages} {1808} (\bibinfo {year}
  {1998})}\BibitemShut {NoStop}%
\bibitem [{\citenamefont {Fisher}\ \textit {et~al.}(1989)\citenamefont {Fisher},
  \citenamefont {Weichman}, \citenamefont {Grinstein},\ and\ \citenamefont
  {Fisher}}]{fisher}%
  \BibitemOpen
  \bibfield  {author} {\bibinfo {author} {\bibfnamefont {M.~P.~A.}\
  \bibnamefont {Fisher}}, \bibinfo {author} {\bibfnamefont {P.~B.}\
  \bibnamefont {Weichman}}, \bibinfo {author} {\bibfnamefont {G.}~\bibnamefont
  {Grinstein}}, \ and\ \bibinfo {author} {\bibfnamefont {D.~S.}\ \bibnamefont
  {Fisher}},\ }\href {\doibase 10.1103/PhysRevB.40.546} {\bibfield  {journal}
  {\bibinfo  {journal} {Phys. Rev. B}\ }\textbf {\bibinfo {volume} {40}},\
  \bibinfo {pages} {546} (\bibinfo {year} {1989})}\BibitemShut {NoStop}%
%
\bibitem [{\citenamefont {Bloch}\ \textit {et~al.}(2008)\citenamefont {Bloch},
  \citenamefont {Dalibard},\ and\ \citenamefont {Zwerger}}]{reviewBloch}%
  \BibitemOpen
  \bibfield  {author} {\bibinfo {author} {\bibfnamefont {I.}~\bibnamefont
  {Bloch}}, \bibinfo {author} {\bibfnamefont {J.}~\bibnamefont {Dalibard}}, \
  and\ \bibinfo {author} {\bibfnamefont {W.}~\bibnamefont {Zwerger}},\
  }\href@noop {} {\bibfield  {journal} {\bibinfo  {journal} {Review Modern
  Physics}\ }\textbf {\bibinfo {volume} {80}},\ \bibinfo {pages} {885}
  (\bibinfo {year} {2008})}\BibitemShut {NoStop}%
%
\bibitem [{\citenamefont {Sierant}\ \textit
  {et~al.}(2017{\natexlab{a}})\citenamefont {Sierant}, \citenamefont
  {Delande},\ and\ \citenamefont {Zakrzewski}}]{Delande_APP17}%
  \BibitemOpen
  \bibfield  {author} {\bibinfo {author} {\bibfnamefont {P.}~\bibnamefont
  {Sierant}}, \bibinfo {author} {\bibfnamefont {D.}~\bibnamefont {Delande}}, \
  and\ \bibinfo {author} {\bibfnamefont {J.}~\bibnamefont {Zakrzewski}},\
  }\href {\doibase 10.12693/APhysPolA.132.1707} {\bibfield  {journal} {\bibinfo
   {journal} {Acta Physica Polonica A}\ }\textbf {\bibinfo {volume} {132}},\
  \bibinfo {pages} {1707} (\bibinfo {year} {2017}{\natexlab{a}})}\BibitemShut
  {NoStop}%
\bibitem [{\citenamefont {Sierant}\ \textit
  {et~al.}(2017{\natexlab{b}})\citenamefont {Sierant}, \citenamefont
  {Delande},\ and\ \citenamefont {Zakrzewski}}]{zak1}%
  \BibitemOpen
  \bibfield  {author} {\bibinfo {author} {\bibfnamefont {P.}~\bibnamefont
  {Sierant}}, \bibinfo {author} {\bibfnamefont {D.}~\bibnamefont {Delande}}, \
  and\ \bibinfo {author} {\bibfnamefont {J.}~\bibnamefont {Zakrzewski}},\
  }\href {\doibase 10.1103/PhysRevA.95.021601} {\bibfield  {journal} {\bibinfo
  {journal} {Phys. Rev. A}\ }\textbf {\bibinfo {volume} {95}},\ \bibinfo
  {pages} {021601} (\bibinfo {year} {2017}{\natexlab{b}})}\BibitemShut
  {NoStop}%
\bibitem [{\citenamefont {Sierant}\ and\ \citenamefont
  {Zakrzewski}(2018)}]{zak2}%
  \BibitemOpen
  \bibfield  {author} {\bibinfo {author} {\bibfnamefont {P.}~\bibnamefont
  {Sierant}}\ and\ \bibinfo {author} {\bibfnamefont {J.}~\bibnamefont
  {Zakrzewski}},\ }\href {\doibase 10.1088/1367-2630/aabb17} {\bibfield
  {journal} {\bibinfo  {journal} {New Journal of Physics}\ }\textbf {\bibinfo
  {volume} {20}},\ \bibinfo {pages} {043032} (\bibinfo {year}
  {2018})}\BibitemShut {NoStop}%
\bibitem [{\citenamefont {Lukin}\ \textit {et~al.}(2019)\citenamefont {Lukin},
  \citenamefont {Rispoli}, \citenamefont {Schittko}, \citenamefont {Tai},
  \citenamefont {Kaufman}, \citenamefont {Choi}, \citenamefont {Khemani},
  \citenamefont {L{\'e}onard},\ and\ \citenamefont {Greiner}}]{Lukin256}%
  \BibitemOpen
  \bibfield  {author} {\bibinfo {author} {\bibfnamefont {A.}~\bibnamefont
  {Lukin}}, \bibinfo {author} {\bibfnamefont {M.}~\bibnamefont {Rispoli}},
  \bibinfo {author} {\bibfnamefont {R.}~\bibnamefont {Schittko}}, \bibinfo
  {author} {\bibfnamefont {M.~E.}\ \bibnamefont {Tai}}, \bibinfo {author}
  {\bibfnamefont {A.~M.}\ \bibnamefont {Kaufman}}, \bibinfo {author}
  {\bibfnamefont {S.}~\bibnamefont {Choi}}, \bibinfo {author} {\bibfnamefont
  {V.}~\bibnamefont {Khemani}}, \bibinfo {author} {\bibfnamefont
  {J.}~\bibnamefont {L{\'e}onard}}, \ and\ \bibinfo {author} {\bibfnamefont
  {M.}~\bibnamefont {Greiner}},\ }\href {\doibase 10.1126/science.aau0818}
  {\bibfield  {journal} {\bibinfo  {journal} {Science}\ }\textbf {\bibinfo
  {volume} {364}},\ \bibinfo {pages} {256} (\bibinfo {year} {2019})},\ \Eprint
  {http://arxiv.org/abs/https://science.sciencemag.org/content/364/6437/256.full.pdf}
  {https://science.sciencemag.org/content/364/6437/256.full.pdf} \BibitemShut
  {NoStop}%
\bibitem [{\citenamefont {Mbeng}(2015)}]{Glen}%
  \BibitemOpen
  \bibfield  {author} {\bibinfo {author} {\bibfnamefont {G.~B.}\ \bibnamefont
  {Mbeng}},\ }\href@noop {} {\enquote {\bibinfo {title} {Localizzazione a molti
  corpi in una catena di bosoni fortemente interagenti},}\ } (\bibinfo {year}
  {2015}),\ \bibinfo {note} {master Thesis, Universit{\`a} di Pisa, available
  at https://etd.adm.unipi.it/t/etd-09252015-163432/}\BibitemShut {NoStop}%
\bibitem [{\citenamefont {Hopjan}\ and\ \citenamefont
  {Heidrich-Meisner}(2019)}]{hopjan2019manybody}%
  \BibitemOpen
  \bibfield  {author} {\bibinfo {author} {\bibfnamefont {M.}~\bibnamefont
  {Hopjan}}\ and\ \bibinfo {author} {\bibfnamefont {F.}~\bibnamefont
  {Heidrich-Meisner}},\ }\href@noop {} {\bibfield  {journal} {\bibinfo  {journal} {Phys. Rev. A}\ }
  \textbf {\bibinfo {volume} {101}},\ \bibinfo {pages} {063617}
  (\bibinfo {year} {2020})}\BibitemShut {NoStop}%
\bibitem [{\citenamefont {Yao}\ and\ \citenamefont
  {Zakrzewski}(2020)}]{yao2020manybody}%
  \BibitemOpen
  \bibfield  {author} {\bibinfo {author} {\bibfnamefont {R.}~\bibnamefont
  {Yao}}\ and\ \bibinfo {author} {\bibfnamefont {J.}~\bibnamefont
  {Zakrzewski}},\ }\href@noop {} {} {\bibfield  {journal} {\bibinfo  {journal} {Phys. Rev. B}\ }
  \textbf {\bibinfo {volume} {102}},\ \bibinfo {pages} {014310}
  (\bibinfo {year} {2020})}\BibitemShut {NoStop}%
\bibitem [{\citenamefont {Kollath}\ \textit {et~al.}(2007)\citenamefont
  {Kollath}, \citenamefont {L\"auchli},\ and\ \citenamefont
  {Altman}}]{kollath}%
  \BibitemOpen
  \bibfield  {author} {\bibinfo {author} {\bibfnamefont {C.}~\bibnamefont
  {Kollath}}, \bibinfo {author} {\bibfnamefont {A.~M.}\ \bibnamefont
  {L\"auchli}}, \ and\ \bibinfo {author} {\bibfnamefont {E.}~\bibnamefont
  {Altman}},\ }\href {\doibase 10.1103/PhysRevLett.98.180601} {\bibfield
  {journal} {\bibinfo  {journal} {Phys. Rev. Lett.}\ }\textbf {\bibinfo
  {volume} {98}},\ \bibinfo {pages} {180601} (\bibinfo {year}
  {2007})}\BibitemShut {NoStop}%
\bibitem [{\citenamefont {Kollath}\ \textit {et~al.}(2010)\citenamefont
  {Kollath}, \citenamefont {Roux}, \citenamefont {Biroli},\ and\ \citenamefont
  {Läuchli}}]{kollath1}%
  \BibitemOpen
  \bibfield  {author} {\bibinfo {author} {\bibfnamefont {C.}~\bibnamefont
  {Kollath}}, \bibinfo {author} {\bibfnamefont {G.}~\bibnamefont {Roux}},
  \bibinfo {author} {\bibfnamefont {G.}~\bibnamefont {Biroli}}, \ and\ \bibinfo
  {author} {\bibfnamefont {A.~M.}\ \bibnamefont {Läuchli}},\ }\href {\doibase
  10.1088/1742-5468/2010/08/p08011} {\bibfield  {journal} {\bibinfo  {journal}
  {Journal of Statistical Mechanics: Theory and Experiment}\ }\textbf {\bibinfo
  {volume} {2010}},\ \bibinfo {pages} {P08011} (\bibinfo {year}
  {2010})}\BibitemShut {NoStop}%
\bibitem [{\citenamefont {Sorg}\ \textit {et~al.}(2014)\citenamefont {Sorg},
  \citenamefont {Vidmar}, \citenamefont {Pollet},\ and\ \citenamefont
  {Heidrich-Meisner}}]{PhysRevA.90.033606}%
  \BibitemOpen
  \bibfield  {author} {\bibinfo {author} {\bibfnamefont {S.}~\bibnamefont
  {Sorg}}, \bibinfo {author} {\bibfnamefont {L.}~\bibnamefont {Vidmar}},
  \bibinfo {author} {\bibfnamefont {L.}~\bibnamefont {Pollet}}, \ and\ \bibinfo
  {author} {\bibfnamefont {F.}~\bibnamefont {Heidrich-Meisner}},\ }\href
  {\doibase 10.1103/PhysRevA.90.033606} {\bibfield  {journal} {\bibinfo
  {journal} {Phys. Rev. A}\ }\textbf {\bibinfo {volume} {90}},\ \bibinfo
  {pages} {033606} (\bibinfo {year} {2014})}\BibitemShut {NoStop}%
\bibitem [{not({\natexlab{a}})}]{note_cruz}%
  \BibitemOpen
  \href@noop {} {}  \bibinfo {note} {the quantum chaotic
  behaviour for $J= 2U$ has been confirmed in~\cite{cruz2020quantum} by means
  of the dynamics of the survival probability.}\BibitemShut {Stop}%
\bibitem [{\citenamefont {Kim}\ \textit {et~al.}(2014)\citenamefont {Kim},
  \citenamefont {Ikeda},\ and\ \citenamefont {Huse}}]{ikeda}%
  \BibitemOpen
  \bibfield  {author} {\bibinfo {author} {\bibfnamefont {H.}~\bibnamefont
  {Kim}}, \bibinfo {author} {\bibfnamefont {T.~N.}~\bibnamefont {Ikeda}}, \ and\
  \bibinfo {author} {\bibfnamefont {D.}~\bibnamefont {Huse}},\ }\href@noop {}
  {\bibfield  {journal} {\bibinfo  {journal} {Phys. Rev. E}\ }\textbf {\bibinfo
  {volume} {90}},\ \bibinfo {pages} {052105} (\bibinfo {year}
  {2014})}\BibitemShut {NoStop}%
\bibitem [{\citenamefont {Santos}\ and\ \citenamefont
  {Rigol}(2010)}]{santos2010localization}%
  \BibitemOpen
  \bibfield  {author} {\bibinfo {author} {\bibfnamefont {L.~F.}\ \bibnamefont
  {Santos}}\ and\ \bibinfo {author} {\bibfnamefont {M.}~\bibnamefont {Rigol}},\
  }\href@noop {} {\bibfield  {journal} {\bibinfo  {journal} {Physical Review
  E}\ }\textbf {\bibinfo {volume} {82}},\ \bibinfo {pages} {031130} (\bibinfo
  {year} {2010})}\BibitemShut {NoStop}%
\bibitem [{\citenamefont {Biroli}\ \textit {et~al.}(2010)\citenamefont {Biroli},
  \citenamefont {Kollath},\ and\ \citenamefont
  {L\"auchli}}]{PhysRevLett.105.250401}%
  \BibitemOpen
  \bibfield  {author} {\bibinfo {author} {\bibfnamefont {G.}~\bibnamefont
  {Biroli}}, \bibinfo {author} {\bibfnamefont {C.}~\bibnamefont {Kollath}}, \
  and\ \bibinfo {author} {\bibfnamefont {A.~M.}\ \bibnamefont {L\"auchli}},\
  }\href {\doibase 10.1103/PhysRevLett.105.250401} {\bibfield  {journal}
  {\bibinfo  {journal} {Phys. Rev. Lett.}\ }\textbf {\bibinfo {volume} {105}},\
  \bibinfo {pages} {250401} (\bibinfo {year} {2010})}\BibitemShut {NoStop}%
\bibitem [{\citenamefont {Pino}\ \textit {et~al.}(2019)\citenamefont {Pino},
  \citenamefont {Tabanera},\ and\ \citenamefont {Serna}}]{pinotto}%
  \BibitemOpen
  \bibfield  {author} {\bibinfo {author} {\bibfnamefont {M.}~\bibnamefont
  {Pino}}, \bibinfo {author} {\bibfnamefont {J.}~\bibnamefont {Tabanera}}, \
  and\ \bibinfo {author} {\bibfnamefont {P.}~\bibnamefont {Serna}},\ }\href
  {\doibase 10.1088/1751-8121/ab4b76} {\bibfield  {journal} {\bibinfo
  {journal} {Journal of Physics A: Mathematical and Theoretical}\ }\textbf
  {\bibinfo {volume} {52}},\ \bibinfo {pages} {475101} (\bibinfo {year}
  {2019})}\BibitemShut {NoStop}%
\bibitem [{\citenamefont {Schreiber}\ \textit {et~al.}(2015)\citenamefont
  {Schreiber}, \citenamefont {Hodgman}, \citenamefont {Bordia}, \citenamefont
  {L{\"u}schen}, \citenamefont {Fischer}, \citenamefont {Vosk}, \citenamefont
  {Altman}, \citenamefont {Schneider},\ and\ \citenamefont
  {Bloch}}]{Schreiber842}%
  \BibitemOpen
  \bibfield  {author} {\bibinfo {author} {\bibfnamefont {M.}~\bibnamefont
  {Schreiber}}, \bibinfo {author} {\bibfnamefont {S.~S.}\ \bibnamefont
  {Hodgman}}, \bibinfo {author} {\bibfnamefont {P.}~\bibnamefont {Bordia}},
  \bibinfo {author} {\bibfnamefont {H.~P.}\ \bibnamefont {L{\"u}schen}},
  \bibinfo {author} {\bibfnamefont {M.~H.}\ \bibnamefont {Fischer}}, \bibinfo
  {author} {\bibfnamefont {R.}~\bibnamefont {Vosk}}, \bibinfo {author}
  {\bibfnamefont {E.}~\bibnamefont {Altman}}, \bibinfo {author} {\bibfnamefont
  {U.}~\bibnamefont {Schneider}}, \ and\ \bibinfo {author} {\bibfnamefont
  {I.}~\bibnamefont {Bloch}},\ }\href {\doibase 10.1126/science.aaa7432}
  {\bibfield  {journal} {\bibinfo  {journal} {Science}\ }\textbf {\bibinfo
  {volume} {349}},\ \bibinfo {pages} {842} (\bibinfo {year} {2015})},\ \Eprint
  {http://arxiv.org/abs/https://science.sciencemag.org/content/349/6250/842.full.pdf}
  {https://science.sciencemag.org/content/349/6250/842.full.pdf} \BibitemShut
  {NoStop}%
\bibitem [{\citenamefont {Sidje}(1998)}]{EXPOKIT}%
  \BibitemOpen
  \bibfield  {author} {\bibinfo {author} {\bibfnamefont {R.~B.}\ \bibnamefont
  {Sidje}},\ }\href@noop {} {\bibfield  {journal} {\bibinfo  {journal} {ACM
  Trans. Math. Softw.}\ }\textbf {\bibinfo {volume} {24}},\ \bibinfo {pages}
  {130} (\bibinfo {year} {1998})}\BibitemShut {NoStop}%
\bibitem [{\citenamefont {Page}(1993)}]{Page_PRL}%
  \BibitemOpen
  \bibfield  {author} {\bibinfo {author} {\bibfnamefont {D.~N.}\ \bibnamefont
  {Page}},\ }\href@noop {} {\bibfield  {journal} {\bibinfo  {journal} {Phys.
  Rev. Lett.}\ }\textbf {\bibinfo {volume} {71}},\ \bibinfo {pages} {1291}
  (\bibinfo {year} {1993})}\BibitemShut {NoStop}%
\bibitem [{Pag()}]{Page_note}%
  \BibitemOpen
  \href@noop {} {}\bibinfo {note} {Strictly speaking, this form of the random
  state is not correct. It does not enforce the energy conservation, being
  random over all the symmetric Hilbert subspace and not only on the
  microcanonical energy shell. Nevertheless, the resulting Page value is always
  the same (as we can see in Fig.~\ref{plots_entro:fig}) and coincides with the
  Page value obtained with a state random in the full Hilbert space. The Page
  value is quite robust and persists even considering highly non-ergodic random
  states~\cite{tomasi2020multifractality}.}\BibitemShut {Stop}%
\bibitem [{\citenamefont {Huang}(2019)}]{Huang_NPB19}%
  \BibitemOpen
  \bibfield  {author} {\bibinfo {author} {\bibfnamefont {Y.}~\bibnamefont
  {Huang}},\ }\href@noop {} {\bibfield  {journal} {\bibinfo  {journal} {Nuclear
  Physics B}\ }\textbf {\bibinfo {volume} {938}},\ \bibinfo {pages} {594}
  (\bibinfo {year} {2019})}\BibitemShut {NoStop}%
\bibitem [{\citenamefont {Luitz}(2016)}]{PhysRevB.93.134201}%
  \BibitemOpen
  \bibfield  {author} {\bibinfo {author} {\bibfnamefont {D.~J.}\ \bibnamefont
  {Luitz}},\ }\href {\doibase 10.1103/PhysRevB.93.134201} {\bibfield  {journal}
  {\bibinfo  {journal} {Phys. Rev. B}\ }\textbf {\bibinfo {volume} {93}},\
  \bibinfo {pages} {134201} (\bibinfo {year} {2016})}\BibitemShut {NoStop}%
\bibitem [{\citenamefont {Haake}(2006)}]{Haake}%
  \BibitemOpen
  \bibfield  {author} {\bibinfo {author} {\bibfnamefont {F.}~\bibnamefont
  {Haake}},\ }\href@noop {} {\textit {\bibinfo {title} {Quantum Signatures of
  Chaos}}}\ (\bibinfo  {publisher} {Springer-Verlag New York, Inc.},\ \bibinfo
  {address} {Secaucus, NJ, USA},\ \bibinfo {year} {2006})\ Chap.~\bibinfo
  {chapter} {7}, pp.\ \bibinfo {pages} {263--274}\BibitemShut {NoStop}%
\bibitem [{\citenamefont {Poilblanc}\ \textit {et~al.}(1993)\citenamefont
  {Poilblanc}, \citenamefont {Ziman}, \citenamefont {Bellisard}, \citenamefont
  {Mila},\ and\ \citenamefont {Montambaux}}]{poilblanc}%
  \BibitemOpen
  \bibfield  {author} {\bibinfo {author} {\bibfnamefont {D.}~\bibnamefont
  {Poilblanc}}, \bibinfo {author} {\bibfnamefont {T.}~\bibnamefont {Ziman}},
  \bibinfo {author} {\bibfnamefont {J.}~\bibnamefont {Bellisard}}, \bibinfo
  {author} {\bibfnamefont {F.}~\bibnamefont {Mila}}, \ and\ \bibinfo {author}
  {\bibfnamefont {G.}~\bibnamefont {Montambaux}},\ }\href@noop {} {\bibfield
  {journal} {\bibinfo  {journal} {Europhys. Lett. 22}\ }\textbf {\bibinfo
  {volume} {537}} (\bibinfo {year} {1993})}\BibitemShut {NoStop}%
\bibitem [{\citenamefont {Berry}(1983)}]{Berry_Les_Houches}%
  \BibitemOpen
  \bibfield  {author} {\bibinfo {author} {\bibfnamefont {M.~V.}\ \bibnamefont
  {Berry}},\ }in\ \href@noop {} {\textit {\bibinfo {booktitle} {Chaotic Behaviour
  of Deterministic Systems}}},\ \bibinfo {series and number} {Les Houches,
  Session XXXVI, 1981},\ \bibinfo {editor} {edited by\ \bibinfo {editor}
  {\bibfnamefont {R.~S.~G.}\ \bibnamefont {Ioos}}, \bibinfo {editor}
  {\bibfnamefont {R.~H.~G.}\ \bibnamefont {Hellemani}}, \ and\ \bibinfo
  {editor} {\bibfnamefont {R.}~\bibnamefont {Stora}}}\ (\bibinfo  {publisher}
  {North-Holland, Amsterdam},\ \bibinfo {year} {1983})\ p.\ \bibinfo {pages}
  {174–271}\BibitemShut {NoStop}%
\bibitem [{\citenamefont {Berry}\ and\ \citenamefont
  {Tabor}(1977)}]{Berry_PRS77}%
  \BibitemOpen
  \bibfield  {author} {\bibinfo {author} {\bibfnamefont {M.~V.}\ \bibnamefont
  {Berry}}\ and\ \bibinfo {author} {\bibfnamefont {M.}~\bibnamefont {Tabor}},\
  }\href@noop {} {\bibfield  {journal} {\bibinfo  {journal} {Proc. Roy. Soc.
  A}\ }\textbf {\bibinfo {volume} {356}},\ \bibinfo {pages} {375} (\bibinfo
  {year} {1977})}\BibitemShut {NoStop}%
\bibitem [{\citenamefont {{ M.\ C.\ Gutzwiller, \textit{Chaos in classical and
  quantum mechanics} (Springer-Verlag, New York, 1990)}}()}]{Gutzwiller90}%
  \BibitemOpen
  \bibfield  {author} {\bibinfo {author} {\bibnamefont {{ M.\ C.\ Gutzwiller,
  \textit{Chaos in classical and quantum mechanics} (Springer-Verlag, New York,
  1990)}}},\ }\href@noop {} {}\BibitemShut {NoStop}%
%
\bibitem [{\citenamefont {Pal}(2010)}]{Palhuse}%
  \BibitemOpen
  \bibfield  {author} {\bibinfo {author} {\bibfnamefont {D.~J.}\ \bibnamefont
  {Pal},\ \bibfnamefont {D.~A.}\ \bibnamefont
  {Huse}}} {\bibfield  {journal}
  {\bibinfo  {journal} {Phys. Rev. B}\ }\textbf {\bibinfo {volume} {82}},\
  \bibinfo {pages} {17441} (\bibinfo {year} {2010})}\BibitemShut {NoStop}%
%
\bibitem [{\citenamefont {Luitz}(2015)}]{Luitz15}%
  \BibitemOpen
  \bibfield  {author} {\bibinfo {author} {\bibfnamefont {D.~J.}\ \bibnamefont
  {Luitz},\ \bibfnamefont {N.~}\ \bibnamefont
  {Laflorencie}\ and\ \bibfnamefont {F.~}\ \bibnamefont
  {Alet}}} {\bibfield  {journal}
  {\bibinfo  {journal} {Phys. Rev. B}\ }\textbf {\bibinfo {volume} {91}},\
  \bibinfo {pages} {081103(R)} (\bibinfo {year} {2015})}\BibitemShut {NoStop}%
%
\bibitem [{\citenamefont {Abanin}\ \textit
  {et~al.}(2019{\natexlab{b}})\citenamefont {Abanin}, \citenamefont
  {Bardarson}, \citenamefont {Tomasi}, \citenamefont {Gopalakrishnan},
  \citenamefont {Khemani}, \citenamefont {Parameswaran}, \citenamefont
  {Pollmann}, \citenamefont {Potter}, \citenamefont {Serbyn},\ and\
  \citenamefont {Vasseur}}]{abanin2019distinguishing}%
  \BibitemOpen
  \bibfield  {author} {\bibinfo {author} {\bibfnamefont {D.~A.}\ \bibnamefont
  {Abanin}}, \bibinfo {author} {\bibfnamefont {J.~H.}\ \bibnamefont
  {Bardarson}}, \bibinfo {author} {\bibfnamefont {G.~D.}\ \bibnamefont
  {Tomasi}}, \bibinfo {author} {\bibfnamefont {S.}~\bibnamefont
  {Gopalakrishnan}}, \bibinfo {author} {\bibfnamefont {V.}~\bibnamefont
  {Khemani}}, \bibinfo {author} {\bibfnamefont {S.~A.}\ \bibnamefont
  {Parameswaran}}, \bibinfo {author} {\bibfnamefont {F.}~\bibnamefont
  {Pollmann}}, \bibinfo {author} {\bibfnamefont {A.~C.}\ \bibnamefont
  {Potter}}, \bibinfo {author} {\bibfnamefont {M.}~\bibnamefont {Serbyn}}, \
  and\ \bibinfo {author} {\bibfnamefont {R.}~\bibnamefont {Vasseur}},\
  }\href@noop {} {\enquote {\bibinfo {title} {Distinguishing localization from
  chaos: challenges in finite-size systems},}\ } (\bibinfo {year}
  {2019}{\natexlab{b}}),\ \Eprint {http://arxiv.org/abs/1911.04501}
  {arXiv:1911.04501 [cond-mat.str-el]} \BibitemShut {NoStop}%
\bibitem [{\citenamefont {L{\"a}uchli}\ and\ \citenamefont
  {Kollath}(2008)}]{lauchli2008spreading}%
  \BibitemOpen
  \bibfield  {author} {\bibinfo {author} {\bibfnamefont {A.~M.}\ \bibnamefont
  {L{\"a}uchli}}\ and\ \bibinfo {author} {\bibfnamefont {C.}~\bibnamefont
  {Kollath}},\ }\href@noop {} {\bibfield  {journal} {\bibinfo  {journal}
  {Journal of Statistical Mechanics: Theory and Experiment}\ }\textbf {\bibinfo
  {volume} {2008}},\ \bibinfo {pages} {P05018} (\bibinfo {year}
  {2008})}\BibitemShut {NoStop}%
\bibitem [{\citenamefont {Edwards}\ and\ \citenamefont
  {Thouless}(1972)}]{Edwards_JPC72}%
  \BibitemOpen
  \bibfield  {author} {\bibinfo {author} {\bibfnamefont {J.~T.}\ \bibnamefont
  {Edwards}}\ and\ \bibinfo {author} {\bibfnamefont {D.~J.}\ \bibnamefont
  {Thouless}},\ }\href@noop {} {\bibfield  {journal} {\bibinfo  {journal} {J.
  Phys. C}\ }\textbf {\bibinfo {volume} {5}},\ \bibinfo {pages} {807} (\bibinfo
  {year} {1972})}\BibitemShut {NoStop}%
\bibitem [{\citenamefont {Wegner}(1980)}]{Wegner}%
  \BibitemOpen
  \bibfield  {author} {\bibinfo {author} {\bibfnamefont {F.}~\bibnamefont
  {Wegner}},\ }\href@noop {} {\bibfield  {journal} {\bibinfo  {journal} {Z.
  Phys. B}\ }\textbf {\bibinfo {volume} {36}},\ \bibinfo {pages} {209}
  (\bibinfo {year} {1980})}\BibitemShut {NoStop}%
\bibitem [{\citenamefont {Ponte}\ \textit {et~al.}(2015)\citenamefont {Ponte},
  \citenamefont {Chandran}, \citenamefont {Papi{\'c}},\ and\ \citenamefont
  {Abanin}}]{ponte2015periodically}%
  \BibitemOpen
  \bibfield  {author} {\bibinfo {author} {\bibfnamefont {P.}~\bibnamefont
  {Ponte}}, \bibinfo {author} {\bibfnamefont {A.}~\bibnamefont {Chandran}},
  \bibinfo {author} {\bibfnamefont {Z.}~\bibnamefont {Papi{\'c}}}, \ and\
  \bibinfo {author} {\bibfnamefont {D.~A.}\ \bibnamefont {Abanin}},\
  }\href@noop {} {\bibfield  {journal} {\bibinfo  {journal} {Annals of
  Physics}\ }\textbf {\bibinfo {volume} {353}},\ \bibinfo {pages} {196}
  (\bibinfo {year} {2015})}\BibitemShut {NoStop}%
\bibitem [{\citenamefont {Serbyn}\ \textit {et~al.}(2017)\citenamefont {Serbyn},
  \citenamefont {Papi\ifmmode~\acute{c}\else \'{c}\fi{}},\ and\ \citenamefont
  {Abanin}}]{PhysRevB.96.104201}%
  \BibitemOpen
  \bibfield  {author} {\bibinfo {author} {\bibfnamefont {M.}~\bibnamefont
  {Serbyn}}, \bibinfo {author} {\bibfnamefont {Z.}~\bibnamefont
  {Papi\ifmmode~\acute{c}\else \'{c}\fi{}}}, \ and\ \bibinfo {author}
  {\bibfnamefont {D.~A.}\ \bibnamefont {Abanin}},\ }\href {\doibase
  10.1103/PhysRevB.96.104201} {\bibfield  {journal} {\bibinfo  {journal} {Phys.
  Rev. B}\ }\textbf {\bibinfo {volume} {96}},\ \bibinfo {pages} {104201}
  (\bibinfo {year} {2017})}\BibitemShut {NoStop}%
\bibitem [{\citenamefont {Rodriguez}\ \textit {et~al.}(2011)\citenamefont
  {Rodriguez}, \citenamefont {Vasquez}, \citenamefont {Slevin},\ and\
  \citenamefont {R\"omer}}]{PhysRevB.84.134209}%
  \BibitemOpen
  \bibfield  {author} {\bibinfo {author} {\bibfnamefont {A.}~\bibnamefont
  {Rodriguez}}, \bibinfo {author} {\bibfnamefont {L.~J.}\ \bibnamefont
  {Vasquez}}, \bibinfo {author} {\bibfnamefont {K.}~\bibnamefont {Slevin}}, \
  and\ \bibinfo {author} {\bibfnamefont {R.~A.}\ \bibnamefont {R\"omer}},\
  }\href {\doibase 10.1103/PhysRevB.84.134209} {\bibfield  {journal} {\bibinfo
  {journal} {Phys. Rev. B}\ }\textbf {\bibinfo {volume} {84}},\ \bibinfo
  {pages} {134209} (\bibinfo {year} {2011})}\BibitemShut {NoStop}%
\bibitem [{\citenamefont {Kravtsov}\ and\ \citenamefont
  {Muttalib}(1997)}]{PhysRevLett.79.1913}%
  \BibitemOpen
  \bibfield  {author} {\bibinfo {author} {\bibfnamefont {V.~E.}\ \bibnamefont
  {Kravtsov}}\ and\ \bibinfo {author} {\bibfnamefont {K.~A.}\ \bibnamefont
  {Muttalib}},\ }\href {\doibase 10.1103/PhysRevLett.79.1913} {\bibfield
  {journal} {\bibinfo  {journal} {Phys. Rev. Lett.}\ }\textbf {\bibinfo
  {volume} {79}},\ \bibinfo {pages} {1913} (\bibinfo {year}
  {1997})}\BibitemShut {NoStop}%
%
\bibitem [{\citenamefont {Alet}(2020)}]{Alet1}%
  \BibitemOpen
  \bibfield  {author} {\bibinfo {author} {\bibfnamefont {N.~}\ \bibnamefont
  {Mac{\'e}}},\ \bibinfo {author} {\bibfnamefont {F.~}\ \bibnamefont
  {Alet}}\ and\ \bibinfo {author} {\bibfnamefont {N.}\ \bibnamefont
  {Laflorencie}},\ } {\bibfield
  {journal} {\bibinfo  {journal} {Phys. Rev. Lett.}\ }\textbf {\bibinfo
  {volume} {123}},\ \bibinfo {pages} {180601} (\bibinfo {year}
  {2019})}\BibitemShut {NoStop}%
%
\bibitem [{\citenamefont {Notarnicola}\ \textit {et~al.}(2020)\citenamefont
  {Notarnicola}, \citenamefont {Silva}, \citenamefont {Fazio},\ and\
  \citenamefont {Russomanno}}]{Notarnicola_2020}%
  \BibitemOpen
  \bibfield  {author} {\bibinfo {author} {\bibfnamefont {S.}~\bibnamefont
  {Notarnicola}}, \bibinfo {author} {\bibfnamefont {A.}~\bibnamefont {Silva}},
  \bibinfo {author} {\bibfnamefont {R.}~\bibnamefont {Fazio}}, \ and\ \bibinfo
  {author} {\bibfnamefont {A.}~\bibnamefont {Russomanno}},\ }\href {\doibase
  10.1088/1742-5468/ab6de4} {\bibfield  {journal} {\bibinfo  {journal} {Journal
  of Statistical Mechanics: Theory and Experiment}\ }\textbf {\bibinfo {volume}
  {2020}},\ \bibinfo {pages} {024008} (\bibinfo {year} {2020})}\BibitemShut
  {NoStop}%
\bibitem [{\citenamefont {de~la Cruz}\ \textit {et~al.}(2020)\citenamefont {de~la
  Cruz}, \citenamefont {Lerma-Hernandez},\ and\ \citenamefont
  {Hirsch}}]{cruz2020quantum}%
  \BibitemOpen
  \bibfield  {author} {\bibinfo {author} {\bibfnamefont {J.}~\bibnamefont
  {de~la Cruz}}, \bibinfo {author} {\bibfnamefont {S.}~\bibnamefont
  {Lerma-Hernandez}}, \ and\ \bibinfo {author} {\bibfnamefont {J.~G.}\
  \bibnamefont {Hirsch}},\ }\href@noop {} {\bibfield  {journal} {\bibinfo  {journal}
  {Phys. Rev. E}\ }\textbf {\bibinfo {volume} {102}},\
  \bibinfo {pages} {032208} (\bibinfo {year} {2020})}\BibitemShut {NoStop}%
\bibitem [{not({\natexlab{b}})}]{nota_freq}%
  \BibitemOpen
  \href@noop {} {} \bibinfo {note} {$\Lambda(\omega)$ has a
  huge peak at vanishing frequency due to the non-vanishing time average of
  $\Lambda(t)$}\BibitemShut {NoStop}%
\bibitem [{not({\natexlab{c}})}]{notev}%
  \BibitemOpen
  \href@noop {} {}  \bibinfo {note} {{Decay of the
  N\'eel order in the disordered symmetric phase has been considered in the
  model Eq.~\eqref{effXXZ1:eqn}
  in~\cite{PhysRevLett.102.130603}.}}\BibitemShut {Stop}%
\bibitem [{\citenamefont {Fazio}\ and\ \citenamefont {van~der
  Zant}(2001)}]{FAZIO2001235}%
  \BibitemOpen
  \bibfield  {author} {\bibinfo {author} {\bibfnamefont {R.}~\bibnamefont
  {Fazio}}\ and\ \bibinfo {author} {\bibfnamefont {H.}~\bibnamefont {van~der
  Zant}},\ }\href {\doibase https://doi.org/10.1016/S0370-1573(01)00022-9}
  {\bibfield  {journal} {\bibinfo  {journal} {Physics Reports}\ }\textbf
  {\bibinfo {volume} {355}},\ \bibinfo {pages} {235 } (\bibinfo {year}
  {2001})}\BibitemShut {NoStop}%
\bibitem [{\citenamefont {Gogolin}\ and\ \citenamefont
  {Eisert}(2016)}]{gogolin2016equilibration}%
  \BibitemOpen
  \bibfield  {author} {\bibinfo {author} {\bibfnamefont {C.}~\bibnamefont
  {Gogolin}}\ and\ \bibinfo {author} {\bibfnamefont {J.}~\bibnamefont
  {Eisert}},\ }\href@noop {} {\bibfield  {journal} {\bibinfo  {journal}
  {Reports on Progress in Physics}\ }\textbf {\bibinfo {volume} {79}},\
  \bibinfo {pages} {056001} (\bibinfo {year} {2016})}\BibitemShut {NoStop}%
\bibitem [{\citenamefont {Essler}\ and\ \citenamefont
  {Fagotti}(2016)}]{essler2016quench}%
  \BibitemOpen
  \bibfield  {author} {\bibinfo {author} {\bibfnamefont {F.~H.}\ \bibnamefont
  {Essler}}\ and\ \bibinfo {author} {\bibfnamefont {M.}~\bibnamefont
  {Fagotti}},\ }\href@noop {} {\bibfield  {journal} {\bibinfo  {journal}
  {Journal of Statistical Mechanics: Theory and Experiment}\ }\textbf {\bibinfo
  {volume} {2016}},\ \bibinfo {pages} {064002} (\bibinfo {year}
  {2016})}\BibitemShut {NoStop}%
\bibitem [{\citenamefont {Essler}\ \textit {et~al.}(2005)\citenamefont {Essler},
  \citenamefont {Frahm}, \citenamefont {G{\"o}hmann}, \citenamefont
  {Kl{\"u}mper},\ and\ \citenamefont {Korepin}}]{essler2005one}%
  \BibitemOpen
  \bibfield  {author} {\bibinfo {author} {\bibfnamefont {F.~H.}\ \bibnamefont
  {Essler}}, \bibinfo {author} {\bibfnamefont {H.}~\bibnamefont {Frahm}},
  \bibinfo {author} {\bibfnamefont {F.}~\bibnamefont {G{\"o}hmann}}, \bibinfo
  {author} {\bibfnamefont {A.}~\bibnamefont {Kl{\"u}mper}}, \ and\ \bibinfo
  {author} {\bibfnamefont {V.~E.}\ \bibnamefont {Korepin}},\ }\href@noop {}
  {\textit {\bibinfo {title} {The one-dimensional Hubbard model}}}\ (\bibinfo
  {publisher} {Cambridge University Press},\ \bibinfo {year}
  {2005})\BibitemShut {NoStop}%
\bibitem [{\citenamefont {Petrosyan}\ \textit {et~al.}(2007)\citenamefont
  {Petrosyan}, \citenamefont {Schmidt}, \citenamefont {Anglin},\ and\
  \citenamefont {Fleischhauer}}]{PhysRevA.76.033606}%
  \BibitemOpen
  \bibfield  {author} {\bibinfo {author} {\bibfnamefont {D.}~\bibnamefont
  {Petrosyan}}, \bibinfo {author} {\bibfnamefont {B.}~\bibnamefont {Schmidt}},
  \bibinfo {author} {\bibfnamefont {J.~R.}\ \bibnamefont {Anglin}}, \ and\
  \bibinfo {author} {\bibfnamefont {M.}~\bibnamefont {Fleischhauer}},\ }\href
  {\doibase 10.1103/PhysRevA.76.033606} {\bibfield  {journal} {\bibinfo
  {journal} {Phys. Rev. A}\ }\textbf {\bibinfo {volume} {76}},\ \bibinfo
  {pages} {033606} (\bibinfo {year} {2007})}\BibitemShut {NoStop}%
\bibitem [{\citenamefont {Rosch}\ \textit {et~al.}(2008)\citenamefont {Rosch},
  \citenamefont {Rasch}, \citenamefont {Binz},\ and\ \citenamefont
  {Vojta}}]{Rosch_PRL08}%
  \BibitemOpen
  \bibfield  {author} {\bibinfo {author} {\bibfnamefont {A.}~\bibnamefont
  {Rosch}}, \bibinfo {author} {\bibfnamefont {D.}~\bibnamefont {Rasch}},
  \bibinfo {author} {\bibfnamefont {B.}~\bibnamefont {Binz}}, \ and\ \bibinfo
  {author} {\bibfnamefont {M.}~\bibnamefont {Vojta}},\ }\href {\doibase
  10.1103/PhysRevLett.101.265301} {\bibfield  {journal} {\bibinfo  {journal}
  {Phys. Rev. Lett.}\ }\textbf {\bibinfo {volume} {101}},\ \bibinfo {pages}
  {265301} (\bibinfo {year} {2008})}\BibitemShut {NoStop}%
\bibitem [{not({\natexlab{d}})}]{note_doublons}%
  \BibitemOpen
  \href@noop {} {} \bibinfo {note} {From the definition of
  $\opbdag{j}$ one finds
  $[\opb{j},\opbdag{j}]=\frac{1}{2}(\opa{j}\opadag{j}+\opadag{j}\opa{j})$ whose
  expectation on every state of the form
  $\ket{\psi}=\alpha\ket{0}+\beta\frac{\opadag{j}}{2}\ket{0}$ is
  1.}\BibitemShut {Stop}%
\bibitem [{not({\natexlab{e}})}]{note_MPS}%
  \BibitemOpen
  \href@noop {} {}  \bibinfo {note} {It is not difficult to
  show that the overlap of the state $\ket{\psi_{02}}$ and any state of the
  symmetry-breaking manifold vanishes for $L\to\infty$. Indeed, both of them
  can be seen as ground states of some non-critical Hamiltonian and then both
  of them can be written as matrix-product states. The overlap of two
  matrix-product states in a system with $L$ sites can be written as the trace
  of some transfer matrix to the power $L$. Because the transfer matrix has
  eigenvalues smaller than 1 for the overlap of two different states, the
  overlap vanishes for $L\to\infty$.}\BibitemShut {Stop}%
\bibitem [{\citenamefont {Tomasi}\ and\ \citenamefont
  {Khaymovich}(2020)}]{tomasi2020multifractality}%
  \BibitemOpen
  \bibfield  {author} {\bibinfo {author} {\bibfnamefont {G.~D.}\ \bibnamefont
  {Tomasi}}\ and\ \bibinfo {author} {\bibfnamefont {I.~M.}\ \bibnamefont
  {Khaymovich}},\ }\href@noop {} {\enquote {\bibinfo {title} {Multifractality
  meets entanglement: relation for non-ergodic extended states},}\ } (\bibinfo
  {year} {2020}),\ \Eprint {http://arxiv.org/abs/2001.03173} {arXiv:2001.03173
  [cond-mat.dis-nn]} \BibitemShut {NoStop}%
\bibitem [{\citenamefont {Barmettler}\ \textit {et~al.}(2009)\citenamefont
  {Barmettler}, \citenamefont {Punk}, \citenamefont {Gritsev}, \citenamefont
  {Demler},\ and\ \citenamefont {Altman}}]{PhysRevLett.102.130603}%
  \BibitemOpen
  \bibfield  {author} {\bibinfo {author} {\bibfnamefont {P.}~\bibnamefont
  {Barmettler}}, \bibinfo {author} {\bibfnamefont {M.}~\bibnamefont {Punk}},
  \bibinfo {author} {\bibfnamefont {V.}~\bibnamefont {Gritsev}}, \bibinfo
  {author} {\bibfnamefont {E.}~\bibnamefont {Demler}}, \ and\ \bibinfo {author}
  {\bibfnamefont {E.}~\bibnamefont {Altman}},\ }\href {\doibase
  10.1103/PhysRevLett.102.130603} {\bibfield  {journal} {\bibinfo  {journal}
  {Phys. Rev. Lett.}\ }\textbf {\bibinfo {volume} {102}},\ \bibinfo {pages}
  {130603} (\bibinfo {year} {2009})}\BibitemShut {NoStop}%
\end{thebibliography}
%

\end{document}